## И. З. ШКУРЧЕНКО

## МЕХАНИКА ЖИДКОСТИ И ГАЗА, ИЛИ МЕХАНИКА БЕЗЫНЕРТНОЙ МАССЫ I.

Монография «Механика жидкости и газа, или механика безынертной массы» адресована специалистам в области теоретической и практической гидродинамики и смежных наук.

Она была написана в 1971 году инженером-конструктором Игорем Захаровичем Шкурченко (Российская Федерация). К сожалению, автор умер в 2003 г. Монографию подготовила к печати дочь автора, которая является редактором всех текстов.

И. З. Шкурченко создал четыре работы по механике жидкости и газа. В 2007 году редактор разместил тексты всех работ И.З. Шкурченко в этом архиве.

Редактор допустил ошибки. Он нашёл их в первой версии редакции. Редактор исправил и дополнил тексты в соответствии с оригиналами и просит читателя использовать только вторую версию (v2) (2022 г.).

Как следует из названия, в монографии «Механика жидкости и газа, или механика безынертной массы» представлен новый взгляд на механическое движение жидкостей и газов, объясняющий явления, связанные с движением вещества в сплошных средах.

Вторая одноименная часть является завершением монографии «Механика жидкости и газа, или механика безынертной массы» и также размещена редактором в этом архиве. Она учитывает свойства реальных жидкостей и газов и дает примеры практического применения теории.

Созданная теория механики жидкости и газа устранит расхождения с практикой и тем самым позволит руководствоваться теорией на практике во всем диапазоне практического использования жидкостей и газов. Это и есть ответ на вопрос, что дает данная работа.

Такое положение позволит не только усовершенствовать любую технику, имеющую дело с жидкостями и газами (в том числе и многострадальные ветряки), но теория будет незаменима для метеорологов, океанологов и многих других специалистов. Автор считал, что теория упругости и пластичности материалов также будет пересмотрена с точки зрения механики безынертной массы.

## I. Z. SHKURCHENKO

## MECHANICS OF LIQUID AND GAS, OR MECHANICS OF THE INERTLESS MASS I.

The monograph "Mechanics of liquid and gas, or mechanics of inertless mass" is addressed to specialists in the field of theoretical and practical hydrodynamics and related sciences. It was written in 1971 by design engineer Igor Zakharovich Shkurchenko (Russian Federation). Unfortunately, the author died in 2003. The monograph was prepared for publication by the author's daughter, who is the editor of all texts. I. Z. Shkurchenko created four works on fluid and gas mechanics. In 2007, the editor posted the texts of all the works of I.Z. Shkurchenko in this archive. The editor made mistakes. He found them in the first version of the edition. The editor has corrected and supplemented the texts in accordance with the originals and asks the reader to use only the second version (v2) of all texts (2022). As the title implies, the monograph " Mechanics of liquid and gas, or mechanics of inertless mass " presents a new look at the mechanical movement of liquids and gases, explaining the phenomena associated with the movement of matter in continuous media. The second part of the same name is the completion of the monograph "Mechanics of liquid and gas, or mechanics of inertless mass" is also placed by the editor in this archive. It takes into account the properties of real liquids and gases and gives examples of the practical application of the theory. The created theory of fluid and gas mechanics will eliminate discrepancies with practice and thus allow the theory to be guided in practice throughout the entire range of practical use of liquids and gases. This is the answer to the question of what this work gives. This situation will not only improve any technique dealing with liquids and gases (including long-suffering windmills), but the theory will also be indispensable for meteorologists, oceanologists and many other specialists. The author believed that the theory of elasticity and plasticity of materials would also be revised from the point of view of the mechanics of inertless mass.

## ПРЕДИСЛОВИЕ РЕДАКТОРА

### *Об авторе*

Шкурченко Игорь Захарович (1935 – 2003) родился и жил в городе Воронеже (РФ), где закончил два вуза по специальностям «промышленное и гражданское строительство» и «двигатели летательных аппаратов». К моменту написания «Механики» работал инженером-конструктором в КБ по проектированию ракетных двигателей. Начало самостоятельному исследованию механического движения среды положило выяснение причин низкочастотных колебаний в двигателях. Судьба этой работы неизвестна, т.к. она была фактически потеряна в какой-то инстанции.

Результаты исследования, в конечном итоге, приобрели общий характер и были изложены автором в виде трёх заявок на предполагаемое открытие *(см. список литературы [2],[3],[4])* . После длительной переписки Комитет по делам изобретений и открытий заявки не принял. Поскольку автор выявил такое свойство массы среды, как безынертность, он понял, что результаты нужно систематизировать, построить на их основе полноценную теорию взаимодействия массы среды с механическими силами, поэтому продолжил исследования. К 1971 году он сформулировал полученные результаты исследования в форме теории механики безинертной массы. Рукопись была озаглавлена «Механика жидкости и газа, или механика безинертной массы».

Автор обращался в разные инстанции, но добиться рассмотрения теории по существу так и не смог. Слово ему не дали. Дело ограничивалось отписками, например:

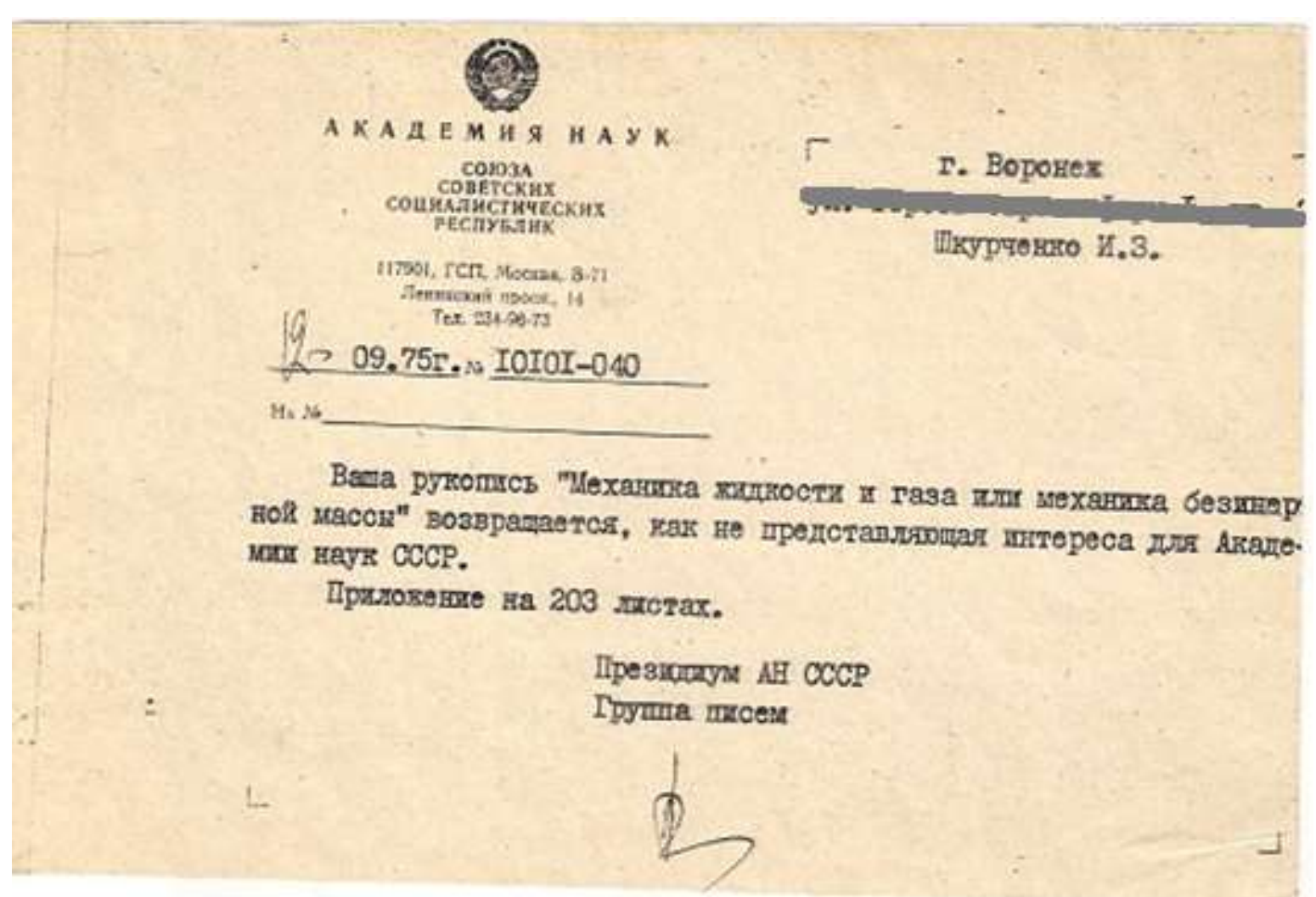

АКАДЕМИЯ НАУК
СОЮЗА СОВЕТСКИХ СОЦИАЛИСТИЧЕСКИХ РЕСПУБЛИК

117901, ГСП, Москва, В-71
Ленинский просп., 14
Тел. 234-96-73

19.09.75г. № IOIOI-040

На №

г. Воронеж

Шкурченко И.З.

Ваша рукопись "Механика жидкости и газа или механика безинерной массы" возвращается, как не представляющая интереса для Академии наук СССР.

Приложение на 203 листах.

Президиум АН СССР
Группа писем

Вскоре, автор был уволен из КБ по сокращению штатов и потерял возможность проводить исследования на практике. Были и другие неприятности. Несмотря на все эти трудности и ограничения в 70-х годах автором были созданы три работы по прикладному применению теории:
«Строение Солнца и планет солнечной системы с точки зрения механики безинертной массы» (1973 -1974),
«Движение твёрдых тел в жидкостях и газах с точки зрения механики безинертной массы (1974 г
«Энергетическая структура Солнца и планет солнечной системы с точки зрения механики безинертной массы» (1977)
Автор предположил, что небесные тела, включая Солнце, имеющие осевое вращение, имеют его потому, что в их недрах организован определённый поток. Эту гипотезу он рассматривает.

Однако, в конечном итоге, вплоть до кончины автора в 2003 году, все работы по теме механики безинертной массы так и остались «в столе».

### *О названии*

Предметом исследования является механическое движение вещества в том случае, когда его масса не имеет определенной количественной величины, то есть является по определению автора *неизолированной массой* (плотность), что и определяет своеобразие её взаимодействия с механическими силами в качестве массы среды. В результате исследования автор выяснил, что масса вещества среды является безинертной. Это значит, *она не имеет свойства запасать механическую энергию в соответствии с первым законом Ньютона*, как это происходит с изолированной массой, т.е. телами. Энергетическое состояние вещества среды в каждый конкретный момент времени определяется характеристиками силового поля или сил давления, действующими на неё в эти конкретные моменты времени.

*О безынертности более подробно см. сноску №3 (стр. 10).*

В настоящее время механика сплошных сред базируется на законах механики твёрдого тела, или механике Ньютона. Цитирую типичный отрывок из учебного пособия для студентов:

«Применительно к среде в целом и к отдельным ее частям постулируется возможность применения общих законов механики: сохранения массы, импульса, момента импульса, механической и общей термодинамической энергии. Из них выводятся основные дифференциальные уравнения механики жидкости и газа и следующие из них общие интегралы или теоремы, к которым добавляются феноменологические законы: уравнение состояния, закон Ньютона для вязкости, закон Фурье – для теплопроводности, Фика – для диффузии и другие».

Ньютон и все его предшественники исследовали взаимодействие механических сил с физическими телами, т.е. изолированной массой. Масса тел имеет свойство инертности и инерциальности, что значит, величина изменения скорости тела зависит от массы тела и тело сохраняет количество движения, пока на него не воздействует другое тело, т.е. механическая сила. Принято считать, что механическая сила действует в центре масс.

Поскольку масса среды безынертна, то она не может подчиняться законам инертной массы. Поэтому в настоящее время, несмотря на усложнение математического аппарата, разрыв между теорией и практикой не сокращается. И не может сократиться. Любое усовершенствование техники осуществляется опытным путём.

Поскольку автор имел соответствующее образование и работал по своей специальности, он видел расхождение теории и практики. В процессе исследования причин низкочастотных колебаний, он выявил механическое свойство жидкостей и газов, которое он назвал безынертностью. Поэтому он понял, в чём корень расхождения теории и практики. Мало выявить свойство. Чтобы устранить причину расхождения, было необходимо создать теорию механики безынертной массы, что он и сделал, несмотря на то, что сначала эта задача показалась ему невыполнимой.

**Созданная теория устранит расхождения с практикой, и тем самым позволит руководствоваться теорией на практике во всём диапазоне практического применения жидкостей и газов.** Это ответ на вопрос, что даёт эта работа.

Такое положение позволит не только усовершенствовать до возможного предела любую технику, имеющую дело с жидкостями и газами (в том числе, многострадальные ветровые электрогенераторы), но она будет незаменима для метеорологов, океанологов и многих других специалистов. Автор полагал, что теория упругости и пластичности материалов тоже будет пересмотрена с точки зрения механики безынертной массы.

Теория механики безынертной массы очень легко проверяется практикой. Автор приводит восемь примеров практического применения зависимостей механики безынертной массы (см. Часть 2, *Приложение*). Этой главной задаче - совмещения теории и практики - посвящены остальные работы.

Так как в главе VI приводится конструктивная схема прибора для измерения динамических сил давления, то, по мнению редактора, самой простой и быстрой проверкой является изготовление этого простейшего прибора. Если он докажет свою эффективность, тогда будет решён вопрос, стоит ли проверять теорию на соответствие действительности в полной мере.

### *О редакторе*

Редактор является родственником автора, и не является специалистом в области механики сплошных сред. Подготовка рукописи «Механики жидкости и газа, или механики безинертной массы» к печати, была начата редактором летом 2002 года, как оказалось, незадолго до болезни и смерти автора, и далась редактору с большим трудом, поскольку пришлось не только осваивать компьютер, но главное, что получить консультации было уже не у кого. Редактор перепечатал на компьютере машинописные тексты в 2003-2004 годах (рукопись «Энергетическая структура Солнца» была обнаружена в архиве позже – в 2008 году).

В 2007 году данном архиве были размещены тексты:

1. *Механика жидкости и газа, или механика безынертной массы* (в двух частях)
   (по условиям публикации текст был искусственно разделён на две части - I и II)
2. *Строение Солнца и планет солнечной системы с точки зрения механики безинертной массы* (в двух частях)

(по условиям публикации текст был искусственно разделён на две части - I и II)

3. *Движение твёрдых тел в жидкостях и газах с точки зрения механики безынертной массы*

4. *Энергетическая структура Солнца и планет солнечной системы с точки зрения механики безынертной массы* (помещён в 2010 году).

К сожалению, в текстах было много опечаток, в том числе смысловых. Были и другие недостатки, в том числе, связанные с разбивкой текстов на части. Всё это, как теперь понятно редактору, негативно влияло на восприятие текстов. В настоящее время редактор проверил все тексты и устранил замеченные недостатки, а также лучше скомпоновал разбитые на части тексты. В 2007 году «*Примеры применения механики безынертной массы*» были помещены в данном архиве отдельным файлом. Теперь они вошли в текст второй части новой редакции «*Механики жидкости и газа, или механики безынертной массы»*, ч. II.

Правила архива, к сожалению, не позволяют удалить предыдущую версию текста. Таким образом, теперь в архиве размещены две версии всех четырёх текстов: v1 – это старая версия, о существовании которой лучше забыть, и v2 – новая, исправленная и дополненная (2022 год).

(см.: https://arxiv.org/search/physics?searchtype=author&query=Shkurchenko%2C+I )

Так как текст «Механика жидкости и газа, или механика безынертной массы» разбит на две части», то они не могут содержать полного оглавления, которое позволяет судить о структуре работы. Поэтому редактор считает нужным привести одним списком содержание обеих частей, которое одновременно будет являться оглавлением к каждой части.

В конце всех размещённых текстов дан адрес электронной почты редактора, по которому можно обращаться по всем вопросам.

***

## ОГЛАВЛЕНИЕ ЧАСТИ I:

### СОДЕРЖАНИЕ ЧАСТИ II:

## ВВЕДЕНИЕ

Механика есть наука об общих законах механического движения и равновесия определённых форм материи. Она изучает состояние покоя, состояние движения конкретно определённых форм материи и действующие соответственно силы. При этом она не вдаётся в физическую, химическую или какую-либо иную сущность изучаемой формы материи. Это положение определяет содержание общей механики.

В окружающем нас мире каждая форма материи конкретно имеет вещественное содержание. В соответствии с этим содержанием делается определённое разграничение по формам материи. Данная работа, «Механика жидкости и газа», обязана своим названием этому разграничению, т. к. предметом её изучения являются жидкости и газы. В более общем смысле, механика жидкости и газа изучает механическое движение и равновесие массы, которая в нашем случае не имеет инерции. По этой причине данная работа имеет второе название – «Механика безынертной массы». Второе название подчеркивает частное различие между предметом изучения механики жидкости и газа и механики твёрдого тела. В принципе, инертность и безынертность являются общими свойствами массы, проявление которых зависит не только от агрегатного состояния вещества, но и от непосредственных условий его механического движения и равновесия. В зависимости от этого к реальной массе применяются либо законы механики твёрдого тела, либо законы механики жидкости и газа.

В основе механики жидкости и газа лежат присущие только ей законы природы, которые объективны и не зависят от нашего сознания. Поэтому задача исследователя упрощается тем, что ему не надо придумывать этих законов, т. к. они проявляют себя как явления природы.

Законы природы, математические зависимости и разъяснения по различным положениям и условиям составляют содержание полностью разработанных наук, вернее, их теоретической части. Законы являются главной частью любой науки, т. к. они выражают сущность природных явлений, поэтому с их помощью человек может определить и качественную, и количественную стороны изучаемого явления. Математические зависимости и различные разъяснения являются лишь следствием этих законов, т. к. они из них вытекают. Законы проверяются экспериментом, прежде чем станут законами, а широкий круг этих экспериментов, связанных с наблюдением их проявления в зависимости от различных условий, позволяет выяснить, насколько полно они отражают сущность многообразных природных явлений и являются ли они общими. Сама экспериментальная проверка становится возможной благодаря тому, что законы выражают качественную и количественную стороны явления природы. При этом всякая наука имеет определённый диапазон своего изучения, например, для механики жидкости и газа он определяется следующей формулировкой: интервалом, или областью её изучения, является сфера действия законов механического движения и равновесия жидкостей и газов.

Конечным доказательством правильности теории служит только эксперимент, практика, а не математические догмы. Зачастую забывают об этом условии. Тогда в науках о явлениях природы законы природы заменяются математическими догмами, которые ничего общего не имеют с этими законами. Именно такое несчастье постигло современную механику жидкости и газа, в которой некоторые математические зависимости возведены в ранг законов природы, хотя они не включают в себя смыслового значения непосредственно самих законов природы.

Правильным же выражением законов природы является словесное выражение, которое включает в себя сущность явлений природы, выраженную через количественные категории. Эти категории затем позволяют выразить законы природы через математические зависимости. По этой причине логика является главной частью закона природы, а математика является лишь одной из частных форм её проявления. Понятно, что частное нельзя поднимать над общим.

По отношению к понятию полностью разработанной науки все ныне существующие отрасли науки можно разделить на две группы:

а) это науки, которые уже имеют полностью разработанное содержание;

б) и науки, которые не имеют полностью разработанного содержания.

К первой группе наук в настоящее время можно отнести только механику твёрдого тела, или механику Ньютона, ко второй группе, пожалуй, все остальные.

В качестве примера рассмотрим содержание теории механики твёрдого тела, или, как её ещё называют, классической механики как полностью разработанной науки. Основное её содержание составляют три закона Ньютона. На базе этих законов были получены и приведены в систему все необходимые математические зависимости, определяющие состояние покоя, движения и равновесия твёрдых тел. Даже система координат этой механики носит название инерционной на основании первого закона Ньютона. Тем самым подчеркивается её частное назначение. Из этого примера следует, что систему знаний этой механики определяют её законы.

Теперь рассмотрим содержание теории современной механики жидкости и газа с точки зрения понятия полностью разработанной науки. Для этой цели приведем в качестве примера некоторые положения из книги [1], точнее – из Введения к этой книги, т. к. в этом разделе отражено не только историческое развитие этой отрасли науки, но и содержатся положения, которые характеризуют всю систему знаний современной механики жидкости и газа.

*Начало научной аэрогидромеханике было положено в 18 столетии трудами академиков Российской Академии наук Леонардо Эйлера (1707 – 1783) и Даниила Бернулли (1700 – 1783). Эйлером были даны общие уравнения движения жидкостей и газов, указаны некоторые интегралы этих уравнений и сформулирован применительно к жидкому телу закон сохранения массы, установленный в общем виде М. В. Ломоносовым. Эйлер исследовал также многие вопросы сопротивления жидкости и применил результаты исследований к практическим задачам кораблестроения и конструирования гидравлических машин. Бернулли, который впервые ввел термин «гидродинамика», последовательно применил в своём труде, имеющим это название, закон сохранения живой силы и установил соотношение между давлением жидкости и её кинетической энергией, известное в настоящее время под именем уравнения Бернулли. Он исследовал также задачу о давлении струи жидкости на пластину.* [1]

Дальше выпишем ещё несколько положений из этого Введения, относящихся непосредственно к развитию механики жидкости и газа и к её оценке.

*Первым известным в настоящее время исследователем проблемы подъё мной силы был знаменитый художник и ученый Леонардо да Винчи. Он предполагал (а заметки по этому вопросу относятся приблизительно к 1505 году), что причиной силы, поддерживающей птицу в воздухе, являются быстрые удары её крыльев, под действием которых воздух под крыльями уплотняется.<...> Долгое время, вплоть до текущего столетия, в науке господствовала другая теория, основанная на представлениях Ньютона (1642 – 1727) о том, что воздух состоит из отдельных, не связанных между собой частиц, которые, двигаясь в потоке, набегающем на препятствие, ударяются о его переднюю стенку и отдают препятствию своё количество движения.* [1]

К этому положению выпишем ещё следующее:

*Лишь при движении в сильно разреженном газе (например, в атмосфере на большой высоте) или при движении газа с очень большой скоростью (значительно большей скорости звука) можно рассматривать среду как состоящую из отдельных молекул, ударяющихся о препятствие. Тогда ударная теория сопротивления среды становится справедливой и в этом отношении.* [1]

Как видно, автор работы [1] признает в конкретных случаях правильность ударной теории сопротивления среды Ньютона.

*Со времени Эйлера стала развиваться и становилась всё более достоверной иная точка зрения на жидкую и газообразную среду, противоположная ньютоновской. Согласно точке зрения Эйлера, жидкость или газ следует рассматривать как непрерывную, легко деформируемую материю. Струйки подходят к препятствию, но не ударяются в него как отдельные (дискретные) частицы, по ударной теории, а отклоняются от препятствия у его передней стороны, плавно со всех сторон обходя (обтекая) его и смыкаясь на задней стороне.* [1]

К этому же положению:

*Эта точка зрения была ближе к действительности (при не очень больших скоростях полета), нежели ударная теория, но так как в жидкости во времена Эйлера не учитывали силы трения, то гипотеза с непрерывной средой приводила, как уже указывалась, к парадоксальному результату об отсутствии сил сопротивления.* [1]

Как видите, автор работы [1] подтверждает также правильность теории Эйлера в определённых случаях, но относит её к разряду гипотез, т. к. она не учитывает сил сопротивления.

*Одно время показалось, что выход из создавшегося положения намечается работами Гельмгольца и Киркофа относящимися к 1868 и 1869 гг. <...> В этой новой точке зрения содержатся в преобразованном виде элементы каждой из этих двух предыдущих.<...> Однако, являясь шагом вперед в определении сопротивления неудобнообтекаемых тел, она оказалась малопригодной для удобнообтекаемых тел.<...> Лишь во второй половине 19 столетия после открытия Рейнольдсем условия перехода ламинарного движения в турбулентное и создания*

*теории подобия многие исследователи в области гидромеханики поняли, что без теории ставящей задачи эксперименту и обобщающей его результаты не может быть научно поставленного эксперимента.* [1]

Из приведенных положений мы видим, что для жидкостей и газов современная механика признает правильной в одних случаях ударную теорию сопротивления среды Ньютона, а в других случаях – точку зрения Эйлера. Хотя это две абсолютно противоположные точки зрения. В конечном итоге наиболее правильной признается теория подобия, основной смысл которой заключается в том, что она даёт возможность переносить характеристики одного эксперимента на множество подобных ему, тем самым предоставляя возможность практике двигаться вперед самостоятельно, без поддержки науки, а так называемая наука просто пожинает её плоды. В результате складывается такая ситуация, что если, например, требуется какая-либо гидравлическая машина с новыми характеристиками, то теория подобия говорит, что пересчитай её со старой. В итоге мы получим так называемый «новый кафтан со старыми дырами». Если хочешь, чтобы этих «дыр» не было, то зашивай их самостоятельно, если конечно сможешь, т.к. теория тебе здесь не поможет. Поэтому получается, что принципиально новые гидравлические или газовые устройства требуют для своего создания большого числа экспериментов и материальных затрат. В настоящее время так и получается, что теория механики жидкости и газа «плетется» за практикой, то есть выполняет несвойственные для себя функции.

Сложившаяся ситуация говорит о том, что в механике жидкости и газа нет своих, присущих только этой отрасли науки закономерностей, которые выражали бы её сущность. Поэтому эта наука относится к разряду наук, которые не имеют полностью разработанного содержания, то есть которые не имеет своей теории. Задача настоящей работы заключается именно в том, чтобы найти общие законы механики жидкости и газа, получить на их основе её теорию и перевести, таким образом, механику жидкости и газа из группы наук с не полностью разработанным содержанием в группу наук с полностью разработанным содержанием. Отметим, что эта задача вполне выполнима, т. к. основные закономерности механики жидкости и газа уже найдены.* Осталось только на их основе создать систему знаний, которая составляет теорию механики жидкости и газа.

* речь идёт о заявках на предполагаемое открытие, см. список литературы [2], [3], [4].

# Г Л А В А I
# ОСНОВНЫЕ ЗАКОНЫ МЕХАНИКИ ЖИДКОСТИ И ГАЗА

Механика жидкости и газа имеет три основных закона, которые полностью определяют её сущность. В последующем тексте мы воспользуемся их содержанием для разработки теории механики жидкости и газа. Количественные зависимости теории будут определяться только положениями этих законов, а описания и исследования будут делаться на основе визуального наблюдения за жидкостями и газами с учётом практического опыта. Поскольку законы являются фундаментом теории, то любое их неправильное истолкование влечёт к неправильному построению теории.

В механике жидкости и газа практика накопила достаточно большой опыт и к настоящему времени намного опережает теорию. В связи с тем, что соответствующая теория должна была бы появиться века на два раньше, т. к. все предпосылки для этого уже были, то в настоящей работе основное внимание будет уделяться построению теории. Что касается экспериментального материала, то каждый из вас всегда сможет найти его и с его помощью проверить правильность положений теории.

## I.1. ПЕРВЫЙ ЗАКОН: ЗАКОН СОХРАНЕНИЯ СОСТОЯНИЯ [2]

***Жидкости и газы сохраняют энергию покоя и установившегося движения только в силовом поле и изменяют её лишь при изменении этого поля.***

Поясним этот закон и с его помощью получим некоторые практические сведения.

Общепринято, что структуру жидкостей и газов составляют молекулы и атомы, которые представляют собой подвижные друг относительно друга мельчайшие частицы материи. Подвижность этих частиц, как принято считать, объясняется отсутствием касательных напряжений между ними. Поскольку молекулы и атомы очень малы, то даже в незначительном е их содержится великое множество.

В окружающем нас мире жидкости и газы занимают большие объёмы, в которых они либо находятся в покое, либо совершают движение. Силовое воздействие на каждую молекулу любого объёма жидкости или газа осуществляется определённым силовым полем. Для пояснения приведем аналогию действия магнитного поля. Если вы приблизите руку к магниту, то ничего не ощутите. Если вы возьмёте какие-нибудь железные предметы и теперь приблизите руки, то ощутите силу притяжения магнита, с которой каждый предмет притягивается к магниту. Подобные силы подобным образом постоянно действуют на каждую молекулу любого объёма независимо от того, находятся ли в нём жидкости и газы в состоянии покоя или установившегося движения. В любом случае жидкости и газы находятся под действием сил силового поля, которое определяет их состояние.

Под действием сил силового поля жидкости и газы могут совершать работу. Это значит, они обладают определённым количеством энергии. Именно об этом состоянии идет речь в законе о сохранении состояния. Находятся ли жидкости и газы в состоянии покоя или установившегося движения, они имеют определённый уровень энергии, который определяется количественно в каждый момент времени силовым полем. Поэтому для изменения количества энергии жидкости и газа надо изменить силовое поле.

Следовательно, этот закон характеризует состояние жидкостей и газов, которое количественно выражается энергией. Поэтому по изменению энергии можно количественно и качественно определить изменение состояния жидкостей и газов. Теперь состояние жидкостей и газов мы можем записать количественными величинами.

Практически энергия состояния для жидкостей и газов измеряется приборами, которые относятся к манометрическому типу. Ниже мы более подробно остановимся на этом вопросе.

Что касается силовых полей, то они могут быть самыми разнообразными. Изучение и классификация их не входят в компетенцию механики жидкости и газа, которой о силовых полях надо знать только то, что они бывают скалярными и векторными, то есть одни из них имеют направленное действие, а другие не имеют направления действия. Примером скалярного силового поля может служить замкнутый объём, под действием границ которого находится содержащийся в нём газ. Примером векторного силового поля может служить поле земного тяготения. Подобное ограничение в позиции силового поля вызвано тем, что механика жидкости и газа есть наука о движении и равновесии, которые определяются характеристиками массы, а по ним можно определить только то, что является ли силовое поле векторным или скалярным и какова сила его воздействия. В механике жидкости и газа нет необходимости знать больше о силовых полях.

Теперь мы можем определить предмет, изучаемый механикой жидкости и газа. Сделаем это, дав определение идеальной жидкости, которая тем отличается от реальных жидкостей и газов, что не обладает сжимаемостью, вязкостью и т.д.

**Идеальной жидкостью называется совокупность молекул или атомов, обладающих абсолютной подвижностью друг относительно друга и составляющих любой объём.**

Таким образом, изучаемая механикой жидкости и газа форма материи получила своё определение.

Совокупность молекул или атомов составляет массу жидкостей и газов в определённых объёмах. Чтобы перейти к пространственному обобщению, введём понятие **среды**, что есть **пространство, заполненное жидкостью**. Это понятие даёт возможность сделать обобщение в отношении масштабности явлений, происходящих в жидкостях и газах, то есть сделать их общими. В этом случае можно выявить количественную величину массы, как количество массы в единице объёма. Эта единица измерения массы носит название плотности $\rho$.

Для сопоставления выпишем некоторые положения современной механики жидкости и газа по этому вопросу:

1. механика не имеет закона сохранения состояния;
2. согласно её положениям, приборами манометрического типа замеряют только давление;
3. определение идеальной жидкости по своей сути не отличается от определения, данного в механике безынертной массы, но оно имеет дополнительное разъяснение такого характера, что «под жидкой частицей в гидродинамике понимают не отдельную молекулу, а малый, по сравнению с характерными размерами потока или тела, элемент объёма, содержащий много молекул» [5].

Вот это дополнение и вносит коренное различие между двумя определениями идеальной жидкости.

## I.2. ВТОРОЙ ЗАКОН: УРАВНЕНИЕ СИЛ РАСХОДНОГО ВИДА ДВИЖЕНИЯ [3]

***Динамические силы давления $P_{дин.}$ равны произведению расхода массы в единицу времени M на линейную скорость W и делённому на площадь сечения потока F:***

$$P_{дин} = \frac{M}{F} W. \qquad (1)$$

Динамическое давление в этой зависимости измеряется силой, которая приходится на единицу площади. Это давление уравновешивается в плоскости сечения потока $F$ расходом массы[1] $M$ с линейной скоростью $W$.

Второй закон даёт нам количественную зависимость между силой и массой при изменении состояния жидкостями и газами. В механике именно при изменении состояния реализуется сила. Согласно первому закону, каждое состояние жидкости и газа определяется количеством энергии. Поэтому переход из одного состояния в другое характеризуется изменением энергии. Непосредственное изменение состояния в уравнении (1) выражается расходом массы в единицу времени $M$.

Это значит, что расход массы в единицу времени является основной характеристикой движения жидкостей и газов. То есть расход массы характеризует любое движение жидкостей и газов, а не ускорение и не линейная скорость, как принято в механике твёрдого тела.

---

[1] В оригинале рукописи *расход массы* обозначен буквой $M$ с точкой наверху. По техническим причинам редактор обозначает *расход массы* в единицу времени буквой $M$ без точки. В тех случаях, когда в формулах будет фигурировать параметр *массы*, он будет обозначен буквой $m$.

Сноски и выделение текста сделаны редактором.

По этой причине определённый вид движения жидкости и газа, описанный в работе [3], был назван расходным, чтобы подчеркнуть необходимость расхода в качестве фундаментальной характеристики механики безынертной массы. Поэтому нам придётся учиться мыслить по новому о движении жидкостей и газов, которое качественно и количественно выражается расходом массы в единицу времени, а не ускорением и не линейной скоростью. В нашем случае линейная скорость *W* является лишь вспомогательной характеристикой[2].

Согласно уравнению сил расходного вида движения, или второму закону механики жидкости и газа, расход массы определяется в неподвижной плоскости (поверхности) сечения потока *F*. Поэтому равновесие между динамическими силами давления и расходом массы в единицу времени сохраняется только в неподвижной плоскости. Это даёт нам право при исследовании движения и равновесия жидкостей и газов пользоваться неподвижной плоскостью сечения *F* потока.

*В уравнении сил расходного вида движения нет ускорения. Это значит, что масса жидкостей и газов не подчиняется первому закону Ньютона, то есть она не обладает инертностью*[3].

---

[2] Движением вещества среды является расход массы в единицу времени на площади сечения потока, которая есть площадь действия сил давления. Все объяснения – в тексте. Пока примите во внимание, о чём идёт речь. Адекватное восприятие текучести должно сформироваться постепенно.

[3] Практика показала, что после этих двух предложений у читателя полностью пропадает интерес к продолжению чтения. Чтобы такого не произошло, редактор решился своими словами, *в меру своего понимания* пояснить, о чём идёт речь в данной работе, т.е. что такое «безынертность». У кого интерес к чтению не пропал, может не читать эту сноску, т.к. весь текст работы и есть пояснение, что это такое и как всё «работает».

Силовое поле действует во всём объёме среды. И т.к. давление по определению не существует без площади действия, то рассматривать действие распределённых сил можно только на плоскости, которая является площадью сечения потока, что является основой метода исследования. В то же время каждая такая плоскость как бы заполнена слоем структурных единиц среды, которые находятся под действием сил, действующих на этой площади. Вещество среды/потока имеет плотность, т.е. количество массы в единице объёма. Поэтому каждая реальная площадь сечения содержит такое количество структурных единиц, масса которых пропорциональна площади сечения, т.е. площади действия сил давления, что значит, каждая единица площади данного сечения должна содержать одинаковое количество массы, с которой взаимодействуют распределённые по этой площади силы. Это значит, что результат взаимодействия поля механических сил определяется плотностью массы среды, а это, в свою очередь, значит, что взаимодействие сил со структурными единицами среды не зависит от массы этих структурных единиц. Для твёрдых тел существует подобная аналогия: их взаимодействие с гравитационным полем Земли не зависит от их массы, что проявляется в одинаковом ускорении свободного падения. Толщина плоскости сечения неизмерима, поэтому имеет значение только её наличие. Она определяется объёмом точки среды, включая в этот объём сферу действия сил давления.

Результатом действия распределённых сил на распределённую по площади их действия массу является движение структурных единиц среды, называемое расходом массы. Оно выражается в том, что распределённая масса как бы выталкивается из одной плоскости действия сил/сечения в следующую. Скорость перехода определяется действующими в данной плоскости силами. Для проталкивания более плотного вещества среды с той же скоростью, что и менее плотного, потребуется большее давление, но при этом во всех случаях скорость выталкивания, или перехода, не зависит от массы конкретной структурной единицы. Все они приобретают одинаковую *скорость* в среде с одинаковой плотностью вещества.

Действующие силы определяют скорость перехода слоя структурных единиц в следующую плоскость. В следующей плоскости действия сил могут действовать такие же силы, тогда скорость расхода (выталкивания слоя массы) не изменяется. Но могут действовать и другие, отличные и по величине и по направлению. Тогда вытолкнутый ими как бы элементарный расход массы, будет вытолкнут уже в следующую плоскость быстрее или медленнее, а также может изменяться направление расхода. Вот такими «шашками» каждая структурная единица объёма среды продвигается в объёме всего потока на определённое расстояние за единицу времени. Это расстояние зависит от скорости, а скорость определяется характеристиками силового поля. Структурные единицы среды не ускоряются и не замедляются от мгновенного силового воздействия. Они каждое мгновение имеют скорость, соответствующую конкретному силовому воздействию на данной площади сечения.

Эти свойства расхода массы как результата взаимодействия массы среды и механических сил, сформулированы в первом и втором законе механики безынертной массы. Третий закон фактически говорит об основном свойстве поля механических сил, которое заключается в том, что распределённые силы могут организовывать расход массы в двух взаимно перпендикулярных плоскостях. Это свойство определяет четыре вида движения среды. В реальности движение среды может представлять комбинацию этих четырёх видов движения. Таким образом, внешне хаотичное движение различных сред, например, даже атмосферы, может быть систематизировано и упорядочено, а также станет предсказуемым и управляемым. В разделе кинематики исследуются все виды движения, т.е. именно расхода массы, в разделе динамики – силовое воздействие их обусловливающее. В примерах практического применения зависимостей механики безынертной массы (см. *Приложение*) и в других прикладных работах с этой точки зрения автором рассмотрены, например, такие явления как сопротивление, обтекаемость, подъёмная сила.

Таким образом, безынертность структурных единиц массы среды, во-первых, выражается в том, что их скорость не зависит от их массы. Во-вторых, они не могут запасать механическую энергию от предыдущего воздействия сил

Поясним примером. Если заполнить какой-либо резервуар жидкостью, то движение этого резервуара происходит в соответствии с законами механики Ньютона. Если мы создадим движение жидкости внутри резервуара, то это движение будет связано с законами механики безынертной массы. Отсюда следует вывод, что молекулы или атомы жидкостей и газов безынертны. Инертным может быть только сам объём, заполненный безынертными молекулами. Поэтому масса в одних случаях должна рассматриваться как инертная масса, а в других как безынертная. По этой причине «Механика жидкости и газа» имеет второе название – «Механика безынертной массы». Ниже мы ещё раз остановимся на втором законе механики безынертной массы, а теперь подведём итоги:

1. второй закон механики безынертной массы даёт **количественную зависимость между силой и расходом массы в единицу времени;**
2. **расход массы** в единицу времени является **основной характеристикой движения** жидкостей и газов;
3. равновесие динамического давления и расхода массы происходит на неподвижной плоскости (поверхности) сечения потока;
4. молекулы, составляющие массу жидкостей и газов, безынертны, и само движение жидкостей и газов реализуется как движение безынертной массы.

---

давления, т.е. в следующей плоскости действия сил они приобретают скорость и направление, соответствующие действующим силам, а имевшаяся скорость и направление как бы обнуляются, т.е. количество движения не сохраняется после прекращения действия сил. Нельзя забывать, автор исследует взаимодействие структурных единиц массы идеальной жидкости с механическими силами, но в реальности существуют термодинамические взаимодействия, которые автор тоже учитывает, когда речь идёт о реальных, а не идеальных, жидкостях и газах.

Наверное, свойство массы структурных единиц среды - не сохранять движение после окончания воздействия механической силы можно назвать безынерциальностью, поэтому первый закон механики безынертной массы противоположен по смыслу первому закону инерциальной массы. Почему же это коренное различие свойств остаётся незамеченным, свойство же должно проявляться так же, как инерционность, в виде природного явления? Оно и проявляется. Например, при уменьшении площадей сечения потока происходит изменение действующих в каждой площади сечения сил - увеличивается динамическая составляющая сил давления. Поэтому в каждой площади сечения величина скорости выталкивания изменяется на бо́льшую величину. Следовательно, увеличение скорости, с которой происходит непосредственно сам расход массы, т.е. выталкивание слоя вещества среды в следующую плоскость действия сил, и так – одновременно по всему объёму потока, для наблюдателя выглядит как ускорение течения (замедление при увеличении площади сечения) потока. Но это, так сказать, мнимое ускорение, поскольку изменение скорости вызывается абсолютно другой причиной, имеет другую природу, которая исследуется в данной работе. Вот таким образом надо понимать фразу «нет ускорения». Его нет не только в уравнении динамических сил, но и в природе. Однако эта фраза выглядит, как отрицание очевидного факта «ускорения», следовательно, редактор должен был устранить данное недоразумение. Автор рассматривает взаимодействие безынертной массы с механическими силами поэтапно, со всех сторон – с кинематической, динамической и энергетической. И только в результате всестороннего рассмотрения может сложиться понимание механического движения массы среды и собственно, само понятие безынертности, как свойства массы среды. Следовательно, данный труд требует вдумчивого чтения, а не сравнения с имеющимися положениями существующей механики сплошных сред, т.к. они естественно, отличаются. Именно поэтому рецензенты указывали на несовместимость с существующими положениями, и на этом основании считали, что данная работа «основана на грубых ошибках». По аналогии, можно сказать, что люди, усвоившие геоцентрическую теорию, точно так же могли указывать автору гелиоцентрической теории на несовместимость с положениями, вместо того, чтобы «посмотреть в телескоп», чтобы выявить, какая теория соответствует действительности.

По определению среда - это не просто пространство, заполненное жидкостью или газом, но эти жидкости или газы должны находиться под действием скалярного и/или векторного силового поля. Вот эта форма материи – безынертная масса - и есть предмет исследования автора с точки зрения её взаимодействия с механическими силами. Возможно, в космическом пространстве существуют скопления молекул или атомов, которые не находятся под действием механических сил, поэтому к ним нельзя применять данную теорию, но на Земле подобной формы материи не существует. Или, автор говорит, что, например, пары` не подчиняются законам механики безынертной массы (см. «Строение Солнца и планет солнечной системы с точки зрения механики безынертной массы»).

Свойство инертности изолированной массы, как явление природы, было известно задолго до Ньютона. Тогда спрашивается, что открыл Ньютон? Он открыл законы механического движения твёрдых тел и дал описание их движения, обусловленного этими законами, что называется теорией. По аналогии, читатель должен понимать, что автор не только сначала выявил свойство безынертности массы среды, но он открыл законы, которым подчиняется безынертная масса при взаимодействии с механическими силами. Весь данный труд является развёрнутой и целой картиной действия этих законов, о чём автором сказано уже во *Введении*. Поэтому заявки в Комитет по делам открытий и изобретений подавались на предполагаемое открытие. Впоследствии на их основе автор создал данную полноценную теорию механики безынертной массы.

Редактор приносит извинения за это пространное пояснение и рассчитывает, что оно подвигнет читателя на продолжение чтения, когда он на первых страницах столкнётся с явной нелепостью с его точки зрения. Другой же точки зрения нет пока ни у кого.

Сопоставим с этими выводами некоторые положения современной механики жидкости и газа:

1. в современной механике жидкости и газа применяется второй закон Ньютона в такой формулировке: сила равна произведению массы на ускорение;
2. масса жидкостей и газов и их структурных единиц принимается инертной в соответствии с первым законом Ньютона;
3. **за основную характеристику движения принимается линейная скорость**.

Как видно, это существенные различия новых и общепринятых положений. Именно эти различия показывают, что мы имеем дело с ранее неизвестным явлением природы[4].

## I.3. ТРЕТИЙ ЗАКОН: ФОРМАЛЬНЫЙ ПРИНЦИП СВЯЗИ ВИДА ДВИЖЕНИЯ С ФОРМОЙ УРАВНЕНИЙ НЕРАЗРЫВНОСТИ И ДВИЖЕНИЯ [4]

***Отсутствие или наличие в уравнении движения параметров пространства и времени определяет его связь с тем или иным видом движения.***

Смысл этого закона заключается в том, что он даёт возможность по количеству сочетаний параметров пространства и времени определить полное число видов движения жидкостей и газов. Таких сочетаний параметров можно составить четыре:

1. отсутствуют параметры пространства и времени;
2. отсутствует параметр времени, присутствует параметр пространства;
3. отсутствует параметр пространства, присутствует параметр времени;
4. присутствуют параметры пространства и времени.

Это значит, что для жидкостей и газов существуют четыре вида движения.

Третий закон был получен из непосредственного наблюдения за движением жидкостей и газов и опыта современной механики жидкости и газа. Поэтому названия для видов движения взяты из существующих прикладных наук и теоретической гидромеханики. Перечислим эти названия:

**1. установившийся вид движения**
(отсутствуют параметры пространства и времени);

**2. плоский установившийся вид движения**
(отсутствует параметр времени, присутствует параметр пространства);

**3. расходный вид движения**
(отсутствует параметр пространства, присутствует параметр времени);

**4. акустический вид движения**
(присутствуют параметры пространства и времени).

Названия этих четырёх видов движения говорят о том, что они должны быть хорошо известны даже не специалистам в этой области. Например, установившееся движение отражает движение жидкостей и газов в трубопроводе при постоянном расходе массы в единицу времени. Примером плоского установившегося движения служит вращательное движение жидкостей и газов типа водоворотов, смерчей, в турбинах и т.п. Расходное движение описывает, например, момент изменения расхода массы во времени в трубопроводе, е и т.п. Акустический вид движения связан, например, с распространение звука в жидкостях и газах.

Как видим, все четыре вида движения относятся к вполне реальным движениям, которые существуют в природе. Так что каждый из вас может легко представить в своём воображении реальную картину любого вида движения.

Отметим, что расходное движение в существующей литературе носило название переменного движения, для которого тоже имелись некоторые зависимости.

Каждый из этих видов движения может существовать самостоятельно и в сочетании с другими видами движения. При этом каждый из них сохраняет свои особенности.

В дополнение к приведённой выше классификации видов движения сделаем существенное замечание:

**каждый из этих видов движения может существовать самостоятельно и в сочетании с другими видами движения. При этом каждый из них сохраняет свои особенности.**

---

[4] т.е. безынертностью массы

*Примечание:*

1. В связи с тем, что формальный принцип связи был получен на основе существующего опыта, то в его названии упоминаются уравнения неразрывности и движения. В данной работе эти уравнения будут называться уравнениями движения и сил, т. к. такие названия выражают их правильный смысл, и сами уравнения будут иметь другой вид. По этой причине его более правильное название будет звучать как " Формальный принцип связи вида движения с формой уравнений движения и сил", а не [с формой] уравнений неразрывности и движения. Об этом надо помнить.

2. Как было показано выше, расход массы в единицу времени является основной характеристикой движения жидкостей и газов, поэтому все изменения в пространстве и времени относятся, прежде всего, к расходу массы в единицу времени.

Теперь посмотрим, в чём данные положения противоречат положениям существующей теории. В ней движение жидкостей и газов определяется общими уравнениями неразрывности и движения, согласно которым жидкости и газы имеют бесчисленное множество различных видов движения. Так что и в этом вопросе у нас имеется коренное различие с положениями существующей механики жидкости и газа.

## I.4. МЕТОД ИССЛЕДОВАНИЯ

Общие законы механики жидкости и газа являются основой, или фундаментом теории, так как они определяют основные понятия и зависимости исследуемых явлений природы. Но теория, прежде всего, является системой знаний и наша задача заключается именно в том, чтобы получить эту систему, то есть свести все основные определения и зависимости к определённой системе. Только в этом случае мы сможем охватить все проявления исследуемых явлений природы в зависимости от различных условий

Основой, или каркасом, системы в нашем случае является метод исследования. Он даёт возможность рассматривать законы механики жидкости и газа при всех условиях движения и равновесия в определённых рамках соответствующим этим законам. Тем самым условия движения и равновесия приводятся в соответствие с основными законами.

Отметим, что метод исследования является активным началом теории. Так как на этапе начального исследования явлений природы (этот этап науки может быть сколь угодно большим по времени) от выбора метода исследования во многом зависит успех всей дальнейшей работы, то есть поиск основных законов. Только он помогает выявить их при анализе результатов эксперимента, опыта.

Конкретно в приложении к теории механики жидкости и газа метод исследования включает в себя определённый комплекс приёмов записи и изображения условий движения и равновесия.

В условия движения и равновесия входят пространство, время и различные факторы, связанные с ними. В комплекс приёмов записи и изображения входят система единиц, система координат, приёмы изображения движения и равновесия. Для механики жидкости и газа принимаем следующие условия движения и равновесия:

1. пространство считается неподвижным и полностью заполненным жидкостью или газом, которое выше определялось как среда;
2. время считается непрерывным и постоянным в своём движении;
3. различные факторы берутся из конкретных условий движения и равновесия.

Комплекс приёмов записи и изображения движения и равновесия:

1. за систему единиц принимаются общепринятые системы единиц для записи пространства, времени, силы, скорости и т.д.;

2. для изображения условий движения и равновесия принимается плоскость (поверхность), для которой записываются конкретные условия движения и равновесия. В соответствии со вторым законом механики жидкости и газа эта плоскость (поверхность) располагается обязательно перпендикулярно направлению движения. Плоскость принимается неподвижной в пространстве.

3. Условия движения и равновесия рассматриваются в одних плоскостях относительно других. И так - для всего пространства. В связи с этой особенностью для пространственной записи движения и равновесия принимается полярная система координат, или неинерциальная система координат.

Эта система координат записывается относительно конкретных условий движения и равновесия. то есть принимают первоначальную плоскость исследования, например, плоскость $S$ в точке $A$ (рис. 1) и относительно этой плоскости располагают систему полярных координат с полюсом $O$, удалённым на расстоянии $r$ от точки $A$. Затем производят исследование в таком порядке: записывают на этой плоскости условия движения и равновесия. Дальше принимают следующую плоскость $S_1$ в точке $A$ на расстоянии $r_1$ от полюса $O$ и углом поворота $\varphi$. Для этой плоскости записывают условия движения и равновесия. О характере движения судят по изменению условий одной плоскости относительно другой.

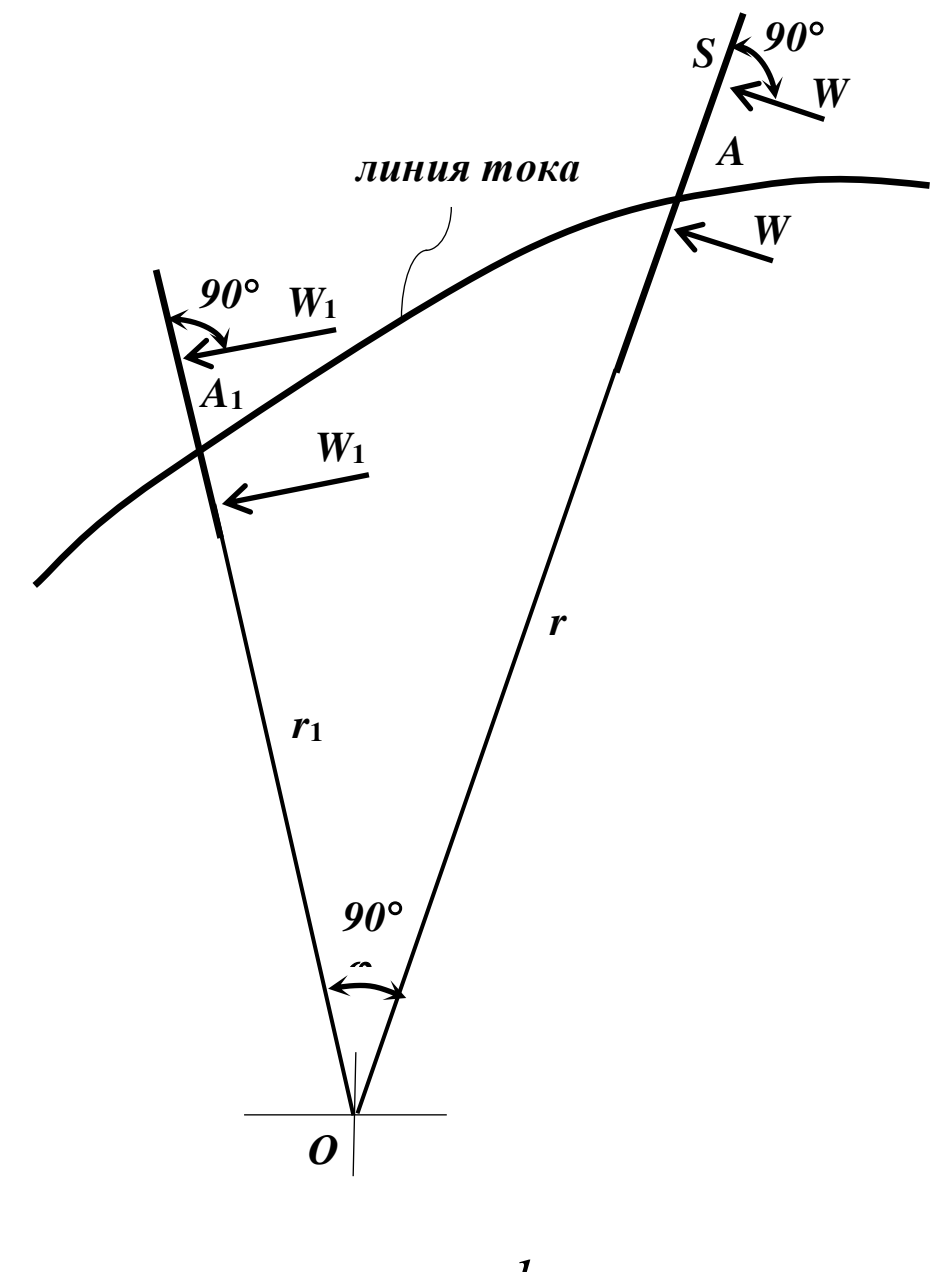

*рис. 1*

Отметим важное следствие, которое вытекает из применения полярной системы координат. Оно заключается в том, что при движении потока жидкости или газа через исследуемые плоскости сечения потока, которые в этом случае расположены на одной прямой, а во втором случае - по окружности с определённым радиусом, характеристики движения остаются постоянными, так как радиус остаётся постоянным как в первом случае, так и во втором. Для прямой линии радиус полярной системы координат равен бесконечности. Это значит, что движение жидкости и газа, линия тока которых является либо прямой линией, либо окружностью, является одним и тем же движением, то есть по отношению к линии тока это два одинаковых движения.

Сопоставим положения современной механики жидкости и газа по этому вопросу с положениями механики безинертной массы:

1. Современная механика жидкости и газа предусматривает два метода исследования: это метод Лагранжа и метод Эйлера. Наиболее близким по отношению к методу исследования механики безынертной массы является метод Эйлера, суть которого заключается в том, что " рассматривается неподвижное пространство, заполненное движущейся жидкостью, и изучается: 1) изменение различных элементов движения в фиксированной точке пространства с течением времени и 2) изменение этих элементов при переходе к другим точкам пространства" [6] .

2. Для обоих методов исследования применяется инерциальная система координат, то есть движение жидкостей и газов рассматривается относительно этой системы.

Следовательно, метод исследования тоже отличается от метода исследования современной механики жидкости и газа.

### I. 5. ИДЕАЛЬНАЯ ЖИДКОСТЬ

Мы уже определили идеальную жидкость как совокупность молекул обладающих абсолютной подвижностью друг относительно друга и составляющих любой объём.

Молекула, с точки зрения механики жидкости и газа, представляет собой минимально возможный для каждой жидкости и газа сгусток массы, который является их структурной единицей. Естественно, что эта структурная единица обладает для каждой жидкости и газа определёнными габаритами, то есть образует определённый объём, который, как утверждают, можно даже замерить. Для идеальной жидкости мы принимаем, что молекулы несжимаемы и не оказывают какого-либо взаимодействия друг на друга, то есть идеальная жидкость представляет собой механическую смесь этих молекул, которые могут заполнить любой объём. Это значит, идеальная жидкость не обладает некоторыми свойствами реальной жидкости, как то – сжимаемостью и вязкостью. Например, такая жидкость, как вода, по своим механическим свойствам близка к свойствам идеальной жидкости.

*Абсолютная подвижность определяется тем, что любая молекула приходит в движение под действием любой силы, как бы ни была мала эта сила, и движется с постоянной скоростью $W$ соответствующей этой силе.* Количественная сторона определяется уравнением сил расходного вида движения. Это значит, что любая молекула, которая является структурной единицей жидкости, не обладает свойством инерции, то есть не подчиняется первому закону Ньютона.

В дальнейшем по отношению к идеальной жидкости будем руководствоваться более общим понятием, таким как среда, которое включает в себя пространство, заполненное идеальной жидкостью. Эта жидкость может находиться либо в состоянии покоя, либо в состоянии движения. Подобное понятие является более универсальным, так как оно включает в себя и пространственное восприятие идеальной жидкости.

## Г Л А В А II. СТАТИКА

Статика включает в себя круг задач, которые изучают равновесие идеальной жидкости в состоянии покоя. В соответствии с законом сохранения состояния жидкость может находиться в состоянии покоя только под действием скалярного или векторного силового поля. Это значит, что жидкость в подобном состоянии находится под действием сил и обладает определённым запасом механической энергии. В разделе статики определим уравнения равновесия для сил скалярного и векторного силового поля. Зависимости для энергии покоя жидкости определим ниже в разделе «Работа и энергия».

Отметим, что состояние покоя для жидкостей можно назвать ещё состоянием застывшего движения. Второе название отражает более правильный смысл, чем первое.

### II. 1. ИССЛЕДОВАНИЕ НЕПОДВИЖНОЙ ТОЧКИ СРЕДЫ

Исследование неподвижной точки среды будет заключаться в том, чтобы найти условия, при которых для неё можно будет применить общий метод исследования. С этой целью изобразим на рисунке 2, *а* среду и возьмём на ней для исследования точку $A$. Согласно общему методу исследования, принятая точка является не объёмной величиной, а плоскостью. То есть в нашем случае точка $A$ есть плоскость, площадь которой равна диаметральной плоскости сечения объёмной точки. Это значит, что точка является бесконечно малой величиной, но всё равно она изображается плоскостью.

Чтобы можно было записать равновесие для этой плоскости, необходимо понять, что в механике жидкости и газа принимают за силу. Понятие силы является одним из основных понятий механики. В настоящее время считают, что сила проявляет себя, когда масса движется с ускорением. Конечно, это верно, но далеко не полно. В механике безынертной массы она проявляется при движении массы с любой скоростью. Например, в поле земного тяготения молекулы жидкости будут двигаться с постоянной скоростью от действия сил этого поля. Ниже мы рассмотрим пример по этому вопросу.

*Сила в механике безынертной массы имеет качественное различие с силой в механике твёрдого тела, то есть она определяется силой отнесённой к площади.* В общем случае она определяется как сила, отнесённая к единице площади. В механике безинертной массы это общее определение имеет название сил давления ($P$) или давления. Практически силы давление, действующие в жидкости, можно замерить только плоскостью, соединённой с фиксатором давления. Плоскость и фиксатор давления в комплексе называются прибором для замера давления. В практике бытует много различных типов подобных приборов.

Отметим что специалисты, пользующиеся этими приборами, не всегда понимают, что они меряют. Так как в настоящее время нет чёткого разграничения между приборами для замера энергии и приборами для замера давления. Здесь надо помнить, что приборы манометрического типа служат для замера энергии, но этими же приборами можно замерять статическое давление, то есть приборы манометрического типа могут иметь две шкалы для одновременной фиксации энергии и статического давления в определённых случаях.

Приборы же для замера давления, то есть приборы, которые воспринимают давление плоскостью, замеряют статическое или динамическое давление в жидкостях.

Вооружившись прибором для замера давления, продолжим исследование неподвижной точки $A$ среды. Для этого представим в своём воображении точку $A$ в увеличенном масштабе, как это показано на рисунке *2,б.*

Согласно этого рисунка, увеличенная площадь $F$ точки $A$ будет иметь форму окружности с радиусом $\Delta r$. Эта плоскость крепится на неподвижном полюсе $O$ и может быть повёрнута относительно него в любом направлении. Площадь $F$ точки $A$ определяется тем, что давление по всей её площади должно быть одинаковым и равным определённой величине $P$.

Дальше совмещаем плоскость прибора для замера давления с плоскостью $F$ точки $A$, Затем начинаем поворачивать совмещённые площади относительно полюса $O$ и в каждом фиксированном положении 1, 2, 3 и т.д., как показано на *рис. 2,в*, производим замеры давления. Мы убедимся, что в любом положении совмещённых плоскостей относительно полюса $O$ прибор замера давления зафиксирует одно и тоже давление $P$. Это значит, что вокруг полюса $O$ на поверхности сферы с бесконечно малым радиусом $\Delta r$ образуется одинаковое во всех её точках определённое давление $P$, см. *рис. 2,г* . то есть неподвижная точка $A$ среды имеет определённую сферу действия определённых сил давления с радиусом $\Delta r$.

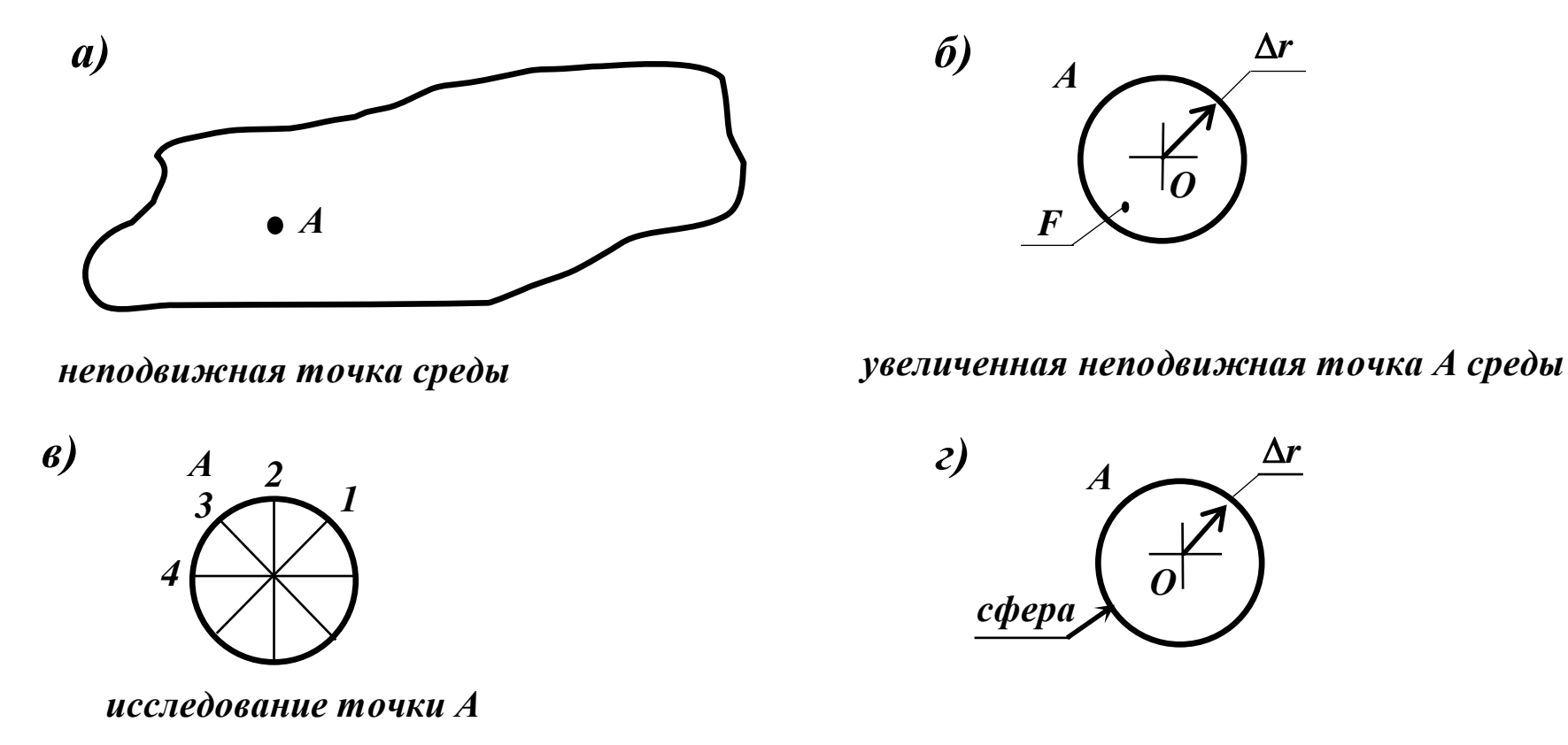

*Рис.2*

Если мы теперь будем подносить с любого направления плоскость прибора для замера давления к сфере действия сил давления точки $A$, то в точке касания сферы с плоскостью прибора сила будет направлена перпендикулярно к этой плоскости. Предположим, что плоскость прибора для замера давления имеет, какой угодно радиус кривизны. В этом случае плоскость прибора будет касаться сферы действия сил давления точки $A$ только в определённой точке касания, и во всех случаях сила давления будет направлена перпендикулярно к этой поверхности. Это, действительно, так и есть, поскольку любая реальная поверхность, как бы мала она ни была, по сравнению с точкой, всё равно, является бесконечно большой величиной, то есть точка, как минимум, на порядок меньше по сравнению с плоскостью, при условии, что и плоскость, и точка являются бесконечно малыми величинами (рис.3).

Из положений *а*, *б* и *в* рисунка 3 следует, что действие сил давления любой точки среды направлено перпендикулярно поверхности независимо от радиуса кривизны этой поверхности.

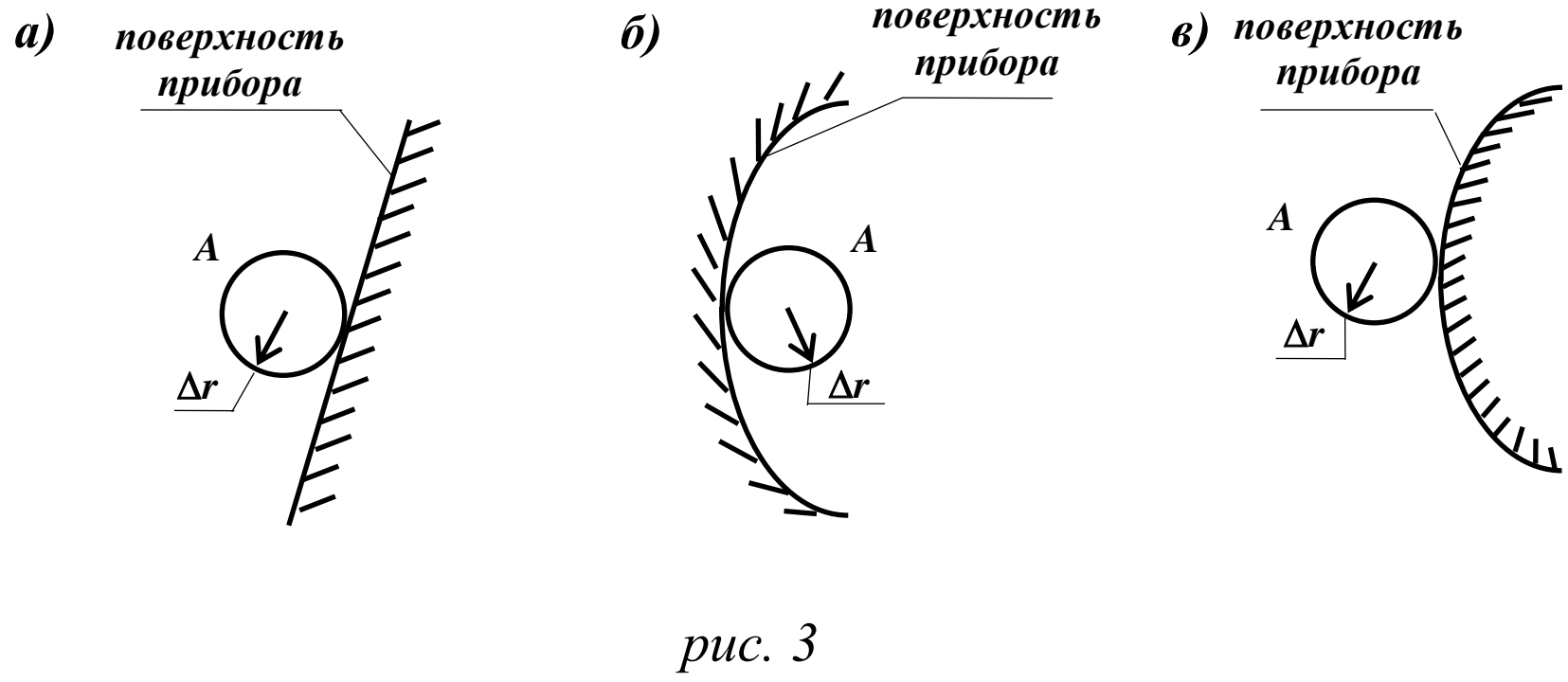

*рис. 3*

Плоскость (поверхность), помещённая в среду, соприкасается непосредственно по всей границе этой поверхности (плоскости) с точками среды. Это значит, что на плоскость (поверхность), помещённую в среду, действуют силы давления, действие которых направлено перпендикулярно к любой точке этой поверхности (плоскости) независимо от формы поверхности (*рис. 4*)

Отсюда следует, что направление действия сил давления можно определить только по направлению действия этих сил давления на поверхность (плоскость) и в зависимости от этого принять положительный или отрицательный знак для них, а не относительно системы координат, как это делается при использовании инерциальной системы координат.

В связи с тем, что действие сил распределено в бесконечно малой сфере любой точки среды, и в пределах этой сферы оно имеет определённое значение, то положения, полученные при исследовании неподвижной точки среды и поверхности любой формы относительно направления сил давления, относятся в одинаковой мере к средам, находящимся под действием, как скалярного, так и векторного силового поля.

Отметим, что общий принцип независимости действия сил верен и для сил давления среды. Если, например, среда находится под действием скалярного и векторного силового поля одновременно, то действия сил этих полей можно определить раздельно, а затем суммировать. Этот принцип сохраняет свою силу и при действии динамических сил давления.

Теперь, используя принцип действия сил давления неподвижной точки среды и общий метод исследования, мы можем приступить непосредственно к исследованию самой среды.

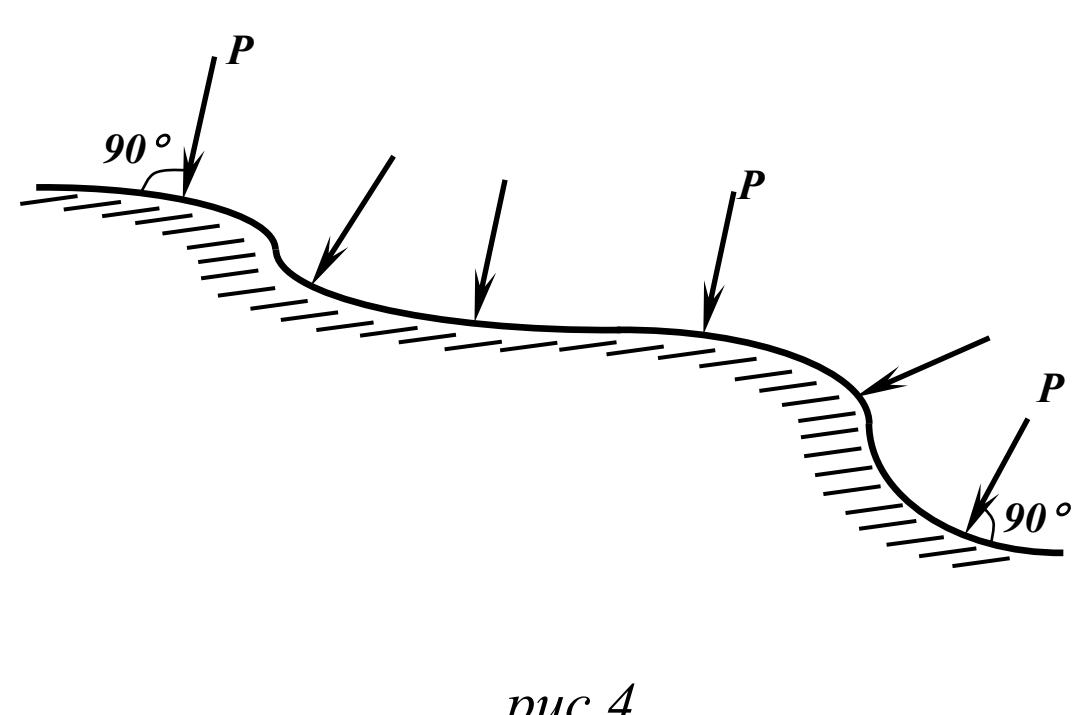

*рис.4*

## II. 2. СОСТОЯНИЕ ПОКОЯ СРЕДЫ В СКАЛЯРНОМ СИЛОВОМ ПОЛЕ

Будем считать, что мы достаточно хорошо ознакомлены с общими характеристиками среды. Теперь нам необходимо применить их к её конкретному состоянию.

Дальше считаем, что действие сил скалярного силового поля во всех точках среды одинаково и равно определённой величине давления (*P*) для каждого конкретного скалярного силового поля. Подобное положение мы принимаем не абстрактно, а на основе практического опыта.

Практически в этом убеждаются следующим образом: берут прибор для замера давления и замеряют им давление в неподвижных точках среды. Таких точек берут по количеству и расположению в определённом минимуме, который даёт возможность убедиться в том, что действительно, давление во всех точках среды есть величина постоянная и равная определённому давлению (*P*). Это предварительное исследование даёт представление о действии скалярного силового поля.

Опять же основываясь на практическом опыте, мы можем сказать, что скалярное силовое поле образуется двумя способами:

первый способ определяется термодинамическими свойствами массы;

второй способ определяется давлением, производимым на среду внешними силами. В соответствии с законом Паскаля давление, производимое на жидкость внешними силами, передаётся ею по всем направлениям равномерно.

Таким образом, мы определили качественно состояние покоя среды в скалярном силовом поле. Количественно же это состояние можно определить зависимостями, которые записываются из условий равновесия.

Чтобы записать условие равновесия, мы должны воспользоваться общим методом исследования, согласно которого первоначально надо взять определённую плоскость или поверхность, расположенную перпендикулярно к скоростям движения, изобразить и записать для неё условия равновесия.

Затем уже относительно неё расположить полярную систему координат и провести общие исследования среды в данном состоянии. Это значит, что должны в первую очередь расположить плоскость или поверхность перпендикулярно скорости движения (*W*) среды. Для этого воспользуемся вторым названием состояния покоя, как состоянием застывшего движения. Оно означает, что жидкость в состоянии покоя сохраняет направленность движения через направление действия сил давления. В скалярном силовом поле среда имеет во всех своих точках одинаковое давление. По этой причине направление скорости движения (*W*) жидкости может быть любым.

Чтобы выразить наиболее общие условие равновесия, выделим из среды некоторый объём *(V) (рис. 5)* с помощью замкнутой криволинейной поверхности *S* . В этом случае направление действия внутренних сил давления *(P)* определяется в соответствии с положениями п. 1 настоящей главы. То есть оно определяется поверхностью *S,* к которой внутренние силы давления действуют перпендикулярно. С этим направлением совпадают скорости движения (*W*) застывшей формы движения.

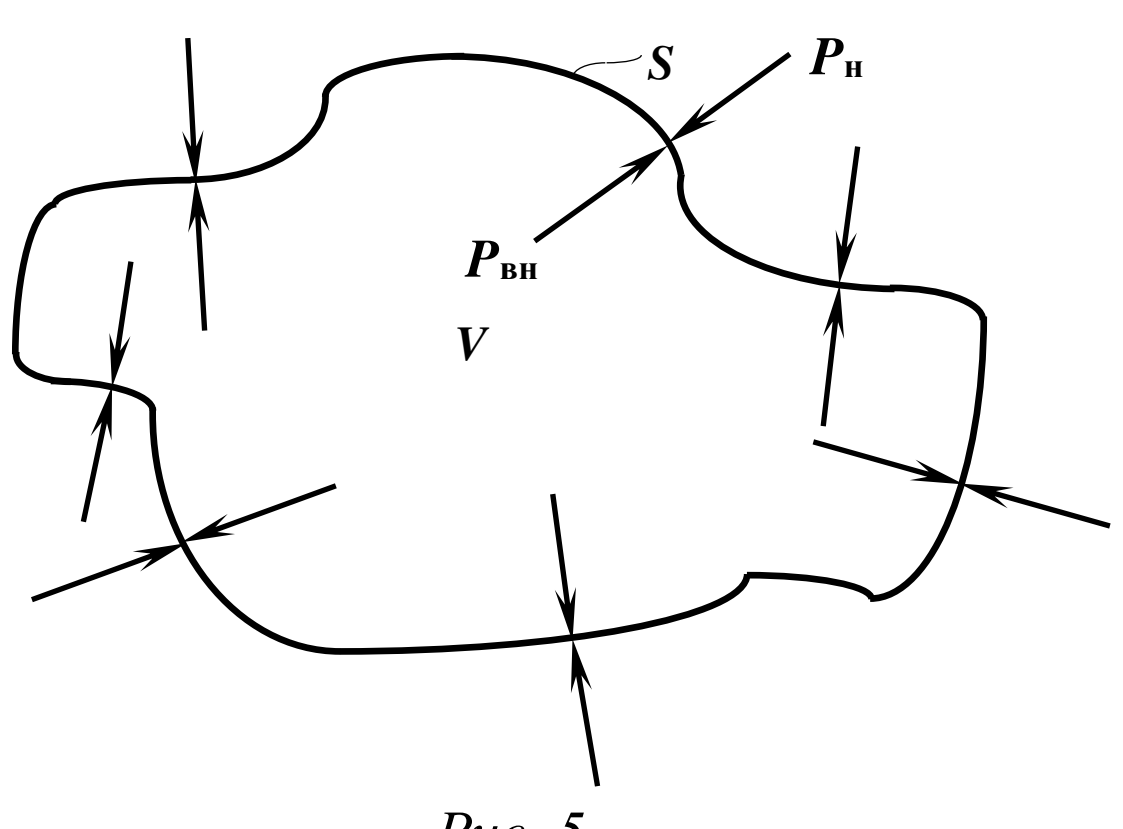

*Рис. 5*

Если не прикладывать к выделенному объёму (*V*) наружных сил давления ($P_н$), то жидкость придёт в движение в соответствии с действующими силами.

Следовательно, чтобы сохранить жидкость в состоянии покоя, мы должны приложить к поверхности (*S*) выделенного объёма наружные силы давления ($P_н$), которые по условию равновесия должны быть равны по величине внутренним силам давления ($P_{вн}$) и иметь противоположное направление.

Тогда условие равновесия запишется следующим равенством:

$$P_н S = P_{вн} S. \qquad (1)$$

Если это равенство отнести к единице площади, то есть разделить его на величину поверхности ($S$), то оно примет вид:

$$P_н = P_{вн}. \qquad (2)$$

Чтобы определить количественно величину внутреннего и наружного давления, необходимо знать зависимость их с массой $m$ выделенного объёма ($V$). Её можно записать как произведение плотности ($\rho$) на выделенный объём, то есть

$$m = \rho V. \qquad (3)$$

Дальше мы должны действовать исходя из двух способов образования скалярного силового поля. В соответствии с первым давление зависит от термодинамического состояния массы выделенного объёма ($V$). Так как мы рассматриваем идеальную жидкость, которая не зависит от термодинамических свойств массы, то мы в данный момент не сможем определить конкретной зависимостью связь давления с массой. Поэтому поступим просто – запишем внутреннее давление ($P_{вн}$) как функцию массы, то есть

$$P_{вн} = f(m). \qquad (4)$$

Для реальных жидкостей и газов, воспользовавшись термодинамическими зависимостями, мы сможем определить зависимость давления от состояния массы. Следовательно, равенство (4) имеет смысл даже в таком общем выражении.

В соответствии со вторым способом по закону Паскаля внутренние силы давления определяются давлением, производимым на среду внешними силами давления. Это значит, что внутренние силы давления $P_{вн}$ зависят от давления производимым на среду внешними силами $P_н$. Тогда мы сможем записать в общем виде зависимость этих давлений, то есть, что внутреннее давление ($P_{вн}$) есть функция давления ($P_н$), которое производят на среду внешние силы. Она будет иметь такой вид:

$$P_{вн} = f(P_н). \qquad (5)$$

Давление ($P_н$) можно всегда определить исходя в каждом случае конкретных условий действия этих сил на среду, то есть исходя из конкретных условий рассматриваемой задачи. Подобных условий может быть очень много, поэтому вид зависимости (5) является общим для них.

Теперь, используя зависимости (4) и (5), мы сможем всегда количественно выразить условие равновесия на поверхности ($S$) выделенного объёма.

Дальше мы должны, согласно общему методу исследования, расположить относительно поверхности исследования ($S$) полярную систему координат и провести исследование для всей среды. Результатом этих исследований должны быть зависимости условий равновесия рассматриваемой среды. В нашем конкретном случае выделенный объём ($V$) может быть как бесконечно малой величиной, так и величиной соизмеримой с объёмом всей исследуемой среды. Всё зависит от того, какой первоначальный объём возьмёт исследователь, так что общий вид зависимостей (1), (2), (4) и (5) при этом остаётся неизменным. Понятие среды включает в себя обобщение явлений в отношении масштаба, что позволяет рассматривать условие равновесия относительно определённых единиц измерения. Отсюда следует, что зависимости (1), (2), (4) и (5) являются общими для состояния среды в скалярном поле сил, то есть они охватывают все условия, связанные с этим состоянием. Следовательно, цель исследования достигнута, так как мы получили общие зависимости. Поэтому в данном случае нет необходимости в полярной системе координат, так как все эти общие уравнения не зависят от пространства.

В заключение отметим, что количественную величину внутренних и наружных сил давления для идеальной жидкости можно определить только от действия внешних сил в соответствии с законом Паскаля, в связи с тем, что идеальная жидкость не зависит от термодинамического состояния своей массы.

## II. 3. СОСТОЯНИЕ ПОКОЯ СРЕДЫ В ВЕКТОРНОМ СИЛОВОМ ПОЛЕ

Здесь имеется в виду силовое поле, например, поле типа гравитационного. Под действием его сил идеальная жидкость находится в состоянии покоя. Для этого состояния среды мы должны получить необходимые зависимости и положения, которые бы характеризовали данное состояние среды как качественно, так и количественно.

Исследование состояния покоя среды в векторном силовом поле начнём с исследования неподвижных точек среды. Для этой цели возьмём прибор для замера сил давления, как в предыдущем параграфе, и замерим им давления во всех интересующих нас направлениях. После исследования таких точек мы придём к следующему выводу:

1. неподвижные точки среды, которые находятся в плоскости, расположенной определённым образом, имеют одинаковое давление;
2. все плоскости среды, в каждой из которых неподвижные точки имеют соответствующее определённое давление, располагаются параллельно относительно друг друга;
3. неподвижные точки среды, расположенные по нормали к этим плоскостям, имеют разные количественные величины сил давления, которые либо уменьшаются, либо увеличиваются по величине в зависимости от направления производимых замеров давления неподвижных точек этой нормали.

Основанием для этих выводов служит практический опыт. Положения данных выводов совместно с общими положениями о среде определяют качественно состояние покоя жидкости в векторном силовом поле. Количественно это состояние можно определить зависимостями, которые записываются из условия равновесия.

Чтобы записать условие равновесия, мы должны воспользоваться общим методом исследования, согласно которому первоначально надо взять определённую плоскость, или поверхность, расположенную перпендикулярно к скорости движения. Затем изобразить и записать для неё условие равновесия. После чего относительно неё расположить полярную систему координат и провести общие исследования среды в данном состоянии.

Далее мы будем действовать в соответствии с общим методом исследования.

1. Нам необходимо выбрать плоскость для первоначального исследования среды. Она должна быть расположена перпендикулярно к скорости движения.

Теперь вспомним, что состояние покоя идеальной жидкости в векторном силовом поле является ещё и состоянием застывшего движения. В этом случае функции скорости движения $W$, связанные с направлением движения, переходят к силам давления, т. к. для данного состояния непосредственно само движение отсутствует. Мы знаем, что направленность действия сил давления проявляется непосредственно через неподвижные точки среды относительно плоскости или поверхности, т. к. сами точки имеют сферу действия сил давления с бесконечно малым радиусом $\Delta r$. Тогда застывшее движение данного состояния можно будет выразить неподвижной плоскостью, все точки которой имеют одинаковую величину сил давления. На рис. 6,*а* изобразим такую плоскость $S$.

2. Теперь, в соответствии с общим методом исследования, мы должны записать и изобразить для неё условия равновесия.

Чтобы окончательно определиться с направлением действия сил давления, выделим из среды определённый объём $V$, основанием которого служит площадь окружности $F$. В этом случае данный объём $V$ можно будет изобразить в виде цилиндра (рис. 6,*б*) высотой $h$. Высота $h$ выделенного цилиндрического объёма жидкости равна по нормали расстоянию между плоскостями $S$ и $S_1$. Плоскость $S_1$ среды характерна тем, что все её неподвижные точки среды имеют давление равное нулю. Отметим, что такое построение цилиндра для выделенного объёма жидкости $V$ не случайно. Смысл его будет разъяснён в дальнейшем описании.

При исследовании среды по неподвижным точкам мы установили, что давление среды в направлении высоты $h$ может изменяться от нуля до бесконечности. Выбор высоты цилиндра, с этой точки зрения, даёт возможность получить в каждом конкретном случае определённую величину сил давления.

Теперь изобразим условие равновесия на площади $F$ плоскости $S$. Для удобства наглядного восприятия изобразим выделенный цилиндр с помощью вида *Б* (рис. 6,*в*). Тогда внутренние силы давления $P_{вн}$ будут действовать перпендикулярно к плоскости $F$, а также непосредственно в плоскости $S$, вернее, на толщине плоскости $S$ по всему контуру площади $F$. Чтобы плоскость оставалась неподвижной, мы должны будем приложить к ней наружные силы давления $P_{н}$, которые будут действовать нормально к площади $F$, но с противоположной стороны этой площади, и в плоскости $S$ по контуру площади $F$ тоже с противоположной стороны. Таким образом, здесь объясняется вопрос о противоположно направленном действии двух скалярных величин. Значит, мы правильно изобразили на рис. 6,*в* условие равновесия. Теперь остаётся записать его.

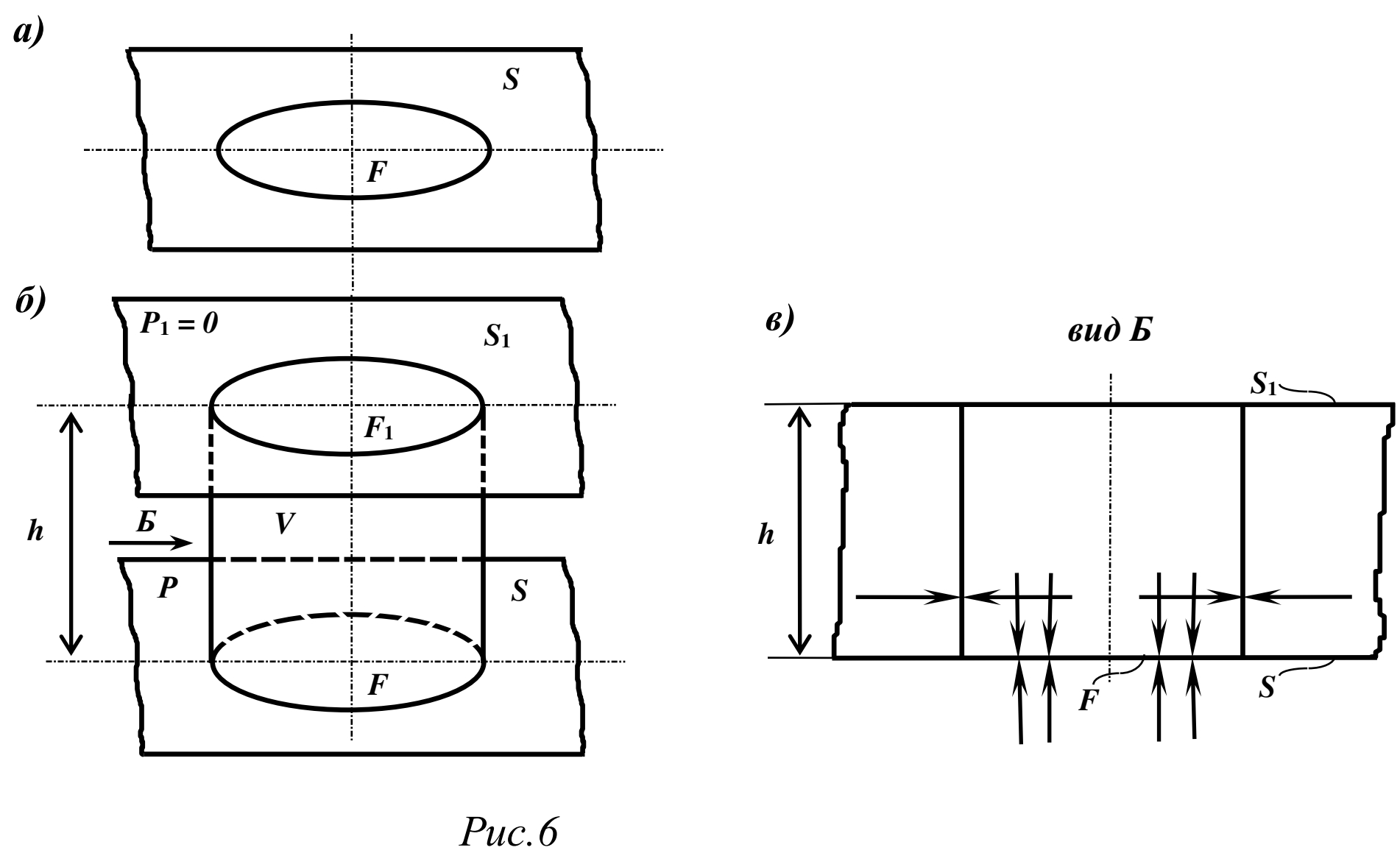

*Рис.6*

Для сил давления, внутренних и наружных, которые действуют в плоскости $S$, толщина этой плоскости есть условная величина, т. к. мы не сможем выразить её какой-либо количественной величиной. Поэтому просто отметим здесь, что величина внутренних и внешних сил давления, которые действуют *в* плоскости $S$, равна величине наружных и внутренних сил давления, которые соответственно действуют перпендикулярно к площади $F$. Можно сказать проще: подобное равенство сохраняется положением о неподвижных точках среды, т. к. они имеют сферу действия сил давления. Это условие в дальнейшем будет необходимо при определении сил давления на поверхности цилиндра или иной поверхности в зависимости от высоты $h$.

Для нормальных сил давления площади $F$ условие равновесия можно записать в таком виде:

$$FP_{вн} = FP_{н}. \qquad (6)$$

Это равенство можно записать также для единицы площади, для чего равенство (6) надо разделить на величину площади $F$. Тогда получим:

$$P_{вн} = P_{н} \; . \qquad (7)$$

В данной зависимости количественную величину как внутренних, так и внешних сил давления мы можем определить только по внутренним силам давления, т. к. при данных условиях равновесия можно записать количественную зависимость только для внутренних сил давления. Чтобы записать её, воспользуемся уравнением сил расходного вида движения, то есть вторым законом механики безынертной массы. Запишем его ещё раз:

$$P = \frac{M}{F} W.$$

Этот закон выражает зависимость сил давления от расхода массы в единицу времени $M$ и скорости движения $W$. У нас на площади $F$ (рис. 6) фактически записано условие застывшего движения. Тогда мы можем записать внутренние силы давления, которые действуют нормально к площади $F$, как

$$P_{\text{вн}} = \frac{M}{F} W. \qquad (8)$$

Теперь остаётся выяснить, как в уравнении (8) надо понимать массовый расход в единицу времени $M$ и скорость $W$.

Массовый расход мы получим следующим образом:

а) запишем сначала массу выделенного объёма $V$ (рис. 6, *б*). Она будет равна произведению этого объёма на плотность:

$$m = \rho V.$$

Если в этом равенстве объём $V$ выразить через произведение площади $F$ на высоту $h$, то получим:

$$m = \rho F h. \qquad (9)$$

б) массовый расход $M$ равен первой производной по времени от количества массы $m$:

$$M = \frac{dm}{dt}. \qquad (10)$$

в) заменим в равенстве (10) массу через уравнение (9), получим:

$$M = \frac{d(\rho F h)}{dt}. \qquad (11)$$

В нашем случае плотность $\rho$ и площадь $F$ являются постоянными величинами. Поэтому их можно вынести, тогда равенство (11) примет вид:

$$M = \rho F \frac{dh}{dt}. \qquad (12)$$

Фактически получается, что высота $h$ тоже является постоянной величиной. В то же время она должна выражать форму застывшего движения. Чтобы совместить оба условия, мы поступим следующим образом: вынесем за знак дифференциала абсолютную величину высоты $h$, то есть считаем её безразмерной, некоторую величину $a$ считаем переменной величиной, тогда равенство (12) примет вид:

$$M = \rho F h \frac{da}{dt}. \qquad (13)$$

Высота $h$ в равенстве (13) определяет количество тех или иных длинновых единиц, но размерности сама не имеет. Первая производная от некоторой длинновой величины $a$ является линейной скоростью $W$, которая зависит от характеристик векторного силового поля. В связи с тем, что мы приняли характеристики этого поля для каждого случая постоянными, то величина скорости $W$ тоже будет постоянной. Тогда равенство (13) примет вид:

$$M = \rho F h W. \qquad (14)$$

Таким способом мы совместили два условия. Теперь подставим равенство (14) в равенство (8), получим:

$$P_{\text{вн}} = \rho h W^2. \qquad (15)$$

С помощью уравнения (15) мы сможем определить величину внутренних сил давления, если будем знать линейную скорость $W$. Для её определения можно воспользоваться, например, таким способом: мы знаем, что выделенный объём (рис. 6, *б*) согласно второму закону Ньютона в поле земного тяготения имеет вес. Второй закон Ньютона имеет такую зависимость:

$$R = mg. \qquad (16)$$

Преобразуем это равенство так, чтобы им можно было пользоваться в нашем случае. Для этого разделим его на площадь $F$, а массу выразим через характеристики объёма $V$ умноженного на плотность $\rho$. Тогда равенство (16) примет вид:

$$P = \rho hg, \qquad (17)$$

где $P$ является распределённой силой веса выделенного объёма на площадь $F$. Распределённая сила $P$ и внутреннее давление $P_{вн}$ должны быть равны по величине, то есть

$$P_{вн} = P.$$

Подставим в это равенство его значения в соответствии с уравнениями (15) и (17) с их единицами измерения, получим:

$$\rho\,[кг*сек^2/м^4]\cdot h\cdot W^2\,[м^2/сек^2] = \rho\,[кг*сек^2/м^4]\cdot h\,[м]\cdot g\,[м/сек^2]$$

После сокращения соответствующих величин и единиц измерения оно примет вид:

$$W^2\,[м^2/сек^2] = g\,[м/сек^2]\cdot[м]$$

В поле земного тяготения для принятой системы единиц величина ускорения $g$ равна 9,81 м/сек$^2$. Тогда скорость будет равна:

$$W = \sqrt{9.81} = 3{,}132\ \ (м/сек).$$

Это значит, что молекулы среды при свободном падении под действием поля земного тяготения движутся с постоянной скоростью, равной 3,132 м/сек. Обозначим эту скорость буквой $w$:

$$w = 3{,}132\ (м/сек).$$

Следовательно, как в своё время экспериментальным путем была получена величина ускорения свободного падения тел, так для жидкостей и газов подобным путем можно получить величину линейной скорости в поле земного тяготения[5]. Таким образом, мы получили полное представление о линейной скорости.

Теперь мы можем пользоваться уравнением (15), то есть мы можем по этой зависимости определить величину внутренних сил давления $P_{вн}$ для исследуемой плоскости. Это значит, что мы можем количественно выразить условия равновесия на площади $F$ поверхности $S$.

3. Далее мы должны относительно плоскости $S$ расположить полярную систему координат и провести исследование всей среды. Результатами этих исследований должны быть общие зависимости условий равновесия рассматриваемой среды.

Уравнение (15) является общим уравнением, которое охватывает все условия состояния среды в векторном силовом поле, т. к. оно включает в себя любые по величине плоскости исследования, которые могут пересекать всю среду одновременно, а общее изменение давления в этих плоскостях в зависимости от высоты $h$ охватывает полный объём среды. Поэтому более общих зависимостей для исследуемой среды, чем уравнение (15), получить нельзя. Следовательно, мы проделали полный комплекс исследований.

[5] В данном случае скорость $w$ движения безынертной массы среды под действием сил земного тяготения получена как характеристика застывшего движения, т.е. как характеристика состояния покоя среды. См. работу «Строение Солнца и планет солнечной системы с точки зрения механики безынертной массы» и «Движение твёрдых тел в жидкостях и газах с точки зрения механики безынертной массы», где эта скорость рассматривается как скорость, в динамике, при движении воздушных масс атмосферы и др.

В заключение отметим, что раздел статики в существующей механике жидкости и газа основан на следующих принципах:

1. за основу берётся инерциальная система координат;
2. при исследовании рассматриваются сосредоточенные силы;
3. в основе всех сил лежит второй закон Ньютона.

Пожалуй, подобные различия остались незамеченными до настоящего времени лишь потому, что абсолютная величина постоянной скорости структурных единиц среды в поле земного тяготения равна корню квадратному из абсолютной величины ускорения для тел, находящихся под действием сил земного тяготения.

## Г Л А В А III. КИНЕМАТИКА

Кинематика является разделом механики жидкости и газа, который изучает движение массы жидкости вне связи с определяющим его силовым взаимодействием. Она рассматривает движение идеальной жидкости в неподвижном пространстве. Главной определяющей характеристикой движения является расход массы в единицу времени. Изменение его в пространстве и во времени определяет вид движения идеальной жидкости. Согласно формальному принципу связи вида движения с формой уравнений неразрывности и движения для жидкости существует всего четыре вида движения:

1. установившийся вид движения;
2. плоский установившийся вид движения;
3. расходный вид движения;
4. акустический вид движения.

Формальный принцип связи значительно упрощает задачи этого раздела механики, т.к. конкретно определяет число видов движения. Каждый вид движения определяется конкретной связью расхода массы в единицу времени с пространством и временем. Изменение расхода массы в пространстве и во времени рассматривается непосредственно в неподвижной относительно линейной скорости движения $W$ плоскости, поверхности, точки. Это изменение движения определяется не вообще изменением в пространстве, а изменением *расхода одного сечения потока относительно другого*. Поэтому в данном разделе применяется полярная система координат, или неинерциальная система координат. Время остаётся временем, и в данном разделе берётся либо как момент времени, либо как промежуток времени. В общем, исследование видов движения проводится в соответствии с общим методом исследования, изложенным в параграфе 4 главы I. Основная задача кинематики заключается в том, чтобы найти уравнение движения для каждого из четырёх видов движения.

### III. 1. УРАВНЕНИЕ ДВИЖЕНИЯ УСТАНОВИВШЕГОСЯ ВИДА ДВИЖЕНИЯ ИДЕАЛЬНОЙ ЖИДКОСТИ

В соответствии с формальным принципом связи вида движения с формой уравнений неразрывности и движения в зависимостях данного вида движения должны отсутствовать параметры пространства и времени. Это значит, что расход массы в единицу времени $M$ является постоянной величиной, которая не зависит ни от параметра пространства, ни от параметра времени, то есть

$$M = f(t, r, \varphi) = \text{const.}$$

Таким образом, мы определились относительно основной характеристики этого вида движения жидкости.

Чтобы получить уравнение движения, мы должны связать расход массы в единицу времени с конкретными характеристиками потока жидкости. Поэтому мы должны знать ещё и характеристики потока жидкости.

Установившийся вид движения мы приняли не абстрактно, а с учётом определённого практического опыта по наблюдению этого вида движения. Руководствуясь им, мы можем дать некоторое описание данного вида движения:

жидкость движется в определённом направлении и образует некоторый объём *V*, который располагается либо симметрично относительно прямой линии, либо симметрично относительно двух взаимно пересекающихся перпендикулярных плоскостей. Линия пересечения этих плоскостей тоже является прямой линией (рис. 7,*б* и 7,*в*). В обоих случаях направление движения, вернее, направление скорости движения *W* жидкости совпадает с направлением этой прямой линии. Назовём её линией тока. Помимо скорости движения *W*, поток жидкости имеет определённую массу *m* с соответствующей плотностью *ρ*. Мы получили описание установившегося вида движения с его характеристиками.

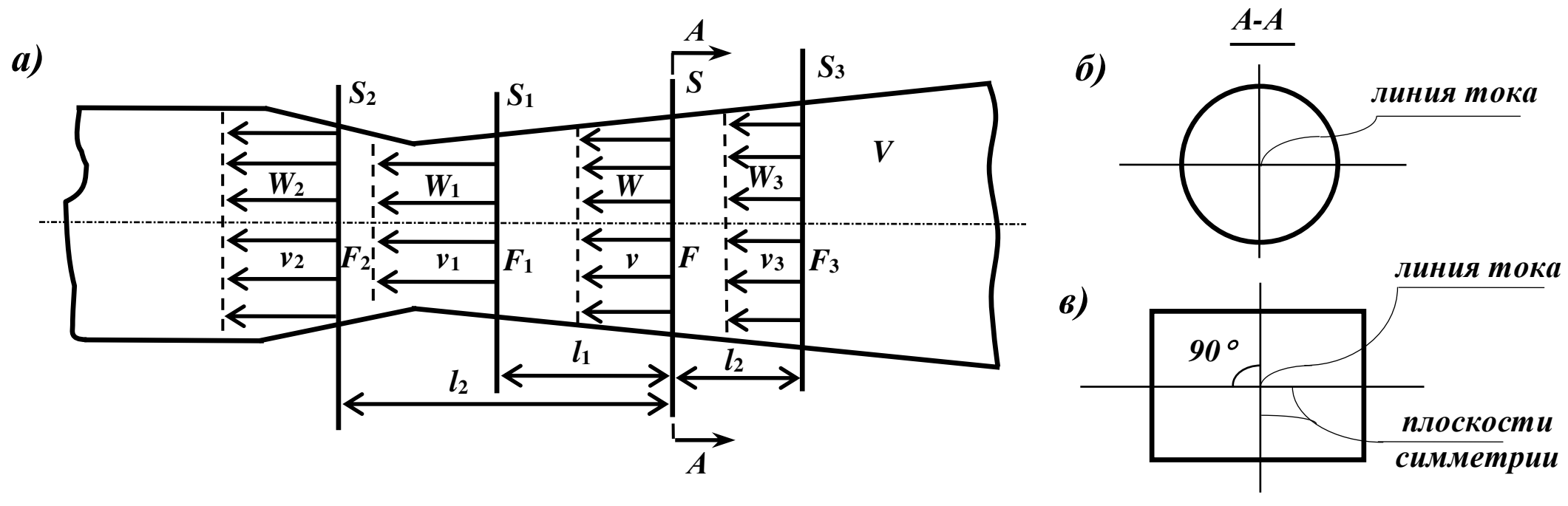

*рис. 7*

Чтобы получить уравнение движения, мы должны воспользоваться общим методом исследования. В соответствии с ним мы должны:

1. выделить неподвижное пространство среды. Это пространство должно включать в себя либо весь поток жидкости, либо его характерную часть. При этом его границы не должны изменять или как-то влиять на движение жидкости. Это значит, что границы среды должны совпадать с границами установившегося потока жидкости.[6] Покажем его на рис. 7,*а*.

2. Расположить или выбрать плоскость *S* для первоначального исследования, которая, как мы знаем, должна располагаться перпендикулярно к скорости движения жидкости *W*. В нашем случае направление скорости движения совпадает с направлением линии тока. Поэтому расположим плоскость *S* для первоначального исследования перпендикулярно линии тока (рис. 7,*а*). Тогда площадь сечения потока будет равна некоторой величине *F*.

3. Изобразить движение жидкости на площади сечения потока *F*.

В соответствии с практическим опытом скорости движения жидкости *в* каждой точке площади сечения *F* одинаковы и равны определённой величине *W* (рис. 7, *а*). Пользуясь характеристиками потока, мы можем записать, что через площадь сечения потока *F* за единицу времени пройдёт некоторый объём жидкости *v*, который будет равняться произведению скорости *W* на площадь сечения потока *F*, то есть

$$v = WF. \tag{18}$$

За ту же единицу времени через сечение потока *F* пройдёт расход массы[7] *M*, т. к. расход является главной характеристикой данного вида движения. Отсюда следует, что масса жидкости,

---

[6] Реальный ламинарный поток имеет слоистую структуру, которая вызвана взаимодействием вещества среды со стенками потока, а затем взаимодействием границ потоков с разными скоростями. Т.к. автор абстрагируется от этого взаимодействия, то рассматриваемый установившийся поток имеет одинаковую скорость по всей площади сечения.

[7] Если сказать, что через площадь сечения, например, прошло 5 кг массы со скоростью 2 м/сек или определённый объём, то эта формулировка будет соответствовать описанию движения тела, т.е. объёма, массой 5 кг со скоростью 2 м/с. Такая формулировка искажает представление о текучести, что невольно заставляет воспринимать поток по аналогии с движением массивных объёмов, пусть и условных.

Редактор считает, что по этой причине здесь и далее автор говорит, что не масса, а расход массы, т.е. суммарная масса структурных единиц, которые прошли через конкретную площадь сечения/исследования за единицу времени, движется с такой-то скоростью или имеет направление, что не масса, а расход массы втекает или вытекает и т.д. Такая формулировка не искажает представление о текучести. В этом случае линейная скорость является только характеристикой процесса расхода, т.е. характеристикой взаимодействия массы с механическими силами, по которой которая содержится в объёме $v$, должна быть равна расходу массы в единицу времени $M$. Так как характеристики потока должны соответствовать характеру движения жидкости, поэтому, умножив объём $v$ на плотность $\rho$ жидкости и приравняв это произведение к расходу массы $M$, получим:

$$M = \rho WF. \tag{19}$$

Уравнение (19) является уравнением *движения* установившегося вида движения на площади сечения $F$ плоскости $S$.

4. Разместить полярную систему координат относительно плоскости $S$ и провести исследование всего потока жидкости относительно движения жидкости в этой плоскости.

В связи с тем, что движение жидкости в установившемся виде движения не зависит от пространства, то есть расход массы в единицу времени является постоянной величиной, то считаем, что полюс $O$ полярной системы координат расположен на бесконечно большом расстоянии от линии тока. При этом характеристики потока не зависят от радиуса $r$ полярной системы координат, что следует из уравнения (19). Тогда исследование всего установившегося потока мы будем производить на площадях сечения потока, которые могут располагаться на любом расстоянии $l$ по линии тока от первоначально принятой для исследования плоскости $S$ (рис. 7,*а*). Далее располагаем плоскости сечения на любом расстоянии $l_1, l_2 ... l_n$ от плоскости $S$. Запишем уравнения движения для каждой площади сечения потока $F_1, F_2 ... F_n$ тем же способом, что и для площади сечения потока $F$ плоскости $S$ (рис. 7, *а*). Мы получим следующие зависимости:

$$\begin{aligned} M &= \rho W_1 F_1, \\ M &= \rho W_2 F_2, \\ &\dots\dots\dots\dots\dots \\ M &= \rho W_n F_n. \end{aligned} \tag{20}$$

5. Сопоставить уравнения движения каждой плоскости друг относительно друга, чтобы выявить общее уравнение движения установившегося вида движения.

Сопоставив уравнения (20) между собой, мы заметим, что от сечения к сечению изменяются линейные скорости $W$ и площадь сечения потока $F$. При этом все изменения происходят в пределах определённой постоянной для них величины расхода массы в единицу времени $M$, то есть всё это можно выразить следующим образом:

$$M = \rho W_1 F_1 = \rho W_2 F_2 = ... = \rho W_n F_n = \text{const}. \tag{21}$$

Это значит, что общим уравнением движения установившегося вида движения жидкости будет уравнение такого вида:

$$M = \rho WF = \text{const}. \tag{22}$$

Таким образом, мы получили уравнение движения установившегося вида движения жидкости − это уравнение (22). Теперь, зная определённое пространственное восприятие этого вида движения жидкости и количественную зависимость, мы сможем уже в какой-то степени принять осознанно этот вид движения.

Отметим, что уравнение (22) было известно в существующей механике жидкости и газа сравнительно давно под названием уравнения неразрывности [7]. Если подходить формально, то различие этих уравнений заключается в их *смысловом восприятии*.

---

можно судить о них. Читатель должен сразу освоиться с терминологией автора, т.к. уже в следующем параграфе автор рассматривает плоский установившийся вид движения, когда расход массы происходит одновременно в двух направлениях, и линия тока не является прямой, что будет невозможно понять, если иметь неправильное представлении о текучести, т.е. собственно расходе массы. Тем более весь остальной материал останется непонятным. Ведь уравнения *движения* являются уравнениями расхода массы, поэтому останется непонятным даже то, о чём идёт речь. Редактор был вынужден дать пояснение терминологии, чтобы этого не произошло. Дальнейший текст не требует пояснений.

В современной механике жидкости и газа это уравнение определяет условие неразрывности, целостности массы жидкости, находящейся в движении. А в данной работе оно определяет непосредственно само движение жидкости, что требует применения уравнения (22) в другом назначении. Поэтому различие заключается именно в назначении. В конечном итоге оно определяет сущность осмысленного понимания установившегося вида движения.

Теперь дополним уже известное описание установившегося вида движения жидкости. Для чего возьмём плоскость *S* первоначального исследования и относительно её линии тока разместим полярную систему координат с определённой величиной радиуса *r* удаления полюса *O* от линии тока (рис. 7`,а)[8].

Затем начнём вращать этот радиус против часовой стрелки и фиксируем его в различных углах $\varphi_1$, $\varphi_2$, $\varphi_3$... $\varphi_n$ плоскости *N*, образованной линией тока и точкой *O* полюса системы координат. В каждом таком фиксированном положении мы получим точки $A_1$, $A_2$, $A_3$... $A_n$ соответственно. В этих точках расположим плоскости $S_1$, $S_2$, $S_3$... $S_n$ таким образом, чтобы соответствующие радиусы разместились бы в этих плоскостях, а сами плоскости были бы перпендикулярными плоскости *N*.

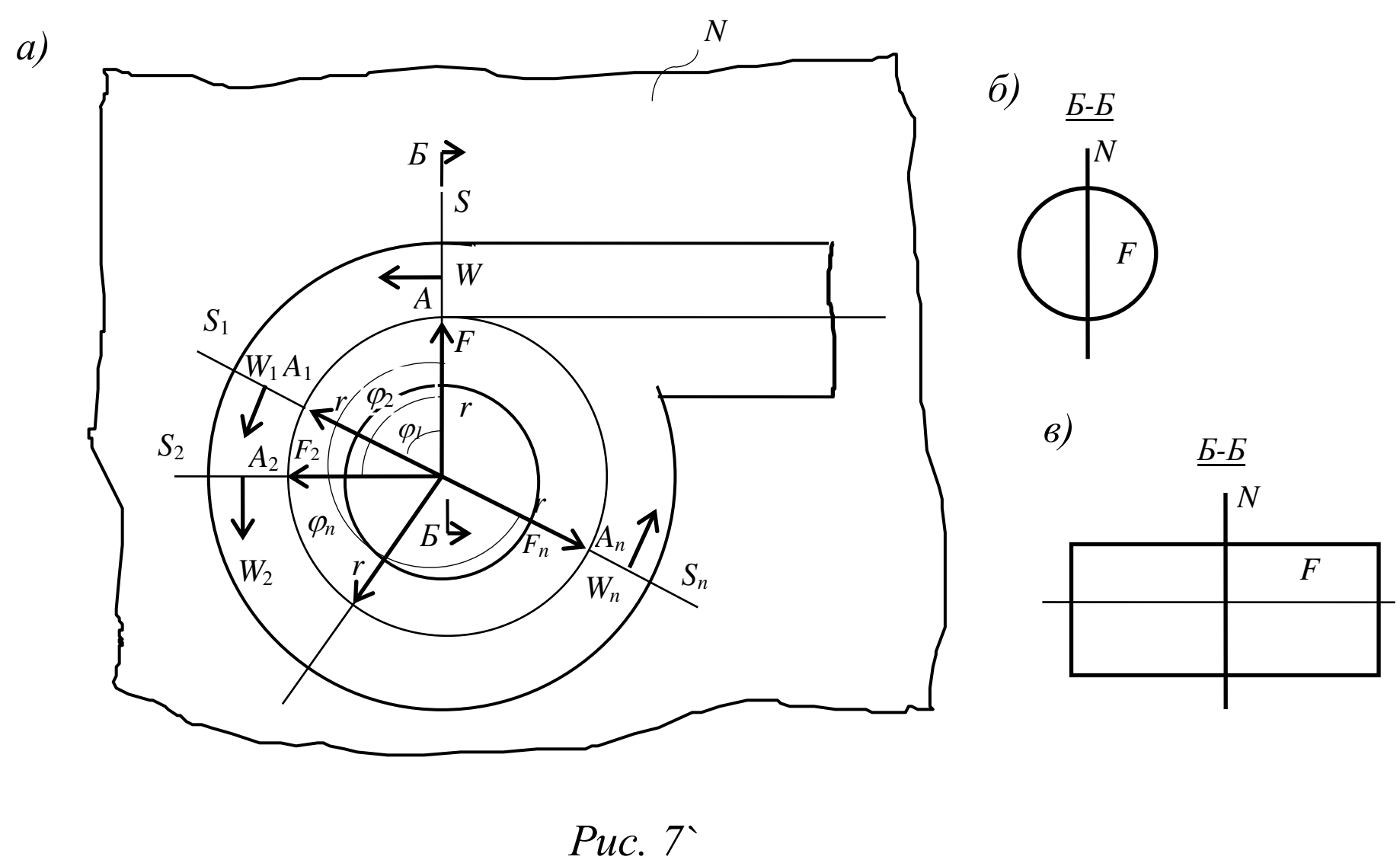

*Рис. 7`*

В этом случае плоскости $S_1$, $S_2$, $S_3$... $S_n$ становятся плоскостями исследования потока жидкости. Соединив все точки $A_1$, $A_2$, $A_3$... $A_n$ линией, мы получим линию тока, к которой все плоскости исследования будут расположены перпендикулярно. Эта линия тока будет иметь форму окружности с постоянным радиусом *r*.

После этого в каждой исследуемой плоскости построим площадь сечения потока $F_1$, $F_2$, $F_3$... $F_n$ , каждая из которых должна быть симметрично расположенной относительно линии тока в соответствующей плоскости $S_1$, $S_2$, $S_3$... $S_n$ исследования (рис. 7`,*б* и 7`,*в*). Соединив все площади $F_1$, $F_2$, $F_3$... $F_n$ по контуру поверхностью, мы получим полный объём потока *V*. Если теперь через этот объём пропустить жидкость с определённым постоянным расходом массы в единицу времени *M*, то мы получим поток установившегося движения жидкости, так как уравнение движения (22) сохраняет своё значение и для этого потока жидкости. Это значит, что линия тока потока жидкости при установившемся виде движения может быть либо прямой линией, либо окружностью. Для данного вида движения это безразлично, что подтверждается практикой.

В том случае, когда линия тока будет образована радиусами различной длины, то есть линия тока не будет ни прямой линией, ни окружностью, мы в этом случае будем иметь дело с другим видом движения жидкости, который называется плоским установившимся. Об этом виде движения речь пойдёт в следующем параграфе.

[8] По техническим причинам редактор пронумеровал этот рисунок как 7`

## III. 2. УРАВНЕНИЯ ДВИЖЕНИЯ ПЛОСКОГО УСТАНОВИВШЕГОСЯ ВИДА ДВИЖЕНИЯ ИДЕАЛЬНОЙ ЖИДКОСТИ

В соответствии с формальным принципом связи вида движения с формой уравнений неразрывности и движения в зависимостях данного вида движения должны отсутствовать параметры времени. Это значит, что расход массы в единицу времени $M$ является величиной, которая зависит от изменения параметров пространства и не зависит от параметров времени, то есть с изменением параметров времени он остаётся неизменным. Это условие можно записать так:

$$M = f(t) = \text{const},$$
$$M = f(r, \varphi).$$

Таким образом, мы определились относительно основной характеристики этого вида движения жидкости. Чтобы получить уравнения движения для него, мы должны связать расход массы в единицу времени с конкретными характеристиками потока жидкости.

Плоский установившийся вид движения мы приняли с учётом определённого практического опыта. Руководствуясь им, мы можем дать некоторое описание данного вида движения следующего содержания:

жидкость движется в определённом направлении и образует некоторый объём $V$. Этот объём имеет форму цилиндра, симметричного относительно оси $O$ (рис. 8).

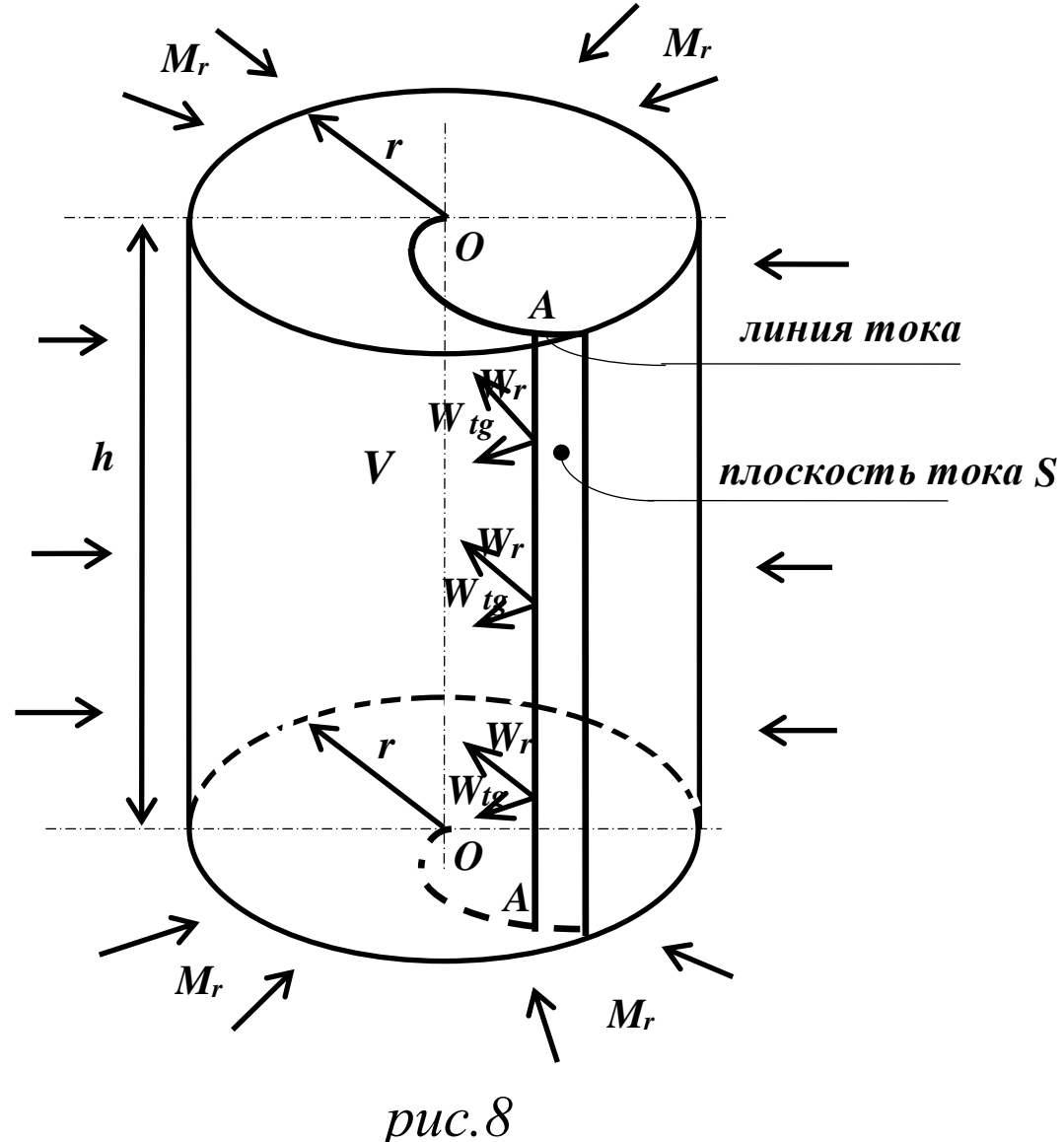

*рис.8*

Жидкость в цилиндре совершает движение в двух направлениях: направление первого движения совпадает с направлением радиуса $r$ цилиндра – назовём его радиальным движением. Направление второго направлено по окружности, то есть перпендикулярно радиусу $r$ цилиндра – назовём его тангенциальным движением. Линейные скорости движения соответственно обозначим: радиальную как $W_r$, тангенциальную как $W_{tg}$. Линейные скорости радиального и тангенциального движения располагаются в плоскостях перпендикулярных оси цилиндра. Собственно, по этой причине данный вид движения жидкости называется *плоским* установившимся видом движения жидкости в отличие от установившегося вида движения жидкости.

Расход массы в единицу времени, который вытекает или притекает к этому цилиндру, движется в радиальном направлении – назовём его радиальным расходом массы в единицу времени и обозначим как $M_r$. При этом он может двигаться как с наружного диаметра цилиндра к его оси, так и от оси цилиндра к его наружному диаметру. Тангенциальная скорость $W_{tg}$ жидкости может быть направлена как по часовой стрелке, так и против неё. Расход массы в единицу времени этой скорости движения располагается по окружности цилиндра в границах объёма этого потока. Расход массы в единицу времени, связанный с тангенциальной скоростью, назовём тангенциальным массовым расходом и обозначим его как $M_{tg}$.

Визуально диаметр цилиндра потока определяется по вращательному движению жидкости, а радиальный расход массы в единицу времени $M_r$ – по непосредственному притоку жидкости к объёму потока. Все точки, расположенные на прямой линии параллельной оси цилиндра, имеют одинаковые тангенциальные и радиальные скорости. Перемещая эту линию от оси цилиндра к его наружной границе или наоборот, мы получим плоскость тока $S$ (рис. 8) для плоского установившегося вида движения. Совокупность этих плоскостей тока составляет цилиндрический объём $V$. В поперечном разрезе цилиндра плоскости тока будут представлены в виде кривых линий (рис. 8) – назовём их линиями тока. Поток в своём объёме $V$ имеет определённую массу $m$ с соответствующей плотностью $\rho$. Таким образом, мы получили описание плоского установившегося вида движения с его характеристиками.

Чтобы получить уравнения движения, мы должны воспользоваться общим методом исследования. В соответствии с этим методом мы должны:

1. выделить неподвижное пространство среды. Это пространство должно включать в себя либо весь поток, либо его характерную часть. При этом границы его не должны изменять или как-то влиять на движение жидкости в потоке. Это значит, что границы среды должны совпадать с границами плоского установившегося потока, то есть границы среды совпадают с границами цилиндра выделенного объёма $V$ потока, показанного на *рис. 8*?

2. Расположить или выбрать плоскость для первоначального исследования.

Для этой цели выделим элемент линии тока (рис. 9, *а*), то есть отрезок одной из линий тока некоторой длины. Выделим на нём точку $A$. В связи с тем, что в плоском установившемся движении жидкость движется в радиальном и тангенциальном направлениях, то есть имеет радиальную $W_r$ и тангенциальную $W_{tg}$ скорости, то для первоначального исследования нам потребуется сразу две плоскости. Поэтому одну плоскость расположим перпендикулярно радиальной скорости $W_r$, другую – перпендикулярно тангенциальной скорости $W_{tg}$.

Радиальная плоскость $S_r$ пройдёт через точку $A$ перпендикулярно радиусу $r$, который проведён от оси цилиндра $O$ до точки $A$. Тангенциальная плоскость $S_{tg}$ пройдёт через точку $A$ и ось цилиндра $O$ (рис.9, *а*). Мы получим две плоскости первоначального исследования: радиальную $S_r$ и тангенциальную $S_{tg}$, которые пересекаются в точке $A$ и располагаются друг относительно друга под углом 90°. Так как в каждой точке потока жидкость движется одновременно с радиальной и тангенциальной скоростями, то нам придётся проводить первоначальное исследование непосредственно в точке потока.

Такой точкой является точка $A$, которая лежит на линии тока и на пересечении радиальной $S_r$ и тангенциальной $S_{tg}$ плоскостей исследования. Поэтому площади исследования будут равны площадям сечения точки $A$ радиальной и тангенциальной плоскостями, то есть радиальная площадь исследования $\Delta F_r$ будет равна тангенциальной площади исследования $\Delta F_{tg}$, т. к. они равны площади сечения точки $A$. Изобразим эти площади в увеличенном масштабе на рисунке 9,*б*.

Здесь надо помнить, что точка $A$ является *прямой линией*, которая располагается параллельно оси цилиндра $O$, согласно рис.8. Длина этой линии равна высоте цилиндра $h$. Таким образом, мы получили характеристики линии тока, вернее, плоскости тока плоского установившегося вида движения.

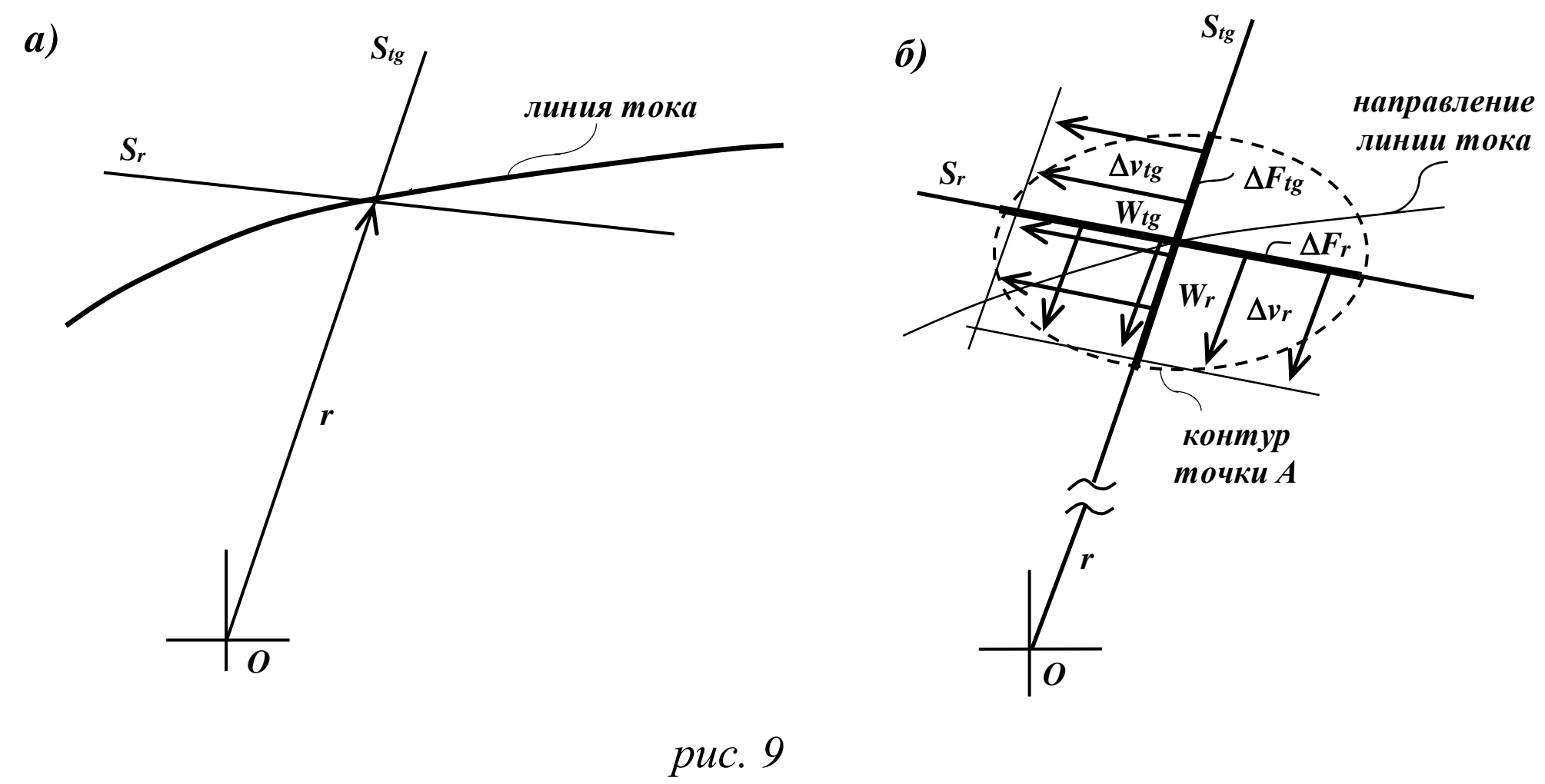

*рис. 9*

3. Изобразить движение жидкости на площадях сечения $\Delta F_r$ и $\Delta F_{tg}$.

В соответствии с практическим опытом скорости движения жидкости в площадях сечения $\Delta F_r$ и $\Delta F_{tg}$ постоянны и равны соответственно определённым величинам $W_r$ и $W_{tg}$ (рис. 9,*б*). Тогда мы можем записать, пользуясь характеристиками линии тока плоского установившегося потока, что через площади сечения $\Delta F_r$ и $\Delta F_{tg}$ за единицу времени пройдут некоторые объёмы жидкости $\Delta v_r$ и $\Delta v_{tg}$, которые будут равны произведениям соответствующих скоростей на соответствующие площади сечения. Получим:

$$\begin{aligned} \Delta v_r &= W_r \Delta F_r, \\ \Delta v_{tg} &= W_{tg} \Delta F_{tg}. \end{aligned} \qquad (23)$$

За ту же единицу времени через эти площади пройдут соответствующие расходы массы. В связи с тем, что расход массы в единицу времени должен равняться расходу массы, выраженному через характеристики потока, то, умножив соответствующие объёмы $\Delta v_r$ и $\Delta v_{tg}$ на плотность жидкости $\rho$ и приравняв их к соответствующим расходам массы в единицу времени, получим:

$$\begin{aligned} \Delta M_r &= \rho W_r \Delta F_r, \\ \Delta M_{tg} &= \rho W_{tg} \Delta F_{tg}. \end{aligned} \qquad (24)$$

Уравнения (24) являются уравнениями движения в точке *A* линии тока для потока жидкости плоского установившегося вида движения.

Теперь представим, что точка *A* имеет некоторый бесконечно малый объём $\Delta V_т$ (рис. 10). Выше мы выяснили, что состояние этого объёма поддерживается двумя расходами массы, которые поступают через радиальную и тангенциальную площади сечения в объёме $\Delta V_т$ точки *A*. Выяснили характер движения радиального и тангенциального расходов массы в единицу времени по отношению к объёму точки *A*.

По отношению к объёму любой расход массы в единицу времени является либо втекающим, либо вытекающим, то есть расход либо поступает в объём, либо выходит из него. За положительный расход массы в единицу времени принимаем расход, который поступает в объём, за отрицательный – вытекающий из него. В дальнейшем положительные и отрицательные значения для любых характеристик механики жидкости и газа мы будем определять непосредственно по отношению к объёму. Подобная классификация расходов вытекает из общих условий безынертной массы и неинерциальной системы координат. Это положение является общим для механики жидкости и газа.

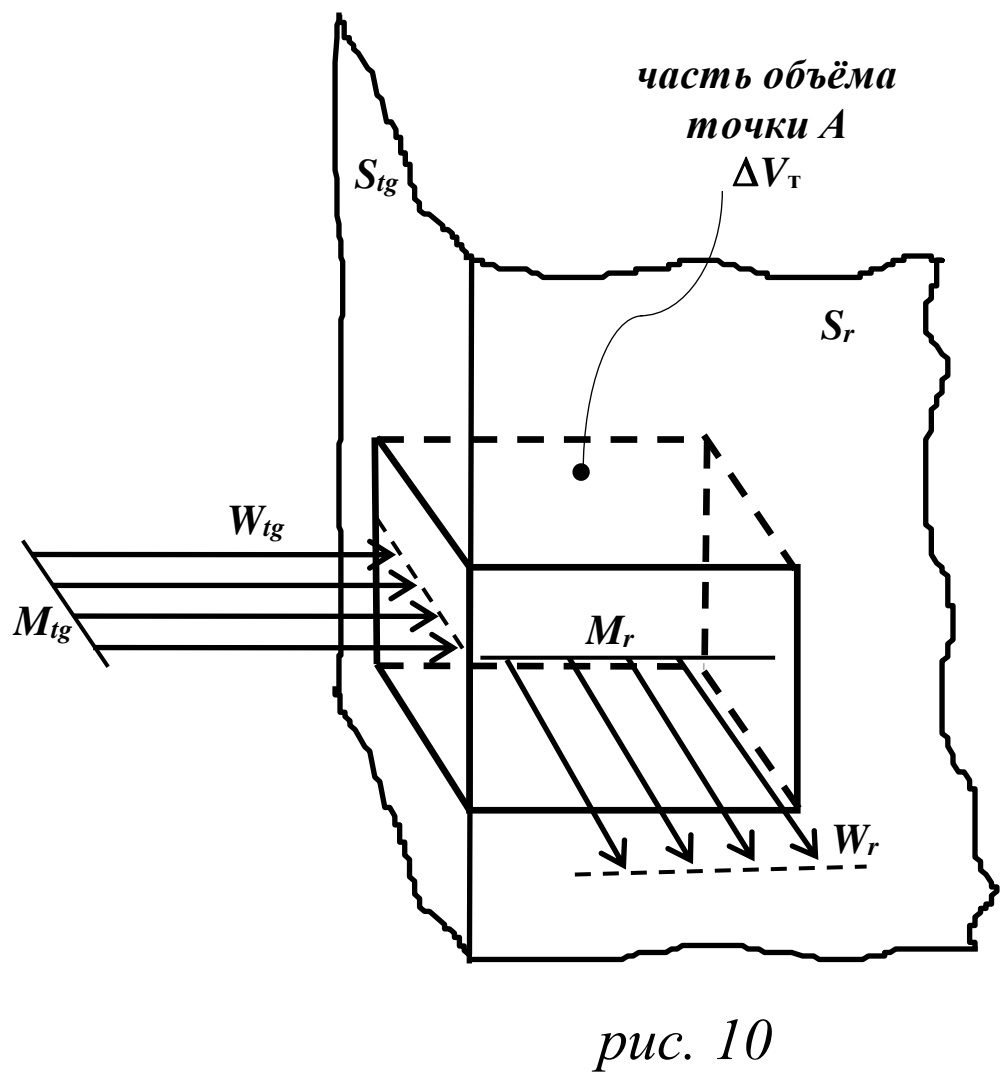

*рис. 10*

В настоящий момент мы знаем, что радиальный и тангенциальный расходы движутся во взаимно перпендикулярных плоскостях, как показано на рисунках 9 и 10.

Теперь определим их относительно объёма точки *A*, то есть определим знаки этих расходов. Сделаем это следующим образом:

предположим, что оба расхода имеют одинаковые знаки. Тогда и радиальный, и тангенциальный расходы массы либо одновременно втекают, либо одновременно вытекают из объёма точки *A*. В этом случае объём точки *A* должен либо уменьшаться, либо увеличиваться, но он остаётся неизменным при движении жидкости. Неизменность объёма можно сохранить при одном условии, что радиальный и тангенциальный расходы массы будут разных знаков, то есть один расход должен втекать в объём точки *A*, а другой вытекать из него. Отсюда следует, что радиальный и тангенциальный расходы массы в единицу времени имеют разные знаки для точки *A*.

Опять же из условия неизменности объёма точки *A* следует, что радиальный расход массы в единицу времени $\Delta M_r$ и тангенциальный расход массы в единицу времени $\Delta M_{tg}$ равны по абсолютной величине. В противном случае  этой точки либо возрастал бы, либо уменьшался с течением времени. В математической форме эти условия можно записать в таком виде:

$$\begin{aligned} &\left|\Delta M_r\right| = \left|\Delta M_{tg}\right|, \text{ или} \\ &\Delta M_r - \Delta M_{tg} = 0. \end{aligned} \tag{25}$$

Из этих условий следует, что и радиальная $W_r$, и тангенциальная $W_{tg}$ линейные скорости в точке *A* тоже равны по абсолютной величине и противоположны по знаку.

4. Разместить полярную систему координат относительно точки *A* и провести исследование всего потока жидкости относительно движения жидкости в этой точке.

В точке *A* мы имеем две плоскости исследования, поэтому можем разместить систему координат в любой из этих плоскостей. В данном случае наиболее рациональным будет размещение системы координат в тангенциальной плоскости исследования $S_{tg}$, в которой располагается ось цилиндра потока и радиус исследуемой точки (рис. 9,*а*). Далее считаем, что полюс системы координат совпадает с осью потока, а радиусы точек линии тока являются координатными радиусами этой системы. Разместив систему координат, проводим исследование всех точек потока плоского установившегося вида движения, чтобы получить общие уравнения для него. В связи с тем, что все точки располагаются на линиях тока, то для каждой из них мы можем записать систему уравнений (24).

5. Сопоставить уравнения движения для каждой точки потока друг относительно друга, чтобы выявить общие уравнения движения плоского установившегося вида движения.

Сделав анализ уравнений движения во всех точках плоского установившегося потока, мы придём к выводу, что для одних точек радиальная и тангенциальная скорости будут иметь одни значения, то есть величины этих скоростей в одной точки будут равны величинам скоростей в другой точке. Для других же точек они имеют другие величины. При этом зависимости (24) остаются неизменными по своей структуре для всех точек потока. Это значит, что уравнения движения являются общими для линий тока в объёме плоского установившегося потока жидкости. Подобный вывод можно получить, основываясь на практическом опыте.

Уравнений (24) и (25) ещё недостаточно, чтобы можно было получить конкретные результаты движения. Для этого необходимо связать общий расход массы в единицу времени всего потока с расходом массы в каждой его точке. Мы знаем из опыта, что в плоском установившемся движении жидкости можно замерить и определить только общий для всего потока радиальный расход массы в единицу времени, т. к. этот расход является определяющим для него. Радиальный расход массы $M_r$ всего потока движется перпендикулярно оси цилиндра (рис. 8). Поверхностью движения для него служат цилиндрические поверхности всего объёма потока. Площадь цилиндрической поверхности определяется следующей зависимостью:

$$F_r = 2\pi rh. \tag{26}$$

Затем, умножив поверхность $F_r$ на плотность жидкости $\rho$ и радиальную скорость $W_r$, мы получим общее уравнение движения для всего потока в радиальном направлении такого вида:

$$M_r = 2\pi rh \cdot \rho W_r. \qquad (27)$$

Теперь определим связь уравнения (27) с уравнениями движения точек потока (24). В первую очередь это уравнение связано с уравнением радиального движения в точке потока. Выпишем его:

$$\Delta M_r = \rho W_r \Delta F_r. \qquad (28)$$

Затем возьмём окружность, вернее, цилиндр в объёме потока на некотором расстоянии $r$ (рис. 11). В связи с тем, что цилиндр образует радиус определённой величины, то радиальные скорости $W_r$ каждой точки этого цилиндра будут одинаковы и равны друг другу. В то же время общий радиальный расход массы $M_r$ при движении через поверхность этого цилиндра образует радиальную скорость $W_r$. При этом плотность жидкости для уравнений (27) и (24) одинакова из условий идеальной жидкости. Сумма поверхностей всех точек, удаленных от полюса $O$ на расстояние радиуса $r$, в пределе равна поверхности цилиндра с радиусом тоже $r$. Это условие можно записать так

$$F_r = \sum_{i=1}^{n} \Delta F_{ri} = 2\pi rh.$$

Общий радиальный расход массы $M_r$ в этом случае должен быть равен сумме радиальных расходов массы каждой точки поверхности цилиндра по условию постоянства объёма потока, то есть

$$M_r = \sum_{i=1}^{n} \Delta M_{ri}.$$

Воспользовавшись всеми этими равенствами для зависимостей (27) и (24), мы получим равенство радиальных скоростей потока, то есть

$$W_r = W_{ri}.$$

Тангенциальный массовый расход и тангенциальная скорость каждой точки цилиндра равны по абсолютной величине радиальному расходу массы и радиальной скорости этих же точек цилиндра. Следовательно, используя зависимости (24) и (27), мы сможем найти различные скорости и расходы массы, как для отдельных точек потока, так и для всего потока в целом.

Остаётся найти зависимости для линий тока. Мы знаем, что для плоского установившегося потока линия тока является кривой линией. Если мы возьмём на наружном цилиндре потока точку $A$ (рис. 11) и начнём перемещать её к центру цилиндра в соответствии с радиальной и тангенциальной скоростями, то получим определённую кривую линию, которая от периферии потока стремится к его оси.

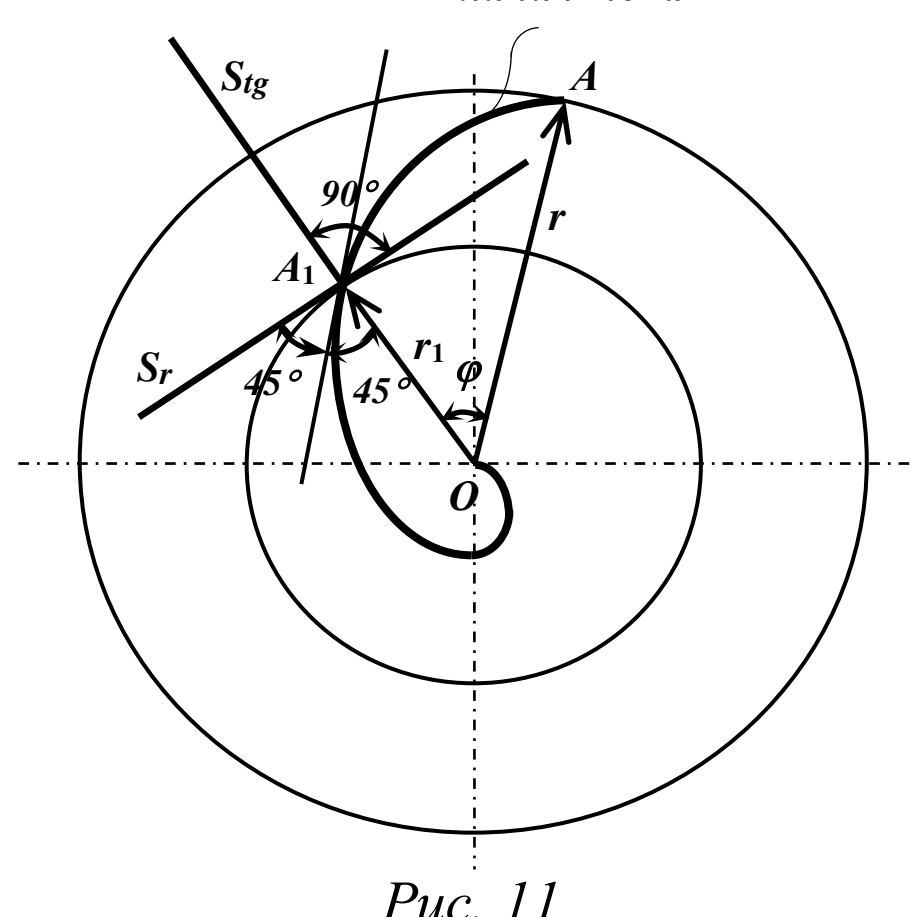

*Рис. 11*

По характеру этой линии можно определить, что она относится к разряду спиралей. Радиальные и тангенциальные скорости движения жидкости в каждой точке цилиндра располагаются во взаимно перпендикулярных плоскостях.

Например, возьмём точку $A_1$ (рис. 11) на линии тока и изобразим в этой точке радиальные и тангенциальные скорости. В связи с тем, что эти скорости равны между собой по абсолютной величине, то касательная линия к линии тока в точке $A_1$ должна располагаться под углом 45° к обеим скоростям движения. Радиальная скорость $W_1$ совпадает с направлением радиуса полярной системы координат. Поэтому касательная к линии тока тоже будет располагаться под углом 45° к радиусу полярной системы координат $r_1$. Так как равенство скоростей сохраняется как для всех точек потока в целом, так и для точек любой линии тока, то касательные в любой точке линии тока должны быть расположены под углом 45° к радиусу. Следовательно, линией тока должна быть такая кривая, которая бы пересекала под одним и тем же углом все лучи, выходящие из одной точки. Такой кривой может быть только логарифмическая спираль. Поэтому для плоского установившегося потока она и будет линией тока. Зависимость для неё возьмём из справочника [10]:

$$r = ae^{k\varphi}, \tag{28}$$

где $\varphi$ – угол полярной системы координат, $a$ – некоторая величина, относящаяся к конкретной спирали, $k = \text{ctg}\,\alpha$, в нашем случае угол $\alpha$ равен 45°, поэтому $\text{ctg}\,\alpha = 1$.
При $\alpha = \pi/2$ кривая превращается в окружность.

Мы получили все зависимости, которые определяют движение жидкости в плоском установившемся потоке.

Отметим, что ось цилиндра для всего объёма потока может быть не только прямой линией, но и цилиндром. Тогда объём всего потока примет форму тора. Все количественные зависимости, полученные для цилиндрического потока с прямолинейной осью, верны без каких - либо изменений и для торового потока.[9]

В итоге мы получили количественные зависимости плоского установившегося вида движения, которые должны быть воспринимаемы в связи с общим восприятием этого вида движения среды. То есть мы должны осознанно воспринимать влияние того или иного параметра на все остальные характеристики движения, чтобы иметь возможность управлять ими.

В связи с тем, что в настоящее время основные положения механики твёрдого тела считаются основными и для механики жидкости и газа, то отсутствие параметров времени и наличие параметров пространства в количественных зависимостях воспринимается как движение с постоянной во времени скоростью и криволинейной траекторией движения. Поэтому всякое изменение параметров в этих зависимостях ассоциируется, прежде всего, с изменением характера кривизны траектории движения, то есть с главным звеном движения, относительно которого уже рассматриваются изменения остальных характеристик движения. По этой причине подобное движение называется просто криволинейным движением.

Хотя плоское установившееся движение жидкости тоже характеризуется отсутствием параметра времени и наличием параметра пространства в зависимостях самого движения, но для него не приемлемо понятие механики твёрдого тела о подобном движении. В соответствии с количественными зависимостями плоского установившегося движения жидкости мы можем получить следующее восприятие этого вида движения:

а) что общий радиальный расход массы определяет объём всего потока, то есть по его величине можно судить о величине этого потока. При этом радиальный расход массы остаётся постоянным во времени;

б) что расход массы, как в отдельных точках потока, так и во всём потоке, движется в двух взаимно перпендикулярных направлениях. При этом радиальные и тангенциальные расходы массы равны друг другу;

[9] Ось потока не является обычной геометрической линией, поскольку механика безынертной массы оперирует объёмными точками, поэтому радиус может только стремиться к нулю.

в) что линейные скорости этих расходов определяются в зависимости от площади тока общего радиального расхода массы. При этом параметр пространства входит непосредственно в зависимость площади для радиального расхода массы;

г) что линией тока этого потока является логарифмическая спираль, которая пересекает все прямые линии, выходящие из одной точки, под углом $45°$ .

Основным звеном в этом восприятии является пункт *б*, который характеризует движение жидкости как движение расхода массы в двух взаимно перпендикулярных направлениях. Поэтому для плоского установившегося вида движения всякое изменение параметра пространства должно ассоциироваться с изменением скоростей радиального и тангенциального расходов массы, относительно которых затем увязываются изменения всех остальных характеристик потока, то есть площади движения общего радиального расхода массы и линия тока. Отсюда следует и название этого вида движения как плоского установившегося.

### III. 3. УРАВНЕНИЕ ДВИЖЕНИЯ РАСХОДНОГО ВИДА ДВИЖЕНИЯ ИДЕАЛЬНОЙ ЖИДКОСТИ

В соответствии с формальным принципом связи вида движения с формой уравнений неразрывности и движения в зависимостях данного вида движения должны отсутствовать параметры пространства. Это значит, что расход массы в единицу времени *M* является величиной, которая зависит от изменения параметров времени и не зависит от параметров пространства. Это условие можно записать в таком виде:

$$M = f(t),$$
$$M = f(r, \varphi) = \text{const}.$$

Таким образом, мы определились относительно основной характеристики этого вида движения жидкости.

Чтобы получить уравнения движения для него, мы должны связать расход массы в единицу времени с конкретными характеристиками потока жидкости. Расходное движение жидкости было выявлено в результате практического наблюдения. Поэтому нам известны некоторые качественные характеристики этого вида движения. Основываясь на них, сделаем его описание:

жидкость в этом виде движения может двигаться во всех направлениях. При этом границы объёма *V* всего потока могут быть либо неподвижными в пространстве, либо перемещаться в нём. Соответственно общий объём потока *V* может быть либо постоянным во времени, либо переменным. Поток жидкости может быть любой формы и конфигурации. В каждый момент времени поток содержит в своём объёме определённое количество массы *m* плотностью $\rho$. Далее мы должны воспользоваться общим методом исследования. В соответствии с ним мы должны:

1. выделить неподвижное пространство среды.

Это пространство должно включать в себя либо весь поток жидкости, либо его характерную часть. При этом его границы не должны изменять или как-то влиять на движение жидкости. Это значит, что границы среды должны совпадать с границами расходного потока жидкости. Исходя из наиболее общих условий движения, мы должны принять границы потока перемещающимися в пространстве с течением времени.

2. Расположить или выбрать плоскость (поверхность) для первоначального исследования. Для чего поступим следующим образом:

а) в определённый фиксированный момент времени *t* принимаем границы потока в пределах границ, которых он достиг к этому времени;

б) в соответствии с общим методом исследования плоскость (поверхность) для первоначального исследования в каждой своей точке должна располагаться перпендикулярно к линейной скорости движения.

Располагая поверхность для первоначального исследования по этому принципу, в общем случае мы получим некоторую замкнутую поверхность $S_t$ какой-то формы для фиксированного момента времени *t* (рис. 12).

Замкнутая поверхность $S_t$ выделяет из среды определённый объём $V_t$, в котором содержится жидкость с плотностью $\rho$. Тогда массу *m* выделенного объёма можно записать как произведение объёма $V_t$ на плотность $\rho$, то есть

$$m_t = V_t \rho. \qquad (29)$$

Будем считать, что мы выполнили работу по второму пункту.

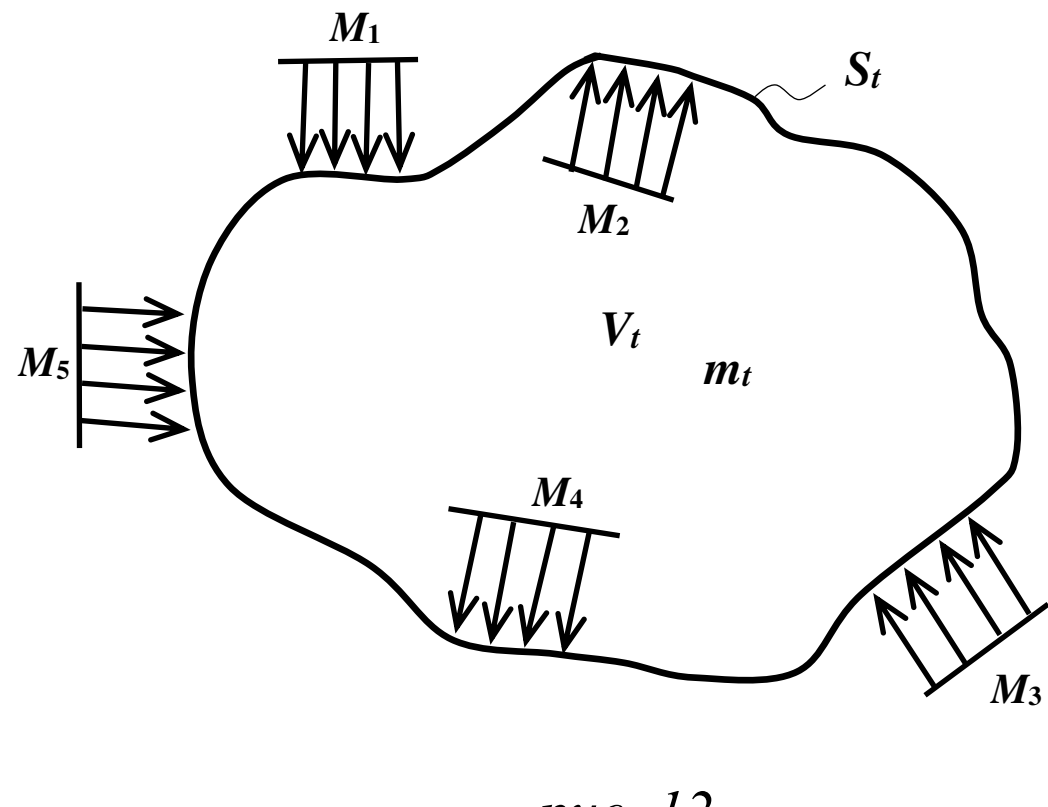

*рис. 12*

3. Изобразить движение жидкости на поверхности $S_t$ потока жидкости и записать для него соответствующие зависимости.

Мы записали некий фиксированный $V_t$ потока жидкости, но от этого поток не прекратил своего движения. Чтобы записать его движение, нам придётся воспользоваться зависимостью (29). При движении жидкости в фиксированном объёме будут изменяться масса жидкости, содержащаяся в этом объёме, и сам фиксированный объём. Плотность жидкости $\rho$ будет оставаться постоянной величиной, т. к. по условию идеальной жидкости она должна оставаться неизменной.

Фиксированный объём, содержащаяся в нём жидкость и плотность жидкости относятся к характеристикам потока, изменение которых должно соответствовать изменению расхода массы в единицу времени $M_t$. Расход массы в единицу времени либо поступает, либо вытекает, либо одновременно поступает и вытекает через поверхность $S_t$. Выше мы установили, что за положительный расход массы принимаем расход, который поступает в объём потока, а за отрицательный – который вытекает из него. Это условие остаётся и для расходного вида движения жидкости.

Из-за непрерывного движения расхода массы через границы объёма $V_t$ количество массы $m$ жидкости в нём будет непрерывно изменяться, а вместе с этой количественной величиной будут изменяться границы объёма и сам объём. Чтобы выразить это условие, возьмём первую производную по времени из уравнения (29), получим:

$$\frac{dm_t}{dt} = \rho \frac{dV_t}{dt}. \tag{30}$$

$dm_t/dt$ в зависимости (30) есть ничто иное, как расход массы в единицу времени $M_t$. Подставим его в это уравнение, получим:

$$M_t = \rho \frac{dV_t}{dt}. \tag{31}$$

Исходя из условия движения жидкости в самом общем случае её движения, то есть когда происходит одновременное поступление и вытекание некоторых расходов массы $M_i$ по различным участкам поверхности $S_t$ (рис. 12), расход массы $M_t$ в уравнении (31) является результирующим, или суммарным расходом, который в каждый момент времени происходит в объёме $V_t$. Это условие можно записать следующим образ

$$M_t = \sum_{i=1}^{n} M_i - \sum_{i=1}^{n} M_j. \tag{32}$$

Подставив это равенство в уравнение (31), мы получим уравнение движения для объёма $V_t$, выраженное через суммы положительных и отрицательных расходов, которые поступают через поверхность $S_t$, то есть

$$\rho \frac{dV_t}{dt} = \sum_{i=1}^{n} M_i - \sum_{j=1}^{n} M_j . \tag{33}$$

Уравнение (33) является общим уравнением движения расходного вида движения жидкости для потока, выделенного первоначальной поверхностью исследования.

4. Разместить полярную систему координат относительно поверхности $S_t$ и провести исследование всего потока жидкости относительно её движения на этой поверхности.

В связи с тем, что в самом общем случае поверхность $S_t$ является переменной во времени и может быть криволинейной, то мы не сможем разместить полярную систему координат относительно этой поверхности. В общем, делать этого не следует, т. к. расходный вид движения не зависит от параметров пространства.

5. Сопоставить уравнения движения каждой поверхности исследования относительно друг друга, чтобы выявить общее уравнение движения расходного вида движения.

Для расходного вида движения объём $V_t$, ограниченный плоскостью $S_t$ первоначального исследования, может занимать как бесконечно малый объём среды, так и всю среду в целом. Как раз в этом выражается независимость уравнения движения расходного вида движения от параметра пространства. Это значит, что уравнение (33) является общим уравнением движения для любого расходного движения.

*Примечание*: при исследовании, например, газов в расходном виде движения плотность $\rho$ является переменной величиной, т. к. газ сжимаем. Если теперь принять переменными величинами и объём $V_t$, и плотность $\rho_t$ для зависимости (29), то после дифференцирования и замены результирующего расхода массы $M_t$ на сумму расходов масс с различными знаками, как для уравнения (33), получим:

$$\rho \frac{dV_t}{dt} + V \frac{d\rho_t}{dt} = \sum_{i=1}^{n} M_i - \sum_{j=1}^{n} M_j . \tag{34}$$

Уравнение движения (34) является более общим уравнением по отношению к уравнению (33), т. к. оно учитывает изменение плотности $\rho_t$. Подобное примечание сделано для того, чтобы не делать специального вывода уравнения (34) при исследовании реальных жидкостей и газов.

В итоге мы получили количественные зависимости расходного вида движения, которые должны быть увязаны с общим восприятием этого вида движения, то есть мы должны осознанно воспринимать влияние того или иного параметра на все остальные характеристики движения, чтобы иметь возможность управлять им.

В связи с тем, что положения механики твёрдого тела в настоящее время считаются основными и для механики жидкости и газа, то отсутствие параметров пространства и наличие параметров времени в количественных зависимостях воспринимается как прямолинейное движение с переменной скоростью. Поэтому всякое изменение параметров времени в этих зависимостях ассоциируется, прежде всего, с изменением скорости, то есть с ускорением, которое является главным звеном этого вида движения в механике твёрдого тела. По отношению к этому звену рассматриваются изменения всех остальных характеристик движения. И по этой причине подобное движение среды называется просто переменным движением.

Хотя расходное движение жидкости тоже характеризуется отсутствием параметров пространства и наличием параметров времени в зависимостях своего движения, но для него неприемлемо понятие механики твёрдого тела о подобном движении. В соответствии с количественными зависимостями расходного вида движения жидкости мы можем получить следующее восприятие этого движения:

что общий суммарный расход массы в единицу времени через поверхность $S_t$ зависит от изменения объёма потока $V_t$ во времени, а при движении газа зависит ещё от изменения плотности $\rho_t$. Основным звеном в этом восприятии является изменение объёма потока, то есть это звено характеризуется движением жидкости в виде изменения объёма потока жидкости. Поэтому для расходного вида движения изменение параметра времени должно ассоциироваться с изменением объёма потока, а при движении газа ещё с изменением плотности, относительно которых затем увязываются изменения всех остальных характеристик потока.

В соответствии с основным звеном движения этот вид движения следовало бы назвать *объёмным* видом движения, а не расходным. Но названием «расходный» подчёркивается особенность этого вида движения, связанная с его восприятием с позиции действия сил и с главной характеристикой движения – расходом массы в единицу времени. Поэтому данный вид движения можно называть *объёмным или расходным видом движения*.

Отметим, что в современной механике уравнение движения (33) известно под названием баланса масс. Его применяли как самостоятельную количественную зависимость при исследовании колебательных движений в проточных объёмах [8], [9]. При этом не применяли соответствующих этой зависимости уравнений сил. Поэтому авторы подобных исследований не могли получить верных результатов.

## III. 4. УРАВНЕНИЯ ДВИЖЕНИЯ АКУСТИЧЕСКОГО ВИДА ДВИЖЕНИЯ ИДЕАЛЬНОЙ ЖИДКОСТИ

В соответствии с формальным принципом связи вида движения с формой уравнений неразрывности и движения в зависимостях данного вида движения должны присутствовать параметры пространства и времени. Это значит, что расход массы в единицу времени $M$ является величиной, которая зависит от изменения параметров пространства и времени. Это условие можно записать в таком виде:

$$M = f(t, r, \varphi).$$

Таким образом, мы определились относительно основной характеристики этого вида движения жидкости.

Чтобы получить уравнения движения, мы должны связать расход массы в единицу времени с конкретными характеристиками потока жидкости.

Акустическое движение жидкости было выявлено в результате практических наблюдений. Поэтому нам известны некоторые качественные характеристики этого вида движения. Основываясь на них, сделаем для него описание.

Местом существования акустического вида движения служит жидкость, которая находится под определённым силовым воздействием, то есть она имеет некоторое давление $P$. Акустическое движение жидкости возникает от действия на эту жидкость определённого источника механического движения. Например, таким источником может быть металлическая пластина, которая периодически совершает возвратно-поступательное движение с определённой частотой. Движение этой пластины направлено перпендикулярно её плоскости. Под воздействием пластины в жидкости возбуждаются определённые возмущения, которые характеризуются периодическим изменением параметров с частотой, совпадающей с частотой колебания пластины.

Эти возмущения образуют фронт, то есть плоскость, точки которой в любой момент времени имеют одинаковые характеристики. Фронт возмущения располагается перпендикулярно направлению движения пластины, то есть параллельно плоскости пластины. Скорость движения фронта возмущения является определённой постоянной величиной для каждой жидкости. Она зависит от физических свойств массы жидкости.

Непосредственно само возмущение образует в жидкости так называемую волну, которая характеризуется тем, что все её точки имеют различные характеристики в любой момент времени. Скорость движения жидкости непосредственно в самой волне отличается по направлению и величине от скорости движения фронта возмущения.

Акустическая волна сохраняет свою форму при движении в пространстве. При своём движении эти волны удаляются от пластины на некоторое расстояние и тем самым охватывают некоторый объём жидкости $V$, которая имеет плотность $\rho$. Этот объём будем считать полным объёмом потока акустического вида движения жидкости.

Отметим, что металлическая пластинка может иметь поверхность отличную от плоскости.

Далее мы должны воспользоваться общим методом исследования. В соответствии с ним мы должны:

1. выделить неподвижное пространство среды. Это пространство должно включать в себя либо весь поток жидкости, либо его характерную часть. При этом границы его не должны изменять или как-то влиять на движение жидкости. Это значит, что границы среды должны совпадать границами акустического потока жидкости. Исходя из этих условий, мы будем действовать следующим образом:

принимаем за источник возмущения среды плоскую пластину с неограниченной площадью, которая совершает возвратно-поступательное движение в пределах длины $l$ (рис. 13).

Считаем, что эта пластина возбуждает возмущения в жидкости с направлением движения по стрелке *Б*. Линия *Б* совпадает с направлением движения пластины и располагается перпендикулярно к её плоскости.

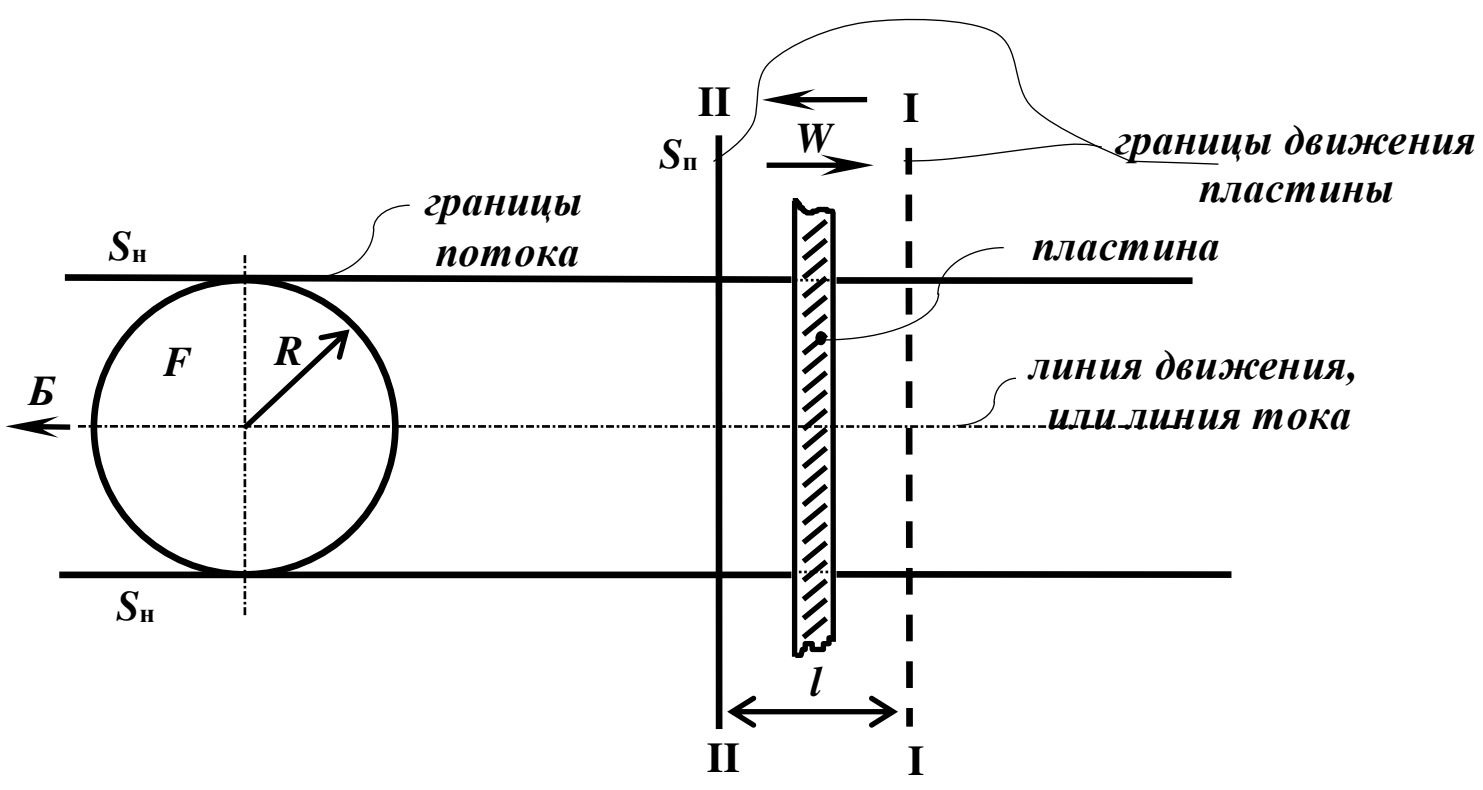

*Рис. 13*

В связи с тем, что пластина не ограничена в своей площади, нам придётся взять в качестве среды характерную часть потока. С этой целью принимаем линию *Б* за линию тока. Это значит, что характерный участок потока должен располагаться симметрично относительно этой линии, как при установившемся виде движения. Поэтому выделенный объём потока жидкости будет иметь форму цилиндра, осью которого является линия *Б*, она же – линия тока. Поперечным сечением цилиндра будет окружность с некоторой площадью $F$. Считаем, что границы среды совпадают с границами этого цилиндра.

2. Расположить или выбрать плоскость (поверхность) для первоначального исследования.

В связи с тем, что движение акустического потока жидкости зависит от параметра пространства, то его расход массы в единицу времени должен двигаться в двух взаимно перпендикулярных направлениях. Это условие следует из плоского установившегося вида движения. Поэтому для рассматриваемого вида движения мы должны иметь две взаимно перпендикулярные плоскости для первоначального исследования. Направление одного расхода массы в единицу времени мы знаем: он совпадает с поступательным движением пластины. Для этого исследования расположим плоскость $S_п$ перпендикулярно скорости поступательного движения пластины (рис. 13). Вторая плоскость должна быть расположена перпендикулярно плоскости $S_п$. Такой поверхностью является поверхность границы среды. Принимаем её за вторую поверхность $S_н$ первоначального исследования. Будем считать, что на этом работы по второму пункту закончены.

3. Изобразить движение жидкости на поверхностях $S_п$ и $S_н$ и записать для них соответствующие зависимости.

Помимо того, что расход массы в единицу времени зависит от параметров пространства, он ещё зависит от параметров времени. Это значит, что акустический вид движения, скажем так, совмещает в себе в определённом сочетании *плоское установившееся движение с расходным объёмным движением.* Чтобы разъяснить это условие, сделаем рисунок 14 с некоторым отступлением от рисунка 13.

Это отступление будет заключаться в том, что в выделенном участке потока акустического движения жидкости вообразим пластину как поршень с площадью $F$, движущийся в цилиндре, длина которого равна длине пути возвратно-поступательного движения $l$ пластины (рис. 14). Теперь рассмотрим ход этого поршня поэтапно.

На первом этапе поршень движется из одного крайнего положения в другое в направлении распространения возмущения жидкости (рис. 14, *а*) с некоторой скоростью $W_{пор}$. При этом движении он будет вытеснять объём жидкости $V_0$, которая, в свою очередь, будет вытекать из этого объёма через плоскость $S_п$. В этом случае расход массы в единицу времени $M_0$ мы можем получить с помощью уравнения движения (33) расходного вида движения как

$$M_0 = \rho \frac{dV_0}{dt}. \qquad (35)$$

Заменим в зависимости (35) объём $V_0$ на произведение площади поршня $F$ и длины хода поршня $l$, тогда получим:

$$M_0 = \rho F \frac{dl}{dt}. \qquad (36)$$

Мы изобразили объём $V_0$ на рисунке 14, *а* чисто символически. В противном случае нам бы пришлось совмещать плоскость $S_п$ с плоскостью поршня и рассматривать движение в каждый определённый фиксированный момент времени. В этом случае мы просто усложнили бы решение задачи, что привело бы к определённым трудностям в понимании акустического вида движения жидкости.

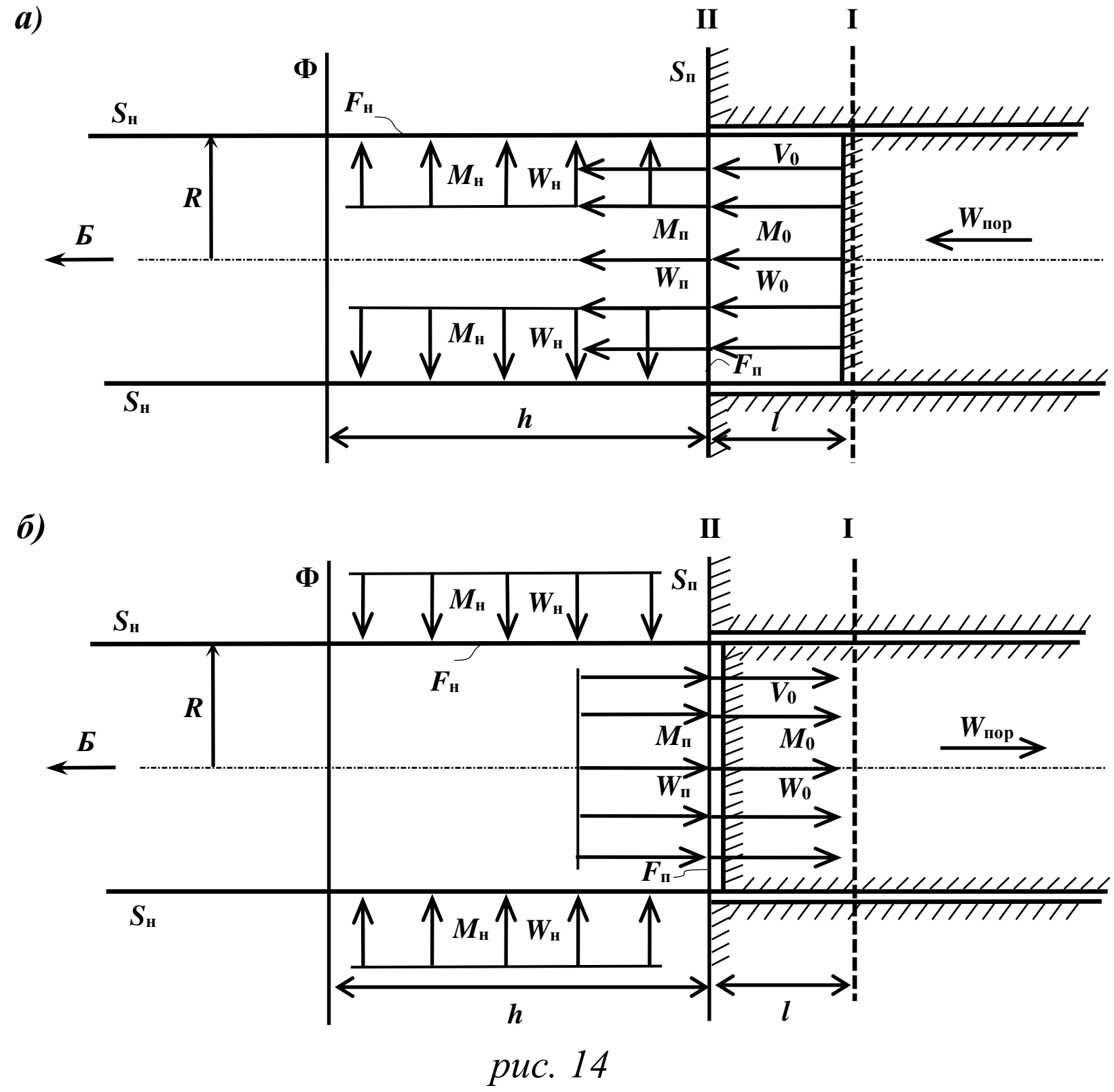

*рис. 14*

В связи с тем, что объём $V_0$ является символическим, то и расход массы в единицу времени $M_0$ тоже является символическим, но из условия рисунка 14,*а* он равен расходу массы поступательного движения $M_п$, то есть

$$M_0 = M_п. \qquad (37)$$

Это значит, что расход массы $M_0$ за плоскостью $S_п$ переходит в равноценный расход массы $M_п$.

К моменту начала движения поршня перед плоскостью $S_п$, как и перед плоскостью поршня, находится жидкость в состоянии покоя. Поэтому расход массы $M_п$, поступающий через площадь $F_п$ плоскости $S_п$, тоже будет вытеснять соответствующее количество жидкости с расходом массы $M_н$. Вытесненный расход массы $M_н$ будет вытекать через поверхность $S_н$ в направлении перпендикулярном скорости движения расхода массы $M_п$. Из условия сохранения постоянства объёма потока при постоянной величине плотности $\rho$, расход массы $M_п$ поступательного движения будет равен расходу массы $M_н$ нормального движения жидкости, то есть

$$M_{\text{п}} = M_{\text{н}}. \tag{38}$$

Теперь выразим нормальный расход массы $M_{\text{н}}$ через характеристики потока. В соответствии с уравнением движения (22) установившегося вида движения он будет равен:

$$M_{\text{н}} = \rho F_{\text{н}} W_{\text{н}}. \tag{39}$$

Нормальная площадь $F_{\text{н}}$ является частью цилиндрической поверхности $S_{\text{н}}$ (рис. 14, *а*). Тогда эта площадь будет равна произведению длины окружности цилиндра на некоторую высоту $h$, то есть

$$F_{\text{н}} = 2\pi R h. \tag{40}$$

Радиус цилиндрической поверхности $R$ мы приняли сами для своих исследований. Поэтому величина его нам известна. Некоторая высота $h$ своё начало берёт от плоскости $S_{\text{п}}$. Полная её величина нам пока неизвестна. Поэтому её требуется найти.

Из практики известно, что возмущения в жидкости распространяются с определённой скоростью $C$, которая зависит от физических свойств массы. Эта скорость называется ещё скоростью звука. С началом движения поршня из положения I от плоскости $S_{\text{п}}$ начнёт распространяться возмущение со скоростью $C$ вдоль линии тока. Начиная с этого момента, путь, пройденный возмущением от плоскости $S_{\text{п}}$, будет увеличиваться с течением времени. В зоне возмущения одновременно происходит поступательное и нормальное движение. По этой причине величина пути возмущения и высота цилиндрической площади $h$ нормального движения жидкости в любой момент времени должны совпадать по величине. Тогда высоту $h$ можно записать как произведение скорости звука $C$ на время $t$:

$$h = C \cdot t. \tag{41}$$

Так как рассматриваемый этап движения у нас ограничивается движением поршня из одного крайнего положения в другое на длине хода поршня равной $l$, то время прохождения этапа $t_{\text{э}}$ определяется скоростью движения поршня $W_{\text{пор}}$. Тогда время будет равно:

$$t_{\text{э}} = \frac{l}{W_{\text{пор}}}. \tag{42}$$

Скорость движения поршня на этапе может быть любой: постоянной, равномерно ускоренной, переменной и т.д. Она непосредственно зависит от самого источника возмущения жидкости. Поэтому в общем случае принимаем, что скорость поршня на этапе движения переменная во времени. Обозначим её как $W_{\text{пор}}(t)$ и подставим в зависимость (42), получим:

$$t_{\text{э}} = \frac{l}{W_{\text{пор}}(t)}. \tag{43}$$

Если мы в уравнение (41) подставим время первого этапа движения поршня, то получим длину $h_{\text{э}}$ возмущённого участка за это время. Запишем её:

$$h_{\text{э}} = C \cdot t_{\text{э}} = C \frac{l}{W_{\text{пор}}(t)}. \tag{44}$$

Затем заменим в уравнении (40) высоту $h$ через зависимость (44) и подставим его в уравнение (39), получим:

$$M_{\text{н}} = \rho W_{\text{н}} \cdot 2\pi R \cdot C \frac{l}{W_{\text{пор}}(t)}. \tag{45}$$

Уравнение (45) есть уравнение движения акустического вида движения жидкости, которое описывает её движение в нормальном направлении. В таком виде оно охватывает движение непосредственно в конце первого этапа, когда поршень находится на бесконечно близком расстоянии от плоскости $S_п$. Если не сделать подобной оговорки, а формально подойти к уравнению (45), то мы заметим, что в этом уравнении стоит полное время первого этапа $t_э$, выраженное через характеристики потока. Это значит, что за это время поршень должен прийти в положение II и остановить своё движение.

При остановке движения поршня из объёма $V_0$ не будет поступать расход массы $M_п$, то есть он превратится в ноль. Поэтому мы были бы обязаны приравнять уравнение (45) к нулю. При оговорке, что поршень находится на бесконечно малом расстоянии от плоскости $S_п$, мы можем оставить уравнение (45) в том виде, в котором оно записано.

Далее мы должны определить знаки для поступательного и нормального расходов массы жидкости. Согласно рисунку 14,*а*, нормальный расход вытекает из объёма потока, а поступательный втекает в него. Тогда уравнения (36) и (45) мы должны записать в таком виде:

$$M_п = \rho F_п \frac{dl}{dt}. \tag{46}$$

$$M_н = -\rho W_н \cdot 2\pi R \cdot C \frac{l}{W_{пор}(t)}. \tag{47}$$

Это окончательные уравнения движения акустического потока на первом этапе движения поршня. Отметим, что в уравнении (46) $dl/dt$ есть ни что иное, как скорость движения поршня, то есть $dl/dt = W_{пор}(t)$.

Теперь рассмотрим движение жидкости непосредственно за весь период движения поршня по участку I – II с длиной хода поршня $l$. Это движение будем рассматривать на примере, когда поршень движется из положения I в положение II с постоянной скоростью $W_{пор} = \text{const}$. Для этого сделаем рисунок 15,*а*.

Изобразим на нём тот же поршень, что и на рисунке 14,*а*. Затем на линии тока от плоскости $S_п$ отложим отрезок прямой, который в соответствующем масштабе равен времени первого этапа движения поршня $t_э$. Получим точку *О*. Нормально к линии тока через точку *О* проведём прямую, которая будет осью скоростей. Время начала движения, или ноль времени и ноль скорости, совпадает с точкой *О*. Изобразим скорости движения жидкости за время первого этапа.

При движении поршня из положения I в положение II с постоянной скоростью, поступательный расход массы в единицу времени тоже будет постоянной величиной, то есть $M_п = \text{const}$. Так как нормальный расход массы $M_н$ равен поступательному расходу массы $M_п$, то он тоже будет равен постоянной величине, то есть $M_н = M_п = \text{const}$. Затем разделим ход поршня $l$ на более мелкие участки величиной $\Delta l$ и рассмотрим движение на каждом из этих участков. Полученные скорости движения в каждый соответствующий момент времени нанесём на оси.

***Поступательная скорость движения***

Поступательный расход потока движется в направлении линии тока. Площадь потока $F_п$ на всём протяжении – постоянная величина. Выше мы приняли объём потока цилиндрическим. Воспользовавшись зависимостью (46), постараемся определить поступательную скорость движения жидкости $W_п$ для каждого участка $\Delta l_i$. В связи с тем, что в этой зависимости расход массы, плотность, площадь потока и скорость движения поршня - постоянные величины, то и поступательная скорость для каждого участка тоже будет постоянной величиной за весь период времени первого этапа. Отложим величину этой скорости на соответствующей оси на рисунке 15,*а*. Отсюда следует, что поступательная скорость $W_п$ акустического потока при постоянной скорости движения поршня есть величина постоянная.

***Нормальная скорость движения***

Нормальный расход массы $M_н$ движется перпендикулярно линии тока. Для определения скорости движения по каждому участку хода поршня воспользуемся зависимостью (47). В данной зависимости нам известны все величины, кроме нормальной скорости. Чтобы определить её, будем подставлять в эту зависимость вместо длины хода поршня $l$ длину его участков в такой последовательности: сначала подставим длину первого участка, затем длину первого и второго участков вместе, затем длину первого, второго и третьего участков вместе и так далее.

Соответственно будем получать нормальные скорости движения для каждого отрезка времени. Нанесём их на оси на рисунке 15, *а*. На первом участке мы получим максимальную скорость, на последнем – минимальную скорость.

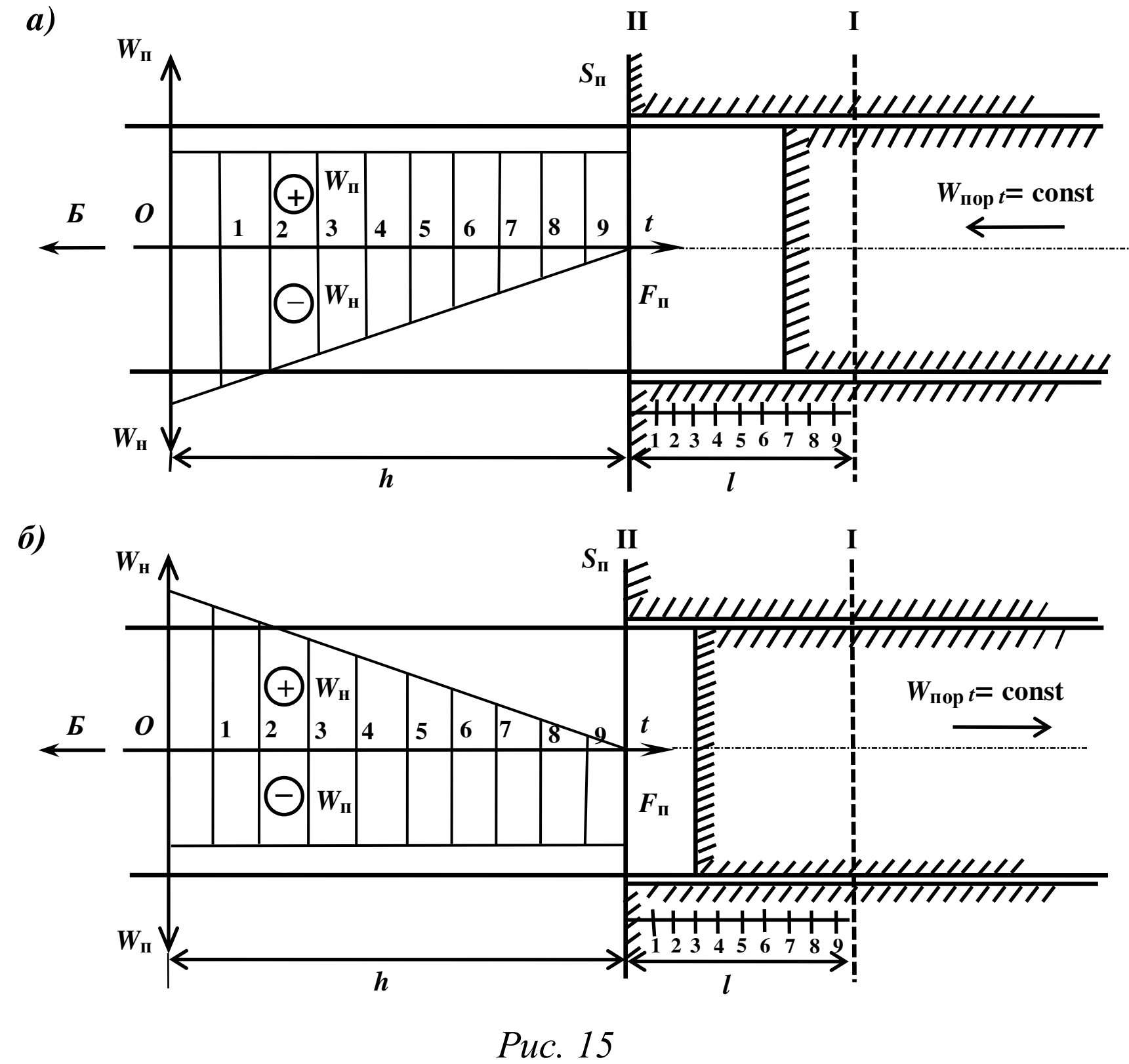

*Рис. 15*

Убывание скоростей по участкам связано с тем, что с ростом числа участков увеличивается площадь проходного сечения для нормального расхода массы, а сам расход остаётся постоянным. Ведь с ростом числа участков увеличивается время движения поршня. И в это же время происходит движение возмущённой жидкости. Поэтому, чем больше число одинаковых участков, тем больше длина возмущённого участка по зависимости (44), а от его длины $h$ зависит величина проходной площади нормального расхода массы $M_н$.

В соответствии с уравнениями (40) и (47) мы получили распределение нормальных скоростей $W_н$ на оси времени рисунка 15,*а*. Как видно, величина нормальной скорости прямо пропорциональна количеству времени движения поршня.

Мы рассмотрели пример, в котором поршень двигался с постоянной скоростью, но в общем случае он может двигаться с различной скоростью на длине $l$ хода поршня. Тогда будет необходимо учесть изменение нормальных и поступательных расходов массы и скоростей в соответствии с его движением. Все эти изменения определяются за время движения поршня на длине $l$ его хода либо аналитическим путем, либо каким-нибудь другим математическим методом.

**Теперь рассмотрим второй этап**. На втором этапе поршень будет двигаться из положения II в положение I (рис. 14,*б*) с некоторой скоростью $W_{пор}$. При этом движении поршня жидкость будет втекать в объём $V_0$ через площадь $F_п$ плоскости $S_п$ под действием наружных сил давления среды, так как при таком своём движении он не может непосредственно воздействовать на жидкость, как на первом этапе движения.

На площади $F_п$ плоскости $S_п$ образуется поступательный расход массы $M_п$, который мы можем получить с помощью уравнения движения (33) объёмного вида движения как:

$$M_п = \rho \frac{dV_0}{dt}. \qquad (48)$$

Заменим в зависимости (48) объём $V_0$ на произведение площади поршня $F_п$ на длину хода поршня $l$, тогда получим:

$$M_{\text{п}} = \rho F_{\text{п}} \frac{dl}{dt}. \quad (49)$$

Поступающий в объём $V_0$ расход массы $M_{\text{п}}$ в объёме потока будет компенсироваться нормальным расходом массы $M_{\text{н}}$, который будет поступать через плоскость $S_{\text{н}}$ перпендикулярно линии тока. Поэтому нормальный расход массы $M_{\text{н}}$ должен быть равен поступательному расходу массы $M_{\text{п}}$, то есть $M_{\text{н}} = M_{\text{п}}$. Так же, как и на первом этапе движения, запишем расход массы $M_{\text{н}}$ через характеристики потока с помощью уравнения (22) установившегося вида движения. Получим:

$$M_{\text{н}} = \rho F_{\text{н}} W_{\text{н}}. \quad (50)$$

Далее мы не будем подробно перечислять всех преобразований, связанных с дальнейшим выводом уравнений движения. Просто отметим, что площадь цилиндра $F_{\text{н}}$ для уравнения (50) получим по уравнению (40), то есть выразим её через радиус цилиндра $R$ и высоту цилиндра $h$, а высоту цилиндра запишем через скорость звука $C$ и время $t$ по уравнению (41). Время $t$ возьмём как время хода поршня из положения II в положение I по уравнению (42). Учитывая, что скорость движения поршня $W_{\text{пор}}$ может быть различной, запишем её как переменную во времени, тогда придём к уравнению (43). Проведя все перечисленные преобразования и подставив их в уравнение (50), получим:

$$M_{\text{н}} = \rho W_{\text{н}} \cdot 2\pi R \cdot \frac{l}{W_{\text{пор}}(t)} C \quad (51)$$

Мы получили уравнение движения для нормального потока жидкости акустического вида движения. В таком виде оно охватывает движение непосредственно в конце второго этапа движения, когда поршень находится на бесконечно малом расстоянии от плоскости положения I. Подобная оговорка, как и для первого этапа движения поршня, необходима для того, чтобы нормальный расход массы не был равен нулю.

Теперь определим знаки уравнений (49) и (51). Согласно рисунку 14,*б*, нормальный расход поступает в объём потока, а поступательный вытекает из него. Поэтому уравнения (49) и (51) второго этапа движения поршня будут отличаться от уравнений движения первого этапа только знаками. Отметим, что в уравнении (49) $dl/dt$ есть скорость движения поршня, то есть $dl/dt = W_{\text{пор}}(t)$.

**Теперь рассмотрим движение жидкости непосредственно за весь период движения поршня по участку II – I длиной *l*.** Это движение будем рассматривать на примере, когда поршень движется из положения II в положение I с постоянной скоростью $W_{\text{пор}} = \text{const}$. Для этого сделаем рисунок 15,*б*.

Изобразим на нём всё тот же поршень, что и на рисунке 14,*б*. Затем на линии тока от плоскости $S_{\text{п}}$ отложим отрезок прямой, который в соответствующем масштабе равен времени второго этапа движения поршня $t_{\text{э}}$. Получим точку *O*. Нормально к линии тока через точку *O* проведём прямую, которая будет осью скоростей. Время начала движения, или ноль времени и ноль скорости, совпадает с точкой *O*. Изобразим скорости движения жидкости во время второго этапа.

При движении поршня из положения II в положение I с постоянной скоростью поступательный расход массы тоже будет постоянной величиной ($M_{\text{п}} = \text{const}$). Так как нормальный расход массы $M_{\text{н}}$ равен поступательному расходу массы $M_{\text{п}}$, то он будет равен постоянной величине, то есть $M_{\text{п}} = M_{\text{н}} = \text{const}$.

Затем разделим ход поршня $l$ на более мелкие равные участки, величиной $\Delta l_i$, и рассмотрим движение на каждом из этих участков. Полученные скорости движения в каждый соответствующий момент времени нанесём на оси.

***Поступательная скорость движения***

Поступательный расход массы потока движется в направлении линии тока. Площадь потока $F_{\text{п}}$ на всём её протяжении постоянна по величине, т. к. элемент потока у нас имеет цилиндрическую форму.

Воспользовавшись зависимостью (49), постараемся определить поступательную скорость движения жидкости $W_п$ для каждого участка хода поршня $\Delta l_i$. В связи с тем, что в этой зависимости расход массы, плотность и площадь потока, а также длина хода поршня и скорость его движения $W_{пор}$ являются постоянными величинами, то поступательная скорость для каждого участка $\Delta l_i$ тоже будет постоянной величиной за весь период времени второго этапа. Отложим величину этой скорости на соответствующей оси на рисунке 15,*б*. Отсюда следует, что поступательная скорость $W_п$ акустического потока при постоянной скорости движения поршня есть величина постоянная.

***Нормальная скорость движения***

Нормальный расход массы движется перпендикулярно линии тока. Для определения скорости движения по каждому участку хода поршня $\Delta l_i$ воспользуемся зависимостью (51). В данной зависимости нам известны все величины, кроме нормальной скорости. Чтобы определить её, будем подставлять в эту зависимость вместо длины хода поршня $l$ длину его участков в такой последовательности: сначала подставим длину первого участка, затем первого и второго вместе, затем длину первого, второго и третьего участков вместе, и так далее. Соответственно будем получать нормальные скорости для каждого участка. Нанесём их на оси на рисунке 15,*б*. На первом участке получим максимальную скорость, на последнем – минимальную. Убывание скорости по участкам, вернее, по времени связано с тем, что с ростом числа участков увеличивается площадь проходного сечения для нормального расхода массы, а сам расход остаётся постоянным. Ведь с ростом числа участков увеличивается время движения поршня. В это же время происходит движение возмущённого участка жидкости. Поэтому чем больше число одинаковых участков, тем больше длина пути возмущённого участка по зависимости (44), а от его длины $h$ зависит величина проходной площади нормального потока $M_н$. В соответствии с уравнениями (40) и (51), мы получаем распределение нормальных скоростей $W_н$ на оси времени рисунка 15,*б*. Как видно, величина нормальной скорости прямо пропорциональна количеству времени движения поршня.

Отметим, что мы рассмотрели пример, где поршень движется с постоянной скоростью, но в общем случае он может двигаться с любой переменной скоростью. Тогда учитывается изменение всех расходов и скоростей с учётом переменной скорости движения поршня на длине его хода.

Мы рассмотрели движение жидкости, заменив колебательное движение пластины источника возмущения акустического движения возвратно-поступательным движением поршня. Чтобы получить полную количественную картину движения жидкости в акустическом потоке, мы должны рассмотреть движение жидкости от непосредственного воздействия пластины источника возбуждения. Для чего мы совместим поступательную плоскость $S_п$ первоначального исследования с плоскостью пластины источника возбуждения (рис. 16), а остальные характеристики акустического потока оставим такими же, какими мы их приняли для рисунков 14 и 15.

Принимаем, что пластина совершает возвратно-поступательное движение в пределах длины $l$ её хода. В акустике подобные движения пластины называются колебательными движениями, так как они характеризуются двумя этапами движения. При перемещении пластины на первом этапе плоскость пластины переместится из положения I в положение II с некоторой скоростью $W_{пор}(t)$ и пройдёт путь длиной $l$ за время соответствующее этому этапу $t_э{}^{I}$. В это время через плоскость исследования $S_п$ будет идти некоторый расход массы в единицу времени. Плоскость $S_п$ совмещена с плоскостью пластины и совершает вместе с ней совместное движение из положения I в положение II. Так как движение поступательного расхода массы $M_п$ происходит на площади $F_п$ плоскости $S_п$, которая равна площади сечения цилиндра выделенного участка потока жидкости, то есть площади окружности цилиндра с радиусом $R$ выделенного объёма потока, то в этом случае движение пластины характеризует изменение а потока, то есть изменение объёма зависит от скорости движения пластины. Тогда в уравнении движения первого этапа (46) мы обязаны заменить $dl/dt$ на скорость движения пластины $W_{пор}(t)$, вернее, отношение $dl/dt$ равнозначно скорости движения пластины, то есть для уравнения (46) мы можем применять любое из этих выражений.

В это время нормальный расход массы $M_н$ будет двигаться через нормальную поверхность $S_н$ потока. Площадь проходного сечения для этого расхода зависит от времени движения пластины и скорости распространения возмущения. В соответствии с уравнением (40), нормальная площадь $F_н = 2\pi Rh$, а высота $h$ равна произведению скорости возмущения $C$ жидкости на время $t$, то есть $h =$ $C\cdot t$. Скорость возмущения $C$ жидкости не зависит непосредственно от скорости движения пластины. Нам известно, что она зависит от физических свойств жидкости и её величина больше скорости движения пластины. Поэтому, когда пластина начнёт своё движение из положения I в положение II, в этот же момент времени возмущение жидкости начнёт распространяться со скоростью $C$ из положения I в направлении движения потока.

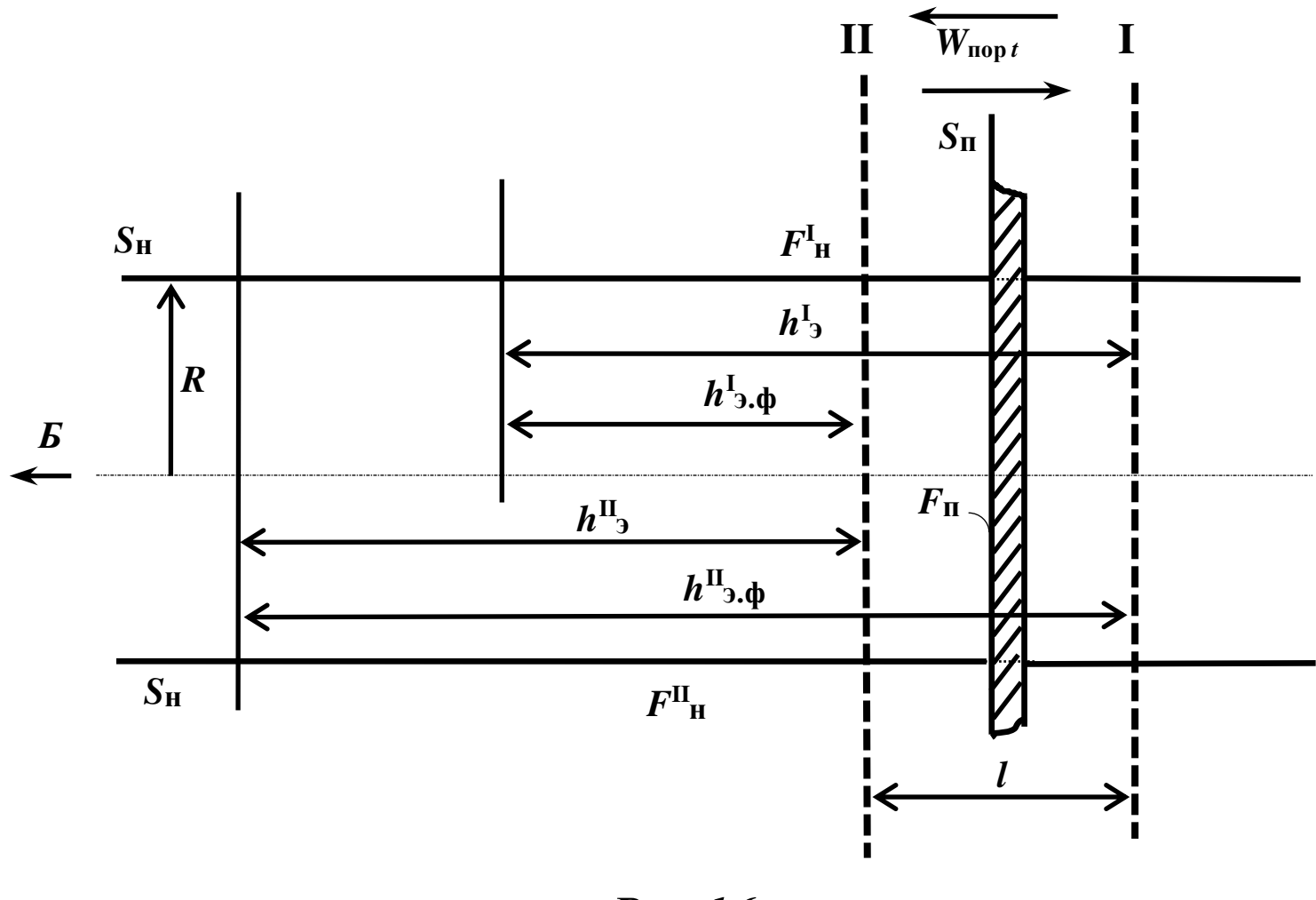

*Рис.16*

За время первого этапа $t^I_э$ возмущение пройдёт путь равный некоторой величине $h^I_э$. За то же время пластина переместится из положения I в положение II и пройдёт путь равный длине хода поршня $l$. В данном случае движение пластины и движение возмущения имеют одинаковое направление. Поэтому фактически высота возмущённого участка будет меньше на величину хода поршня $l$, то есть

$$h^I_{э.ф} = C\cdot t^I_э - l, \tag{52}$$

т. к. фактическая высота возмущения отсчитывается от истинного положения плоскости первоначального исследования $S_п$. Поэтому мы должны внести эту поправку в уравнение движения (47) первого этапа движения. Оно примет вид:

$$M_н = -\rho W_н \cdot 2\pi R\left(C\frac{l}{W_{пор}(t)} - l\right). \tag{53}$$

При непосредственном исследовании движения жидкости на любой длине хода поршня первого этапа движения в уравнение (53) подставляется фактическая длина отрезка хода пластины $\Delta l_i$ и фактическая скорость движения пластины. При этом условии мы всегда получим интересующие нас результаты.

Уравнения движения (46) и (53) являются окончательными уравнениями движения для акустического потока жидкости в первоначальных плоскостях исследования.

При движении пластины из положения II в положение I на втором этапе движения (рис. 16) поступательный расход массы $M_п$ тоже движется через площадь $F_п$ плоскости $S_п$, которая совмещена с плоскостью пластины и совершает с ней совместное движение. Поэтому изменение объёма в уравнении (48) зависит от скорости движения пластины $W_{пор}(t)$. Следовательно, мы должны в уравнении (49) вместо отношения $dl/dt$ подставить скорость движения пластины $W_{пор}(t)$. Это значит, что отношение $dl/dt$ и скорость движения пластины $W_{пор}(t)$ - равнозначные выражения, то есть $dl/dt = W_{пор}(t)$.

Для нормального расхода массы $M_н$ второго этапа движения площадь проходного сечения $F_н$ зависит от скорости движения возмущения и времени. Движение возмущения со скоростью $C$ начнётся из положения II в направлении *Б* линии тока в момент начала движения пластины из положения II в положение I. За время второго этапа $t^{II}_э$ возмущение пройдёт путь равный некоторой высоте $h_{э}^{II}$. За то же время пластина переместится на длину хода поршня $l$ в направлении противоположном направлению распространения возмущения. Поэтому фактическая высота возмущения участка $h_{э.ф}^{II}$ будет больше на величину хода поршня $l$, то есть

$$h_{э.ф}^{II} = h_{э}^{II} + l.$$

Следовательно, мы должны внести эту поправку в уравнение движения нормального расхода массы второго этапа (51). Тогда оно примет вид:

$$M_{н} = \rho W_{н} \cdot 2\pi R\left(C\frac{l}{W_{пор}(t)} + l\right). \qquad (54)$$

При непосредственном исследовании движения жидкости на любой длине хода пластины второго этапа в уравнение (54) подставляется фактическая скорость движения пластины. При этом условии мы всегда получим интересующие нас результаты.

Уравнения движения (49), (54) являются окончательными уравнениями движения акустического потока жидкости в первоначальных плоскостях исследования. Таким образом, мы получили количественные зависимости для движения жидкости на первом и втором этапах движения.

Отметим, что на рис. 16 высоты возмущённых участков $h_{э}^{I}$ и $h_{э}^{II}$ изображены разной длины. Этим подчеркивается то положение, что скорости движения пластины могут быть разными на первом и втором этапах движения. Поэтому величины времени первого и второго этапа движения тоже могут быть разными. Следовательно, из-за разницы времени этапов при одинаковой скорости движения возмущения $C$ высоты участков возмущения $h_{э}^{I}$ и $h_{э}^{II}$ могут иметь различные величины.

**Теперь определим качественно движение потока за два этапа.** Это значит, что мы должны связать движение акустического потока с определённым зрительным восприятием и определить основное звено этого движения, относительно которого происходит изменение всех параметров потока.

Принимаем в начальный момент времени среду в состоянии покоя. Считаем, что пластина источника возбуждения акустического движения жидкости не колеблется. Перпендикулярно к пластине проводим линию тока в сторону основного движения акустического потока (рис.17). Затем размещаем плоскости первоначального исследования потока (рис. 17).

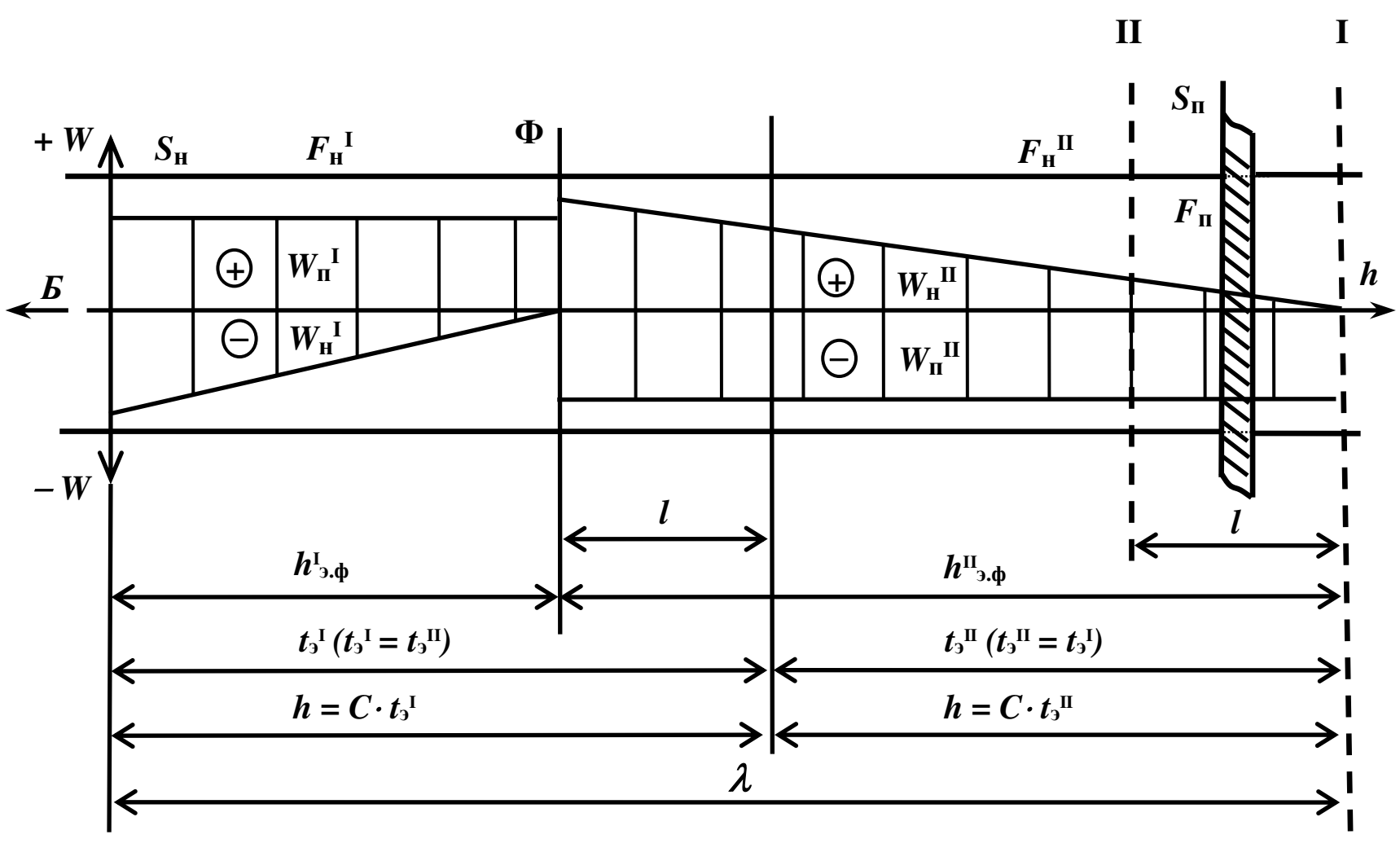

*Рис. 17*

Для фиксации происходящих изменений скоростей акустического потока принимаем плоскость, которая проходит через линию тока и считаем, что она может двигаться со скоростью возмущения потока.

Полагаем, что пластина совершила поступательное движение в два этапа Одновременно с пластиной двигалась плоскость фиксации скоростей, на которой фиксировались скорости [в] поступательной и нормальной ($S_п$ и $S_н$) плоскостях. Результаты этого движения изображены на рис. 17. Для данного рисунка было принято, что пластина двигалась с постоянной скоростью на обоих этапах. Подобное допущение было принято для удобства понимания движения.

С началом движения пластины жидкость, которая находилась в состоянии покоя, тоже начнёт движение. Движение её образуется за счёт поступательного и нормального расходов массы. Жидкость вовлекается в движение не одновременно во всём потоке, а образует в начальный момент зону движущейся жидкости. Назовём её зоной возмущения.

Границы зоны возмущения не остаются постоянными. Поэтому объём возмущённой жидкости начинает непрерывно увеличиваться. После образования зоны возмущения в начальный момент движения пластины, прирост её объёма происходит за счёт перемещения площади $F_п$ в направлении движения пластины. Эта площадь образует фронт движения, перемещающийся с определённой скоростью С, которую мы назвали скоростью возмущения. Перед началом движения плоскость фронта движения совпадает с поступательной плоскостью и плоскостью пластины. В момент начала движения пластины она тоже начинает двигаться со скоростью возмущения С. В данном случае скорость движения возмущения превышает скорость движения пластины.

Совместим начальное положение плоскости фиксации с плоскостью фронта движения. Тогда эта плоскость при движении будет находиться в пределах границ зоны возмущения. На её поверхности будут фиксироваться скорости движения жидкости через нормальную и поступательную плоскости ($S_п$ и $S_н$) фактические границы зоны возмущения. Результаты движения этой плоскости даны на рис. 17.

При первом этапе движения поступательная и нормальная скорости движения сохраняют постоянное направление. Поступательная скорость $W_п^{I}$ является постоянной величиной за всё время первого этапа, т.к. мы приняли постоянной величиной скорость движения пластины.

Нормальная скорость движения $W_н^{I}$ на начальном этапе движения имеет некую максимальную величину, которая к концу первого этапа становится равной нулю. К концу первого этапа пластина приходит в положение II и как бы останавливается на какое-то мгновение. В этот период расход массы равен нулю. Дальше должна снова идти как бы невозмущённая зона жидкости, но в этот период пластина начинает двигаться из положения II в положение I, то есть начинается второй этап движения пластины.

По отношению к общему потоку жидкости второй этап движения начинается как бы в покоящейся жидкости. Поэтому на втором этапе движения тоже происходит образование зоны возмущения потока, фронт которой движется со скоростью возмущения *C* в направлении движения фронта первого этапа движения. Тем самым сохраняется общая направленность движения возмущённой зоны.[10]

Нормальная и поступательная скорости движения меняют своё направление на противоположное относительно скоростей движения первого этапа. Поступательная скорость $W_п^{II}$ сохраняет свою величину постоянной за время второго этапа. Её величина равна величине поступательной скорости первого этапа движения $W_п^{I}$, так как скорости движения пластины равны по абсолютной величине. Нормальная скорость движения $W_н^{II}$ второго этапа движения тоже имеет максимальную величину на начальном этапе движения, которая в конце второго этапа переходит в ноль. Пластина достигает положения I и останавливается, и жидкость переходит в состояние покоя.

При сравнении фактических высот зон возмущения первого и второго этапов движения мы заметим, что фактическая высота зоны возмущения первого этапа $h^{I}_{э.ф}$ меньше фактической высоты возмущения второго этапа движения $h^{II}_{э.ф}$ на две длины хода пластины (2*l*), то есть

$$h^{I}_{э.ф} = h^{II}_{э.ф} - 2l.$$

Подобное соотношение сохраняется только при условии постоянной скорости движения пластины на обоих этапах. В общем случае оно может быть другим.

[10] Обратите внимание, что акустический поток как бы квантован на полуволны первого и второго этапа

Если теперь пластина продолжит свои колебания, то мы получим поэтапное чередование характеристик потока в первоначальных плоскостях исследования.

*Определяющим звеном этого движения является скорость движения пластины и скорость распространения возмущения С.*

Таким образом, мы получили количественные и качественные зависимости акустического потока жидкости для первоначальных плоскостей исследования.

4. Разместить полярную систему координат относительно поступательной поверхности $S_п$ и провести исследование всего потока жидкости относительно её движения на этой поверхности.

Особенностями акустического вида движения является то, что пространственное движение жидкости в двух взаимно перпендикулярных плоскостях требует прямолинейной линии тока, т. к. оно возбуждается возвратно-поступательным движением пластины, то есть является расходным объёмным видом движения жидкости, который не зависит от параметров пространства. По отношению к прямой линии тока радиус полярной системы координат является бесконечно большой величиной. Поэтому нет смысла размещать эту систему относительно поступательной плоскости исследования $S_п$.

Чтобы провести исследование всего потока, разместим необходимое число поступательных плоскостей исследования $S_п$ на разных расстояниях друг от друга по линии тока. Нормальную поверхность исследования $S_н$ продолжим на всю длину потока (рис.18). Затем проведём исследование движения жидкости в этих плоскостях с помощью плоскости фиксации, которая может двигаться со скоростью возмущения *С*.

В какой-то момент времени *t* произведём двухэтапное движение пластины. С этого момента плоскость фиксации, границей которой служит фронт возмущения, начнёт перемещаться со скоростью возмущения *С* в направлении *Б*. Перемещаясь таким образом, она достигнет первой поступательной плоскости исследования $S_{п1}$ за время $t_1$, которое будет равным частному от деления расстояния $l_1$ на скорость возмущения *С*. Последующие плоскости исследования она достигает за соответствующее время, которое для каждой плоскости будет равным частному от деления расстояния $l_i$ на скорость возмущения *С*.

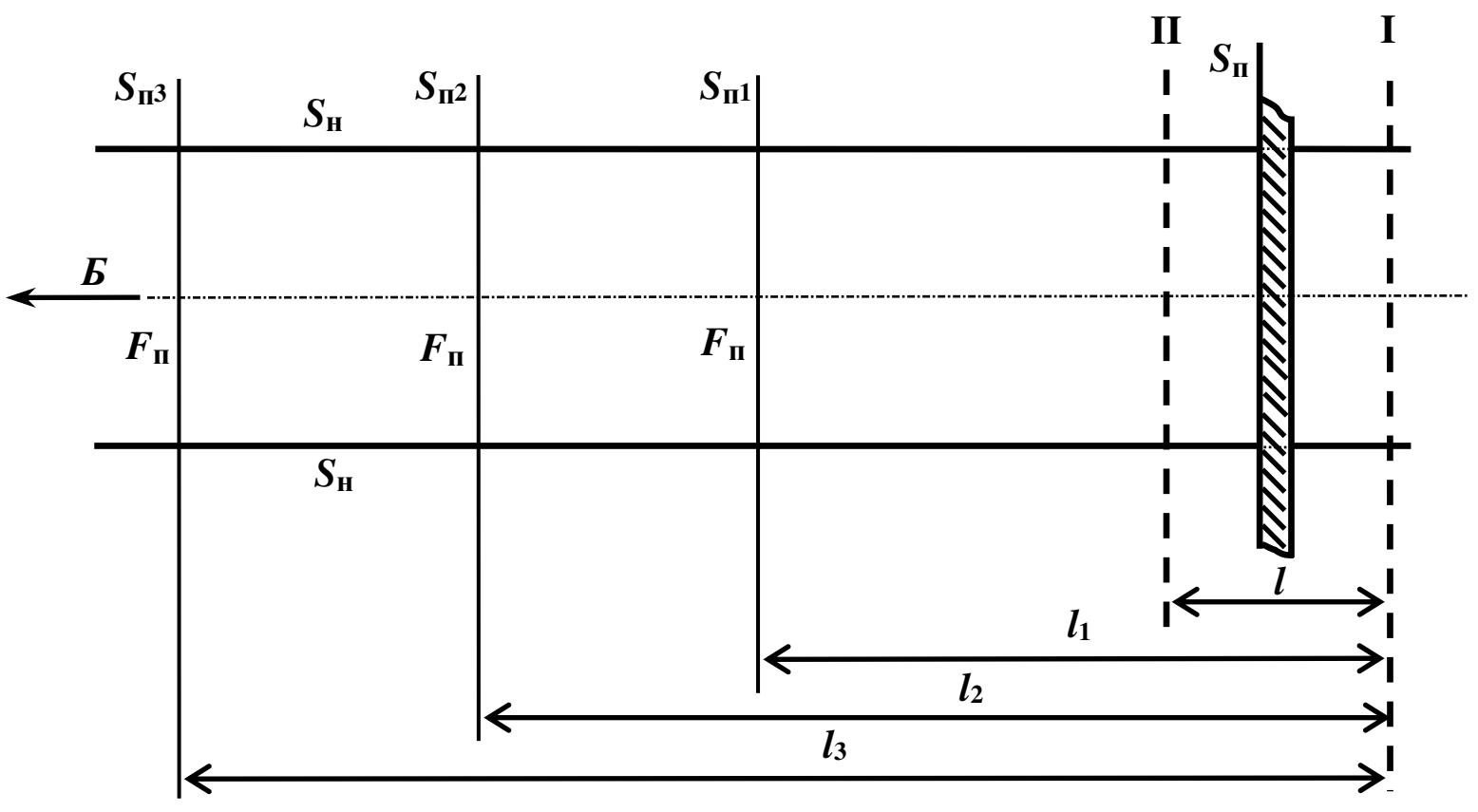

*Рис. 18*

При этом для каждой плоскости исследования на плоскости фиксации зарегистрируется картина движения жидкости, изображенная на рисунке 17, для первоначальных плоскостей исследования, то есть для каждой из всех плоскостей через определённый промежуток времени зафиксируется картина возмущения жидкости при двухэтапном движении пластины.

После прохождения возмущения жидкости двухэтапным движением пластины через все плоскости исследования на плоскости фиксации больше не регистрируются другие возмущения, то есть после возмущения жидкость снова приходит в состояние покоя. Это значит, что волна возмущения жидкости сохраняет свою форму на протяжении всего потока и движется со скоростью равной скорости возмущения *С*.

Следовательно, уравнения движения (46), (53), (49), (54) действительны для любых плоскостей исследования расположенных в объёме потока.

Следуя условиям рисунка 17, полная длина волны возмущения λ будет равна сумме фактических высот возмущённого потока для двух этапов движения пластины, то есть

$$\lambda = h^{\mathrm{I}}_{\text{э.ф}} + h^{\mathrm{II}}_{\text{э.ф}}. \qquad (55)$$

Высоты возмущённого потока вычисляются с помощью уравнений движения (46), (53), (49) и (54). Следовательно, эти уравнения определяют характеристики волны возмущения жидкости для двухэтапного движения пластины.

Если теперь предположить, что пластина колеблется непрерывно, то поток превращается в цепь перемещающихся волн возмущения.

Отметим, что в поступательных плоскостях исследования $S_{\text{п}}$ и связанных с ними участков нормальной поверхности $S_{\text{н}}$ в любой фиксированный момент времени <параметры> возмущённого потока будут одинаковыми, если между поступательными плоскостями по линии тока укладывается целое число длин волн λ при условии, что характер колебаний пластины остаётся неизменным.

Таким образом, мы выяснили, что уравнения движения (46), (53), (49), (54) применимы как для любых плоскостей исследования, так и для всего потока в целом, и каким образом они объединяют поток жидкости.

5. Сопоставить уравнения движения каждой поверхности исследования относительно друг друга, чтобы выявить общие уравнения движения акустического вида движения.

Исследования, связанные с данным пунктом, мы провели в пункте 4, то есть получили там общие уравнения движения акустического потока жидкости. Поэтому здесь не будем повторяться, а сделаем ссылку на пункт 4 данного параграфа.

Отметим:

а) при выводе уравнений движения (46), (49), когда проводится исследование акустического движения газа, необходимо пользоваться уравнением движения (34) объёмного вида движения, а не уравнением (33) .

б) если линия тока акустического потока жидкости представляет собой окружность с некоторым постоянным радиусом, то все уравнения движения и все положения, связанные с акустическим потоком с прямолинейной линией тока, полностью приемлемы для неё.

Пройдемся по положениям существующей механики жидкости и газа, относящихся к данному виду движения. В настоящее время акустика существует как самостоятельный её раздел. Её уравнения получены на основе общих зависимостей механики жидкости и газа, то есть

1. используется инерциальная система координат;

2. используются уравнения движения такого вида [10]:

$$\rho \frac{\partial \dot{\xi}}{\partial t} = -\frac{\partial P}{\partial x},$$

$$\rho \frac{\partial \dot{\eta}}{\partial t} = -\frac{\partial P}{\partial y},$$

$$\rho \frac{\partial \dot{\nu}}{\partial t} = -\frac{\partial P}{\partial z},$$

где $\rho$ – плотность, $\xi, \eta, \nu$ – скорости по соответствующим осям координат, $P$ – давление;

3. используется потенциал скорости [10]:

$$\int_0^t \frac{P}{\rho} dt + \Phi_0 = \Phi(x, y, z, t);$$

4. используется уравнение неразрывности [10]:

$$\mathrm{div}V = -\frac{\partial\left(\frac{\partial P}{\rho_0}\right)}{\partial t};$$

5. общее волновое уравнение имеет такой вид [10]:

$$C^2\Delta\Phi = \partial^2\Phi/\partial t^2.$$

Функция так называемого потенциала скоростей позволяет использовать практически любую математическую функцию в волновом уравнении.

Все эти положения и количественные зависимости коренным образом отличаются от положений и количественных зависимостей, изложенных в настоящей работе для акустического движения потока.

## III. 5. ОСНОВНЫЕ ПОЛОЖЕНИЯ СУЩЕСТВУЮЩЕЙ МЕХАНИКИ ЖИДКОСТИ И ГАЗА ПО РАЗДЕЛУ КИНЕМАТИКИ

В данном параграфе сопоставим положения кинематики существующей механики жидкости и газа с вновь полученными положениями. Для чего возьмём работу [1], в которой, в отличие от других работ, есть раздел кинематики, и рассмотрим её основные положения по данному вопросу.

«Основной задачей кинематики жидкости является определение скоростей частиц» [1]. Для этого используются два метода: метод Лагранжа и метод Эйлера. Метод Лагранжа: «Для каждой частицы жидкости должна быть определена её траектория, то есть координаты этой частицы должны быть определены как функция времени:

$$\left.\begin{aligned} x &= F_1(t,a,b,c),\\ y &= F_2(t,a,b,c),\\ z &= F_3(t,a,b,c) \end{aligned}\right\}\text{» [1].}$$

«Этим практическим вопросам отвечает метод Эйлера, который в том, как раз, и заключается, что фиксируется не частица (как в методе Лагранжа), а точка в пространстве с координатами *x*, *y*, *z*, и исследуется изменение скорости в этой точке с течением времени:

$$\left.\begin{aligned} v_x &= f_1(x,y,z,t),\\ v_y &= f_2(x,y,z,t),\\ v_z &= f_3(x,y,z,t) \end{aligned}\right\}\text{» [1].}$$

В данных положениях имеются следующие противоречия с вновь полученными положениями:

а) за основную характеристику движения принимается линейная скорость, а во вновь полученных основной характеристикой движения является расход массы в единицу времени;

б) здесь применяется инерциальная система координат, то есть относительно этой системы рассматривается движение. Во вновь полученных положениях принимается неинерциальная, которая размещается относительно движущегося потока жидкости.

в) во вновь полученных положениях движение жидкости исследуется в плоскостях и поверхностях.

Эти противоречия являются основными противоречиями, которые определяют коренное различие между положениями ныне существующей кинематики и вновь полученной.

Далее, в существующих положениях принимается так называемый потенциал скоростей, функция которого получается из уравнения неразрывности при условии безвихревого движения. Эта условная функция является чисто надуманной, полученной исключительно с помощью математических манипуляций. Поэтому она не имеет ничего общего с практикой. Она даёт лишь возможность записать какую-то внешнюю форму явления, то есть сделать математическую аппроксимацию явления природы, не вникая в его сущность. Пожалуй, функция потенциала скоростей является одной из главных причин того, что до настоящего времени не были выявлены неточности в количественных зависимостях движения жидкости, т. к. она давала возможность свободно манипулировать математическими функциями вне зависимости от явлений природы.

Теперь обратимся к другому разделу книги [1], который называется «Основные законы аэродинамики». Эта глава начинается с закона сохранения массы. Такой закон был открыт М. В. Ломоносовым, но он относится к изолированной массе, которая заключена в определённом объёме. По этой причине этот закон не может быть применим в механике жидкости и газа, которая рассматривает движение изолированной и неизолированной массы, а в общем случае её зависимости относятся к неизолированной массе жидкости. Уравнения движения (33) и (34) расходного вида движения во вновь полученных положениях непосредственно указывают на *неприменимость* закона сохранения массы в положениях и зависимостях механики жидкости и газа.

## Г Л А В А IV. ДИНАМИКА

Динамика есть часть теоретической механики, в которой устанавливается связь между движением определённых форм материи и действующими силами. Из этого определения следует, что в динамике изучается движение материи от действия на неё определённых сил.

Основные задачи динамики состоят в следующем:

1. зная движение жидкости, найти силы, действующие на эту жидкость;
2. зная силы, действующие на жидкость, найти её движение.

Обе задачи сводятся к определению связи действующих сил с характеристиками потока.

При решении этих задач динамика устанавливает общие количественные соотношения между различными физическими величинами, теснейшим образом связанными с движением жидкости. В основе динамики лежат основные законы механики жидкости и газа. К ним относятся:

1. закон сохранения состояния;
2. уравнение сил расходного вида движения;
3. формальный принцип связи вида движения с формой уравнений неразрывности и движения.

Для получения количественных зависимостей динамики воспользуемся общим методом исследования. В соответствии с этим методом мы должны для определённого потока жидкости проделать пять действий, в результате чего мы получим общие количественные зависимости. Вот эти действия:

1. выделить неподвижное пространство, которое включает в себя либо весь поток жидкости, либо его характерную часть;
2. выбирается и располагается плоскость, поверхность для первоначального исследования, расположение которой зависит от направления линейной скорости $W$;
3. изображается движение и силы давления в принятых плоскостях или поверхностях исследования, и записываются для них соответствующие количественные зависимости;
4. размещается полярная система координат, вернее – неинерциальная система координат, относительно принятой плоскости исследования и проводится исследование всего потока жидкости;
5. сопоставляются уравнения сил, полученные в различных плоскостях потока жидкости относительно друг друга, и таким образом выявляют общие уравнения сил для всего потока жидкости.

На этом исследование движения жидкости заканчивается. Количественные зависимости для плоскостей исследования записываются из условий равновесия.

В данном разделе будут получены уравнения сил для четырёх видов движения идеальной жидкости.

### IV. 1. УРАВНЕНИЕ СИЛ УСТАНОВИВШЕГОСЯ ВИДА ДВИЖЕНИЯ ИДЕАЛЬНОЙ ЖИДКОСТИ

С характеристиками установившегося вида движения жидкости мы ознакомились в предыдущей главе. Поэтому не будем делать излишних описаний этого вида движения, а приступим к его непосредственному исследованию в соответствии с порядком действий общего метода исследования.

***Действие первое.*** Здесь мы должны выделить неподвижное пространство среды. Как это делается, см. п.1, гл. III. В соответствии с этим параграфом мы должны изобразить объём установившегося потока. Покажем его на рисунке 19.

***Действие второе.*** Здесь мы должны выбрать и расположить плоскость $S$ для первоначального исследования. Для этой цели возьмём плоскость $S$ и расположим её перпендикулярно линии тока (рис. 19). Часть этой плоскости, которая лежит в пределах сечения потока, является непосредственной площадью исследование $F$ плоскости $S$ движущегося потока жидкости.

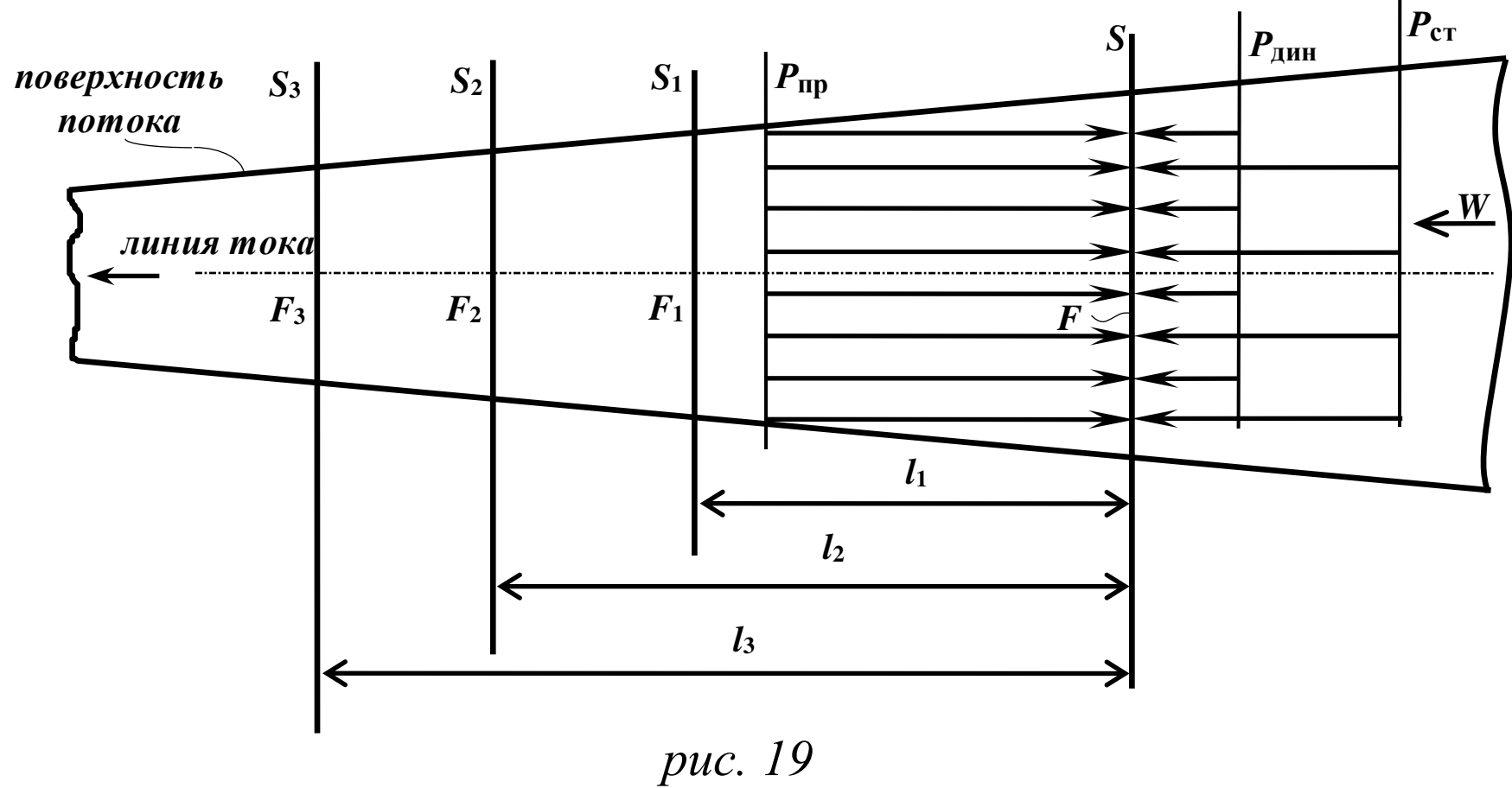

*рис. 19*

***Действие третье.*** Здесь мы должны изобразить действие сил давления на площади $F$ и записать для неё уравнение сил. На площади исследования $F$ плоскости $S$ будут действовать определённые силы давления. Изобразим их действие на площадь $F$ со стороны движения потока жидкости (рис. 19). Из практического опыта известно, что в этом случае на площадь $F$ будут действовать два типа сил давления: это статические силы давления $P_{ст}$ и динамические силы давления $P_{дин}$.

Направление действия статических сил давления $P_{ст}$ определяется расположением поверхности, т. к. их действие всегда направлено перпендикулярно к поверхности. Так и покажем их на рис.19. Динамические силы давления тоже относятся к разряду скалярных величин. Поэтому направление их действия определяется направлением линейной скорости $W$ движения жидкости в потоке. Наша плоскость исследования $S$ расположена перпендикулярно к линейной скорости $W$, значит, динамические силы давления $P_{дин}$ будут направлены перпендикулярно к плоскости $S$. Изобразим их на рисунке 19.

Под действием статических и динамических сил давления плоскость $S$ должна будет смещаться в направлении движения потока. Согласно общему методу исследования эта плоскость должна быть неподвижной. Чтобы выполнить это условие, мы должны приложить с противоположной стороны площади $F$ соответствующие силы давления $P_{пр}$, величины которых должны быть равны сумме статических и динамических сил давления. Это требование следует из условия равновесия. Тогда величину приложенных, или принятых, сил давления $P_{пр}$ можно записать следующим образом:

$$FP_{пр} = FP_{ст} + FP_{дин}. \qquad (1)$$

Фактически уравнение (1) уже является уравнением сил потока жидкости в исследуемой плоскости $S$. Нам остаётся только выразить силы давления потока через его характеристики.

Статические силы давления $P_{ст}$ потока зависят от физических свойств жидкости и от внешних сил. Поэтому их нельзя выразить непосредственно через характеристики потока. Динамические силы давления согласно уравнению сил расходного вида движения равны произведению расхода массы в единицу времени $M$ на скорость $W$ и делённому на площадь сечения потока $F$, то есть

$$P_{дин} = \frac{M}{F} W.$$

Подставим значение динамических сил давления в уравнение (1), получим:

$$FP_{пр} = FP_{ст} + MW. \qquad (2)$$

Уравнение (2) является уравнением сил потока жидкости в плоскости *S*. Если теперь в уравнении (2) заменить расход массы по уравнению движения (III - 19) и отнести силы давления к единице площади, то есть разделить правую и левую часть уравнения (2) на площадь *F*, то получим:

$$P_{\text{пр}} = P_{\text{ст}} + \rho W^2. \tag{3}$$

Уравнение (3) есть уравнение сил единицы площади потока.

***Действие четвёртое.*** Здесь мы должны разместить полярную систему координат относительно плоскости *S* и провести исследование всего потока жидкости.

В данном случае нет необходимости пользоваться полярной системой координат (см. гл. III, п. 1). Поэтому расположим плоскости исследования $S_1, S_2, S_3, \ldots S_n$ в объёме всего потока от плоскости первоначального исследования *S* на различных расстояниях $l_1, l_2, l_3, \ldots l_n$. Затем для каждой плоскости исследования запишем уравнение сил в соответствии с действующими на них статическими и динамическими силами давления потока, получим:

$$\begin{aligned}
F_1 P_{\text{пр}1} &= F_1 P_{\text{ст}1} + MW_1, \\
F_2 P_{\text{пр}2} &= F_2 P_{\text{ст}2} + MW_2, \\
F_3 P_{\text{пр}3} &= F_3 P_{\text{ст}3} + MW_3, \\
&\ldots\ldots \\
F_n P_{\text{пр}\,n} &= F_n P_{\text{ст}\,n} + MW_n.
\end{aligned} \tag{(4)}$$

***Действие пятое.*** Здесь мы должны сопоставить уравнения сил каждой плоскости друг относительно друга, чтобы выявить общее уравнение сил установившегося вида движения.

Сопоставляя уравнения (4) между собой, мы заметим, что величины каждого члена уравнения в соответствующих плоскостях исследования равны между собой, если площади сечения потока в этих плоскостях одинаковы, то есть $F_1 = F_2 = F_3 \ldots = F_n$. И различны, если площади сечения различны. Сопоставив таким образом уравнения сил различных плоскостей исследования, мы придём к выводу, что уравнения (2) и (3) являются общими уравнениями сил для потока жидкости установившегося вида движения.

Отметим, что уравнения (2) и (3) в одинаковой степени пригодны как для потоков установившегося вида движения с прямолинейной линией тока, так и для потоков, линией тока которых является окружность.

В ныне существующей механике жидкости и газа, например в книге [7], содержится уравнение (3), но там оно называется уравнением количества движения, так как оно было получено в результате так называемого «преобразования» уравнения количества движения в гидродинамическую форму.

## IV. 2. УРАВНЕНИЯ СИЛ ПЛОСКОГО УСТАНОВИВШЕГОСЯ ВИДА ДВИЖЕНИЯ ИДЕАЛЬНОЙ ЖИДКОСТИ

Общие характеристики движения плоского установившегося вида движения даны в гл. III, п. 2. Поэтому мы не будем повторять их в данном разделе, а приступим сразу к его основным задачам.

***Действие первое.*** Здесь мы должны выделить неподвижное пространство среды.

Это пространство показано на рисунке 8 с соответствующим описанием, которое дано в гл. III, п. 2. Рисунок и описание плоского установившегося вида движения действительны и для данного раздела.

***Действие второе.*** Здесь мы должны выбрать и расположить плоскости первоначального исследования.

В потоке плоского установившегося движения жидкость движется одновременно в двух взаимно перпендикулярных направлениях. Линиями тока его служат логарифмические спирали, которые, прилегая друг к другу, создают в сумме объём потока. Поэтому нам придётся провести первоначальные исследования в некоторой точке потока. Для этой цели выделим элемент одной из линий тока и возьмём на нём точку *A* (рис. 20,*а*). В этой точке *A* и проведём исследование.

Поэтому расположим в ней радиальную $S_r$ и тангенциальную $S_{tg}$ плоскости первоначального исследования. Непосредственное движение и исследование жидкости будет проходить на площадях $\Delta F_r$ и $\Delta F_{tg}$, которые образуются сечением точки $A$ радиальной и тангенциальной плоскостями. Поэтому площади $\Delta F_r$ и $\Delta F_{tg}$ являются площадями сечения точки $A$ соответствующими плоскостями исследования. Изобразим эти площади в увеличенном масштабе на рисунке 20, *б*.

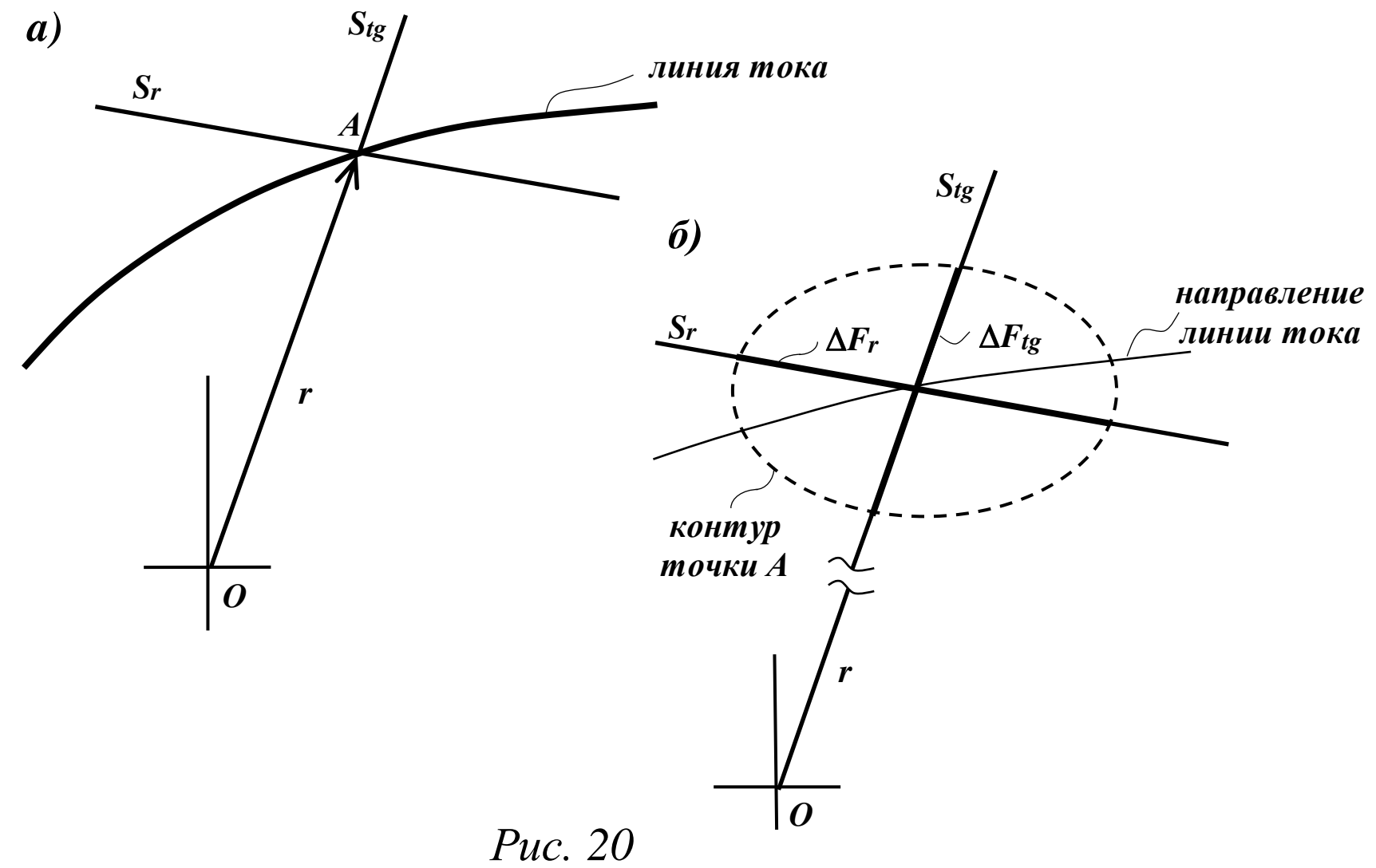

*Рис. 20*

***Действие третье.*** Здесь мы должны изобразить действующие силы давления на площадях сечения $\Delta F_r$ и $\Delta F_{tg}$ и записать для них уравнения сил.

Движущаяся жидкость будет оказывать определённое силовое воздействие на выделенные площади. Это воздействие будет складываться из статических и динамических сил давления. Изобразим действие этих сил на рисунке 21.

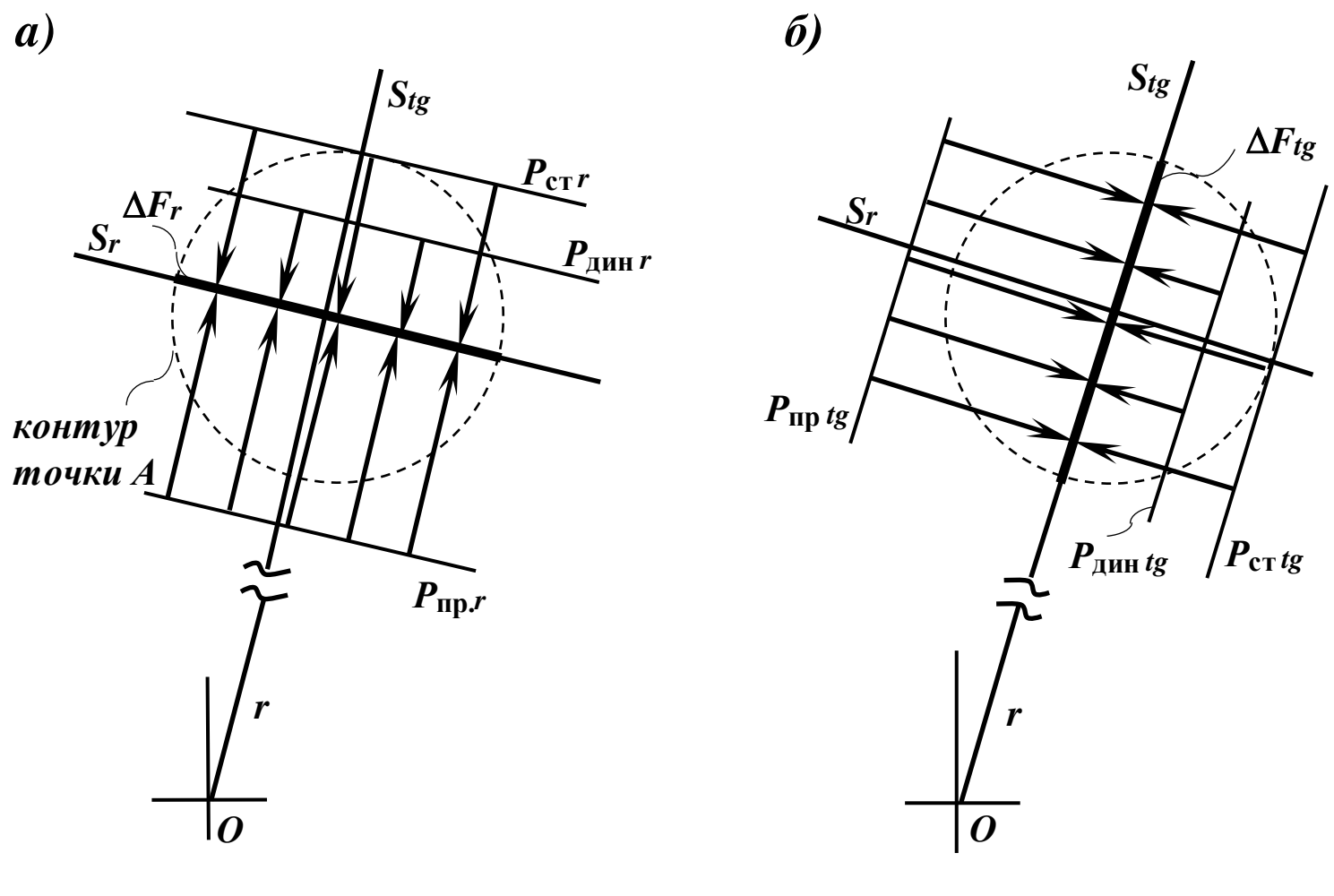

*рис. 21*

На радиальную площадь со стороны потока будут действовать статические радиальные силы $P_{ст\ r}$ и динамические радиальные силы $P_{дин\ r}$ давления (рис. 21,*а*). Под действием этих сил площадь $\Delta F_r$ будет перемещаться. Чтобы удержать её в определённом положении, приложим к ней некоторые силы давления с противоположной стороны (рис.21,*а*), так называемые радиальные принятые силы давления $P_{пр.r}$. Из условия равновесия величина суммы радиальных сил давления потока должна быть равна величине радиальных принятых сил давления. Запишем это равенство:

$$\Delta F_r P_{пр.r} = \Delta F_r P_{ст.r} + \Delta F_r P_{дин.r}. \qquad (5)$$

Теперь мы должны выразить силы давления потока через его характеристики. Радиальные статические силы давления останутся в том виде, в котором они записаны в уравнении (5), т. к. их нельзя выразить через характеристики потока.

Радиальные динамические силы давления с помощью уравнения сил расходного вида движения представим как произведение радиального расхода массы в единицу времени на радиальную скорость и делённую на радиальную площадь $\Delta F_r$, то есть

$$P_{дин.r} = \frac{\Delta M_r}{\Delta F_r} W_r.$$

Подставим его в уравнение (5), получим:

$$\Delta F_r P_{пр.r} = \Delta F_r P_{ст.r} + \Delta M_r W_r. \quad (6)$$

Мы получили уравнение сил для радиальной составляющей движения потока в точке *A*.

Затем для тангенциальной площади $\Delta F_{tg}$ изобразим действующие силы потока и уравновесим их тангенциальной принятой силой давления $P_{пр.tg}$ (рис. 21,*б*). Запишем для них условия равновесия:

$$\Delta F_{tg} P_{пр.tg} = \Delta F_{tg} P_{ст.tg} + \Delta F_{tg} P_{дин.tg}. \quad (7)$$

Заменим динамическую составляющую тангенциальных сил давления с помощью уравнения сил расходного вида движения, получим:

$$\Delta F_{tg} P_{пр.tg} = \Delta F_{tg} P_{ст.tg} + \Delta M_{tg} W_{tg}. \quad (8)$$

Мы получили уравнение сил для тангенциальной составляющей движения потока в точке *A*.

Сопоставим радиальное уравнение сил (6) с тангенциальным уравнением сил (8).

Согласно уравнению (III - 25), радиальный расход массы $\Delta M_r$ и радиальная скорость($W_r$) равны по абсолютной величине тангенциальному расходу массы $\Delta M_{tg}$ и тангенциальной скорости ($W_{tg}$) соответственно, но имеют противоположный знак.

Радиальная сила давления $P_{ст.r}$ равна тангенциальной силе давления $P_{ст.tg}$, т. к. статическое давление в одной точке не может быть различным. При этом статические давления не могут быть отрицательными в одной точке.

Определить, какой знак имеют динамические составляющие сил, в общей форме невозможно, т. к. знак их зависит непосредственно от конкретных условий движения. Поэтому перед динамическими составляющими обоих уравнений поставим знак «±». Тогда уравнения (6) и (8) примут вид:

$$\Delta F_r P_{пр.r} = \Delta F_r P_{ст.r} \pm \Delta M_r W_r, \quad (9)$$

$$\Delta F_{tg} P_{пр.tg} = \Delta F_{tg} P_{ст.tg} \pm \Delta M_{tg} W_{tg}. \quad (10)$$

Это окончательная форма уравнений сил движущейся жидкости в точке *A*. Подстановка знаков в этих уравнениях делается для конкретных условий движения. Уравнения сил (9) и (10) можно отнести к единице площади, то есть разделить их на соответствующие площади, тогда они примут вид:

$$P_{пр.r} = P_{ст.r} \pm \rho W_r^2, \quad (11)$$

$$P_{пр.tg} = P_{ст.tg} \pm \rho W_{tg}^2. \quad (12)$$

Это будут уравнения сил единицы площади потока для точки *A*.

***Действие четвёртое.*** Здесь мы должны разместить полярную систему координат относительно точки $A$ и провести исследование всего потока жидкости.

Разместим полярную систему координат в тангенциальной плоскости $S_{tg}$ исследования точки $A$. В этом случае ось полярной системы координат совместится с осью потока, как это сделано в п. 2, гл. III. В этом случае мы можем для любой точки потока записать систему уравнений (9), (10) или (11), (12). Предположим, что мы действительно записали эти уравнения для всех точек потока и получили множество уравнений. Теперь остаётся разобраться с ними, что мы сделаем в следующем действии.

***Действие пятое.*** Здесь мы должны сопоставить уравнения сил каждой точки потока относительно друг друга, чтобы выявить общие уравнения сил.

Сопоставление уравнений сил всех точек потока в настоящий момент мы можем произвести фактически только для динамических составляющих сил давления. Из раздела кинематики мы знаем, что радиальный и тангенциальный расход массы в единицу времени есть величина постоянная для всего потока. Поэтому изменение динамических составляющих сил давления будет зависеть от изменения радиальных и тангенциальных скоростей движения жидкости.

Статические составляющие сил давления мы не можем определить в данный момент, т. к. для этого необходимо иметь уравнения энергии, которые будут получены в следующей главе данной работы. При этом они тоже зависят от изменения динамических составляющих сил давления. Поэтому проведём сопоставление уравнений сил по динамическим составляющим сил давления.

Сопоставив уравнения, мы получим, что во всех точках любой цилиндрической поверхности, расположенной в объёме потока, динамические и статические силы давления одинаковы по величине и направлению, что во всех точках линии тока динамические и статические силы давления имеют разные величины. Из этого сопоставления следует, что уравнения сил точки $A$ - (9), (10), (11), (12), - являются общими уравнениями сил для потока жидкости плоского установившегося вида движения.

## IV. 3. УРАВНЕНИЕ СИЛ РАСХОДНОГО ВИДА ДВИЖЕНИЯ ИДЕАЛЬНОЙ ЖИДКОСТИ

В разделе кинематики мы получили необходимые количественные и качественные характеристики этого вида движения. Поэтому перейдем непосредственно к задачам, связанным с динамикой расходного вида движения.

***Действие первое.*** Для расходного потока жидкости принимается неподвижное пространство, границы которого перемещаются или изменяются с течением времени. В этом заключается своеобразие этого вида движения, то есть выделенный объём потока не перемещается в пространстве, а изменяет свои границы. Такой объём можно легко себе представить.

***Действие второе.*** Здесь мы должны выбрать и расположить плоскость (поверхность) для первоначального исследования.

Учитывая особенности расходного потока жидкости, мы должны поступить следующим образом:

а) в определённый фиксированный момент времени $t$ принять границы потока в пределах тех границ, которых он достиг к этому времени;

б) в соответствии с общим методом исследования должны расположить поверхность первоначального исследования так, чтобы она в каждой своей точке была бы расположена перпендикулярно скорости движения.

Располагая поверхность первоначального исследования по этим принципам, мы получим некоторую замкнутую поверхность $S_t$ какой-то формы для фиксированного момента времени $t$ (рис. 22).

Замкнутая поверхность $S_t$ выделит из среды объём $V_t$, в котором содержится жидкость с плотностью $\rho$. В этот же фиксированный момент времени $t$ мы будем иметь некоторый фиксированный расход массы $M_t$, который поступал или вытекал в этот момент времени из фиксированного объёма. Таким образом, мы получили поверхность $S_t$ для первоначального исследования и необходимые характеристики потока жидкости.

***Действие третье.*** Здесь мы должны показать действующие на поверхности $S_t$ силы давления и записать для них уравнения сил.

В фиксированном объёме $V_t$ в этот момент времени жидкость будет находиться под действием некоторых сил давления $P_v$. Эти силы давления одинаковы по величине во всех точках объёма. Поэтому они будут действовать на всю поверхность $S_t$ тоже силой давления $P_v$. Изобразим эти силы на рисунке 22.

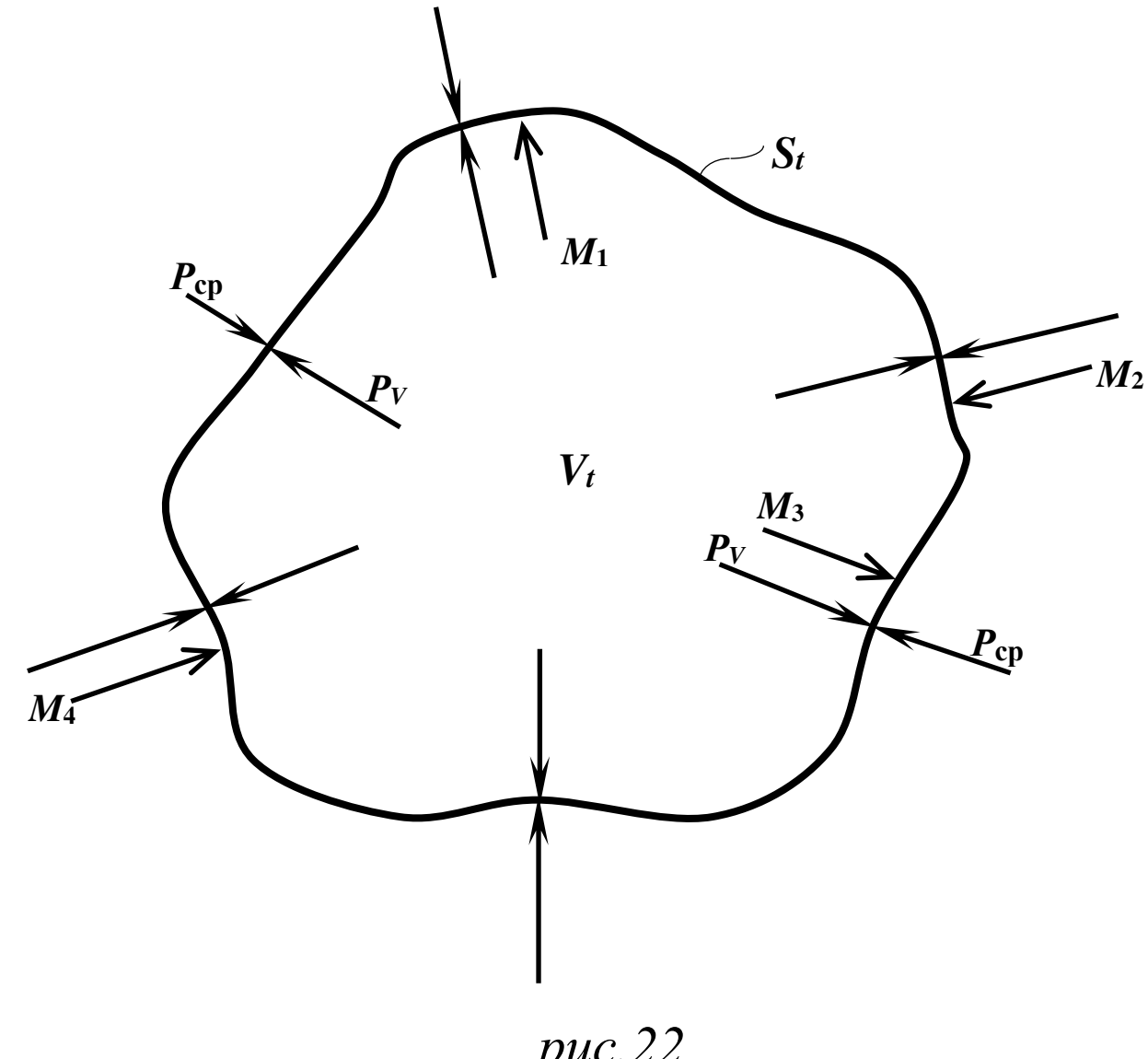

*рис.22*

В этот фиксированный момент времени среда тоже имеет некоторую величину сил давления $P_{ср}$, которые, в свою очередь, действуют на всю поверхность $S_t$ с внешней стороны. Изобразим их на рисунке 22. В это же время от действия объёмных сил давления $P_v$ и наружних сил давления $P_{ср.}$ по всей поверхности $S_t$ будет проходить некоторый расход массы в единицу времени $M_t$.

Дальше полагаем, что силы давления среды $P_{ср}$ равны по величине, то есть по всей поверхности $S_t$ они имеют равную величину, что силы давления $P_v$ выделенного объёма меньше, чем силы давления среды $P_{ср}$, то есть $P_{ср}>P_v$. Так как расход массы поступает через поверхность $S_t$ первоначального исследования, то эта же поверхность одновременно выполняет функцию площади потока $F_t$, то есть в данном случае поверхность исследования и площадь потока равнозначны: $S_t = F_t$. Теперь мы можем записать условие равновесия для выделенного объёма потока по его границе в фиксированный момент времени $t$. В математической форме оно будет иметь такой вид:

$$F_t P_{ср} = F_t P_v + M_t W_t. \tag{13}$$

Произведение площади потока $F_t$ на силы давления среды $P_{ср}$ выражает действие внешних сил на выделенный объём . Эти силы уравновешиваются объёмными силами давления $P_v$, которые в уравнении (13) записаны как произведение площади $F_t$ на объёмные силы давления $P_v$. В связи с тем, что объёмные силы давления меньше, чем силы давления среды, то эта разница покрывается расходом массы в единицу времени $M_t$, который поступает через площадь $F_t$ в выделенный объём потока со скоростью $W_t$.

Величина расхода массы и скорость движения жидкости должны соответствовать разности сил давления. Поэтому, прибавив к объёмным силам в уравнении (13) произведение расхода массы на скорость, мы тем самым заменили знак неравенства между силами на знак равенства. Следовательно, уравнение (13) выражает условие равновесия.

В задачу данного раздела входит определение связи сил давления с характеристиками потока. Поэтому перенесём члены уравнения, содержащие силы давления, в левую часть уравнения (13), а остальные оставим в правой, тогда получим:

$$F_t (P_{ср} - P_v) = M_t W_t. \tag{14}$$

В этом уравнении нас интересуют давления, которые затачиваются на проталкивание расхода массы в единицу времени. Поэтому обозначим разность сил давления в уравнении (14) просто как динамическую силу давления $P_{\text{дин}}$, то есть $P_{\text{дин}} = P_{\text{ср}} - P_v$, и подставим её в уравнение (14), получим:

$$F_t P_{\text{дин}} = M_t W_t. \tag{15}$$

Уравнение (15) выражает силы давления через характеристики движущейся массы. Это уравнение будет уже уравнением сил расходного вида движения, но пока мы не можем сказать, что оно общее, т. к. мы получили его при определённых ограничениях. Поэтому сделаем исследование уравнения (15) на предмет общих условий движения.

Для чего предположим, что объёмные силы давления больше, чем силы давления среды, то есть $P_v > P_{\text{ср}}$. Тогда расход массы будет иметь противоположное направление, то есть он будет вытекать из выделенного объёма. Это значит, что правый член уравнения (15) может быть отрицательным. Следовательно, он должен иметь и положительный, и отрицательный знак, то есть «±», чтобы учесть и это условие движения. Тогда уравнение (15) примет вид:

$$F_t P_{\text{дин}} = \pm M_t W_t. \tag{16}$$

Уравнение (16) будет более общим, чем уравнение (15).

Теперь полагаем, что на отдельных участках площади $F_t$ силы давления среды $P_{\text{ср}}$ имеют разные величины. Объёмные силы давления $P_v$ в пределах выделенного объёма имеют одинаковое значение по закону Паскаля. Тогда в общем случае на этих участках одновременно могут быть и положительные, и отрицательные расходы масс в единицу времени и при одинаковых знаках иметь большее или меньшее значение. Для отдельного участка выделенного объёма уравнение равновесия можно записать с помощью уравнения (15) как

$$\Delta F_{ti} P_{\text{дин}\,tj} = M_{ti} W_{tj}. \tag{17}$$

Тогда общее уравнение равновесия для выделенного объёма будет выражаться как сумма уравнений всех этих участков, то есть

$$\sum_{1}^{n} F_{ti} P_{\text{дин}\,tj} = \sum_{1}^{n} M_{ti} W_{tj} - \sum_{1}^{n} M_{ti} W_{tj}. \tag{18}$$

В самом общем случае для первоначальной поверхности исследования, фиксированной в определённый момент времени, уравнение сил будет иметь вид уравнения (18). Если теперь принять условие, что границы объёма непрерывно меняются во времени, то в уравнении (18) придётся учесть изменение площади сечения потока $F_t$ по времени, то есть принять площадь сечения потока $F(t)$ как функцию времени и подставить её в уравнение (18). В остальном уравнение (18) останется без изменений.

***Действие четвёртое.*** Здесь мы должны разместить полярную систему координат относительно поверхности $S_t$ и провести исследование всего потока жидкости.

Так как поверхность первоначального исследования $S_t$ является криволинейной, то мы не сможем разместить полярную систему координат относительно неё. И, в общем, делать этого не нужно, т. к. расходный вид движения не зависит от параметров пространства. В этом случае мы можем получить ряд значений уравнения (18) для других фиксированных моментов времени, полагая, что мы получили такие значения для различных поверхностей исследования.

***Действие пятое.*** Здесь мы должны сопоставить уравнения сил каждой поверхности исследования друг относительно друга, чтобы выявить общее уравнение сил расходного вида движения.

Сопоставив уравнения сил поверхностей исследования, мы придём к выводу, что формы уравнений (15), (16) и (18) являются общими формами уравнений сил для всего потока жидкости расходного вида движения. Уравнение (18) является общим уравнением сил для любого потока жидкости расходного вида движения.

Отметим, что в современной механике жидкости и газа существует уравнение так называемой реактивной силы $R$ [7], которое имеет такой вид:

$$R = MW.$$

Это уравнение является почти точной копией уравнения (15), которое имеет такое же значение для механики безынертной массы, как произведение массы на ускорение для механики твёрдого тела. В то же время, в современной механике жидкости и газа произведение массы на ускорение принято основным понятием силы. Подобный курьёз, пожалуй, связан с тем, что в силу сложившихся привычек люди иногда не считаются даже с фактами. Например, уравнение (15) и уравнение реактивной тяги определяют массу как безынертную, что является фактом, и в то же время почему-то считают инертность основным свойством массы.

***<u>От автора</u>:*** Например, если подойти формально ко второму закону Ньютона, который говорит, что сила равна первой производной по времени от количества движения ( $K = mW$ ), тогда можно записать следующие зависимости:

$$R = m\frac{dW}{dt}, \tag{19}$$

$$R = W\frac{dm}{dt}. \tag{20}$$

Оба эти уравнения, согласно вышеизложенной формулировке второго закона Ньютона, выражают силу. Различие между ними заключается лишь в том, что уравнение (19) получено при условии, что масса является постоянной величиной, а уравнение (20) получено при условии постоянства скорости движения. Теперь в уравнении (20) заменим силу $R$ на произведение площади сечения потока $F$ и сил давления $P$, получим:

$$FP = WM. \tag{21}$$

Уравнение (21) одинаково с уравнением сил (15) и с уравнением реактивной силы. Это значит, что второй закон механики Ньютона является общим, как для механики твёрдого тела, так и для механики жидкости и газа. Что подтверждает правильность уравнений сил расходного вида движения (15), (16) и (18).

В то же время уравнения (19) и (20) являются зависимостями, которые выражают сущность определённых явлений природы, поэтому они называются законами природы. Первый и третий законы Ньютона конкретно связывают уравнение (19) с определённой формой материи, которая, согласно этим законам, является массой, обладающей свойством инертности и неизменным объёмом. Изменение состояния этой массы происходит при действии силы за счёт изменения скорости её объёма, то есть за счёт ускорения при действии силы. Свойство инертности массы сделали основным свойством массы. При этом не учли, что оно проявляется только при определённом её состоянии, поэтому является частным свойством <массы>. По этой причине решили упростить второй закон Ньютона, который в учебной литературе чаще трактуется в таком виде, что сила равна произведению массы на ускорение.

Всё это, конечно, верно, если представлять себе, что в природе не существует каких-либо иных явлений. Но, «к сожалению», в других состояниях масса не имеет свойства инертности. Когда она находится в жидком и газообразном состоянии и совершает какое-то перемещение, то она обладает ещё вторым свойством - свойством безынертности. Это свойство определило иное явление природы, которое изучает механика жидкости и газа. Поэтому уравнение (20) потребовало для выражения этой конкретной формы материи ещё две закономерности: это *закон сохранения состояния* и *формальный принцип связи вида движения с формой уравнений неразрывности и движения,* которые полностью выразили сущность изучаемого явления природы и закрепили качественное различие между механикой твёрдого тела и механикой жидкости и газа.[11] Свойство безынертности массы было выявлено при выводе уравнения сил расходного вида

[11] массой называется способность вещества взаимодействовать с механическими силами, т.е., масса является характеристикой механических свойств вещества. Поэтому, строго говоря, только само вещество может находиться в различных агрегатных состояниях, а не его масса. Согласно второму закону Ньютона механическая сила одна, но имеет

движения. Только после этого стали понятными качественные различия между уравнениями (19) и (20) .

Цель данной работы заключается именно в том, чтобы показать различие свойств массы в механике твёрдого тела и в механике жидкости и газа и дать описание новому явлению природы, то есть дать описание механического состояния жидкостей и газов. Так или иначе, это новое для людей восприятие явления природы. Задача автора заключается лишь в том, чтобы разъяснить, как можно понятнее, это новое в зависимости от своих субъективных и объективных возможностей. Право судить об этом новом предоставляется другим.

Что касается авторитета Ньютона, то он должен возрасти, так как данная работа является продолжением классической механики, фундамент которой был заложен Галилеем, Ньютоном и другими учёными более поздних времён.

### IV. 4. УРАВНЕНИЯ СИЛ АКУСТИЧЕСКОГО ВИДА ДВИЖЕНИЯ ИДЕАЛЬНОЙ ЖИДКОСТИ

Характеристики потока этого вида движения были получены в разделе кинематики. Поэтому сразу перейдем к непосредственным задачам данного параграфа. Определим уравнение сил акустического вида движения с помощью общего метода исследования.

***Действие первое.*** Здесь мы должны выделить неподвижное пространство среды. Как это делается, изложено в п. 4 гл. III (рис. 13). В данном действии мы принимаем полностью этот акустический поток жидкости.

***Действие второе.*** Здесь мы должны выбрать и расположить плоскости первоначального исследования.

Для первоначального исследования принимаем две взаимно перпендикулярные поверхности:

для поступательного движения жидкости принимаем поступательную плоскость $S_{\text{п}}$ (рис. 13), расположенную перпендикулярно к линии тока в положении II пластины возмущения;

для нормального движения жидкости принимаем цилиндрическую поверхность $S_{\text{н}}$, расположенную симметрично относительно линии тока (рис. 13).

***Действие третье.*** Здесь мы должны изобразить действующие силы на поверхностях исследования $S_{\text{п}}$ и $S_{\text{н}}$ и получить для них уравнения сил.

Акустическое возмущение среды происходит за счёт возвратно-поступательного движения пластины. Поэтому исследование движения акустического потока будет проводиться в два этапа. Изобразим на рис. 23 поэтапный ход пластины и, как на рис. 14, покажем её в виде поршня.

На первом этапе движения поршень движется из положения I в положение II со скоростью пластины. Вытесняемая жидкость движется через поступательную площадь сечения потока $F_{\text{п}}$ плоскости $S_{\text{п}}$ и оказывает на неё определённое силовое воздействие. Силовое воздействие будет складываться из статических поступательных сил давления $P_{\text{п.ст.}}$, покажем их на рис. 23,*а*, и динамических сил давления $P_{\text{п.дин.}}$, которые тоже покажем на рис. 23,*а*. Чтобы плоскость не двигалась, приложим к ней с противоположной стороны принятые поступательные силы давления $P_{\text{пр.п.}}$, покажем их на рис. 23,*а*. В этом случае плоскость $S_{\text{п}}$ будет находиться в состоянии равновесия. Запишем условие равновесия:

$$F_{\text{п}} P_{\text{пр.п.}} = F_{\text{п}} P_{\text{ст.п}} + F_{\text{п}} P_{\text{дин.п}}. \tag{22}$$

Заменим с помощью уравнения сил расходного вида движения в уравнении (22) динамические силы давления характеристиками потока, получим:

$$F_{\text{п}} P_{\text{пр.п.}} = F_{\text{п}} P_{\text{ст.п}} + M_{\text{п}} W_{\text{п}}. \tag{23}$$

---

две формы взаимодействия с веществом - как сила и как распределённая сила. Для сил не имеет значения агрегатное состояние вещества. Масса вещества делится для них на изолированную и неизолированную, т.е. инертную и безынертную. В соответствии с этим, силы принимают соответствующую форму. Для изолированной массы они могут принимать и ту и другую форму, поскольку им приходится взаимодействовать и с веществом тела, которое для них является неизолированной массой. Редактор считает, что именно по этой причине автор механики безынертной массы говорил, что теория упругости и пластичности материалов должна быть пересмотрена с точки зрения механики безынертной массы.

Уравнение (23) является уравнением сил поступательного движения жидкости на первом этапе. Все силы давления в уравнении (23) изменяются во времени в зависимости от скорости движения поршня.

При движении поршня на первом этапе нормальный поток жидкости будет оказывать силовое воздействие на нормальную поверхность $S_{н}$ на площади сечения потока $F_{н}$. Площадь потока является переменной величиной. По уравнению (III-40) её величина равна длине окружности цилиндра умноженной на высоту $h$, а высота изменяется во времени со скоростью возмущения $C$. Поэтому на первом этапе движения поршня принимаем некоторый фиксированный момент времени $t$ и рассмотрим действие сил потока на нормальную площадь потока $F_{н}$ нормальной поверхности $S_{н}$.

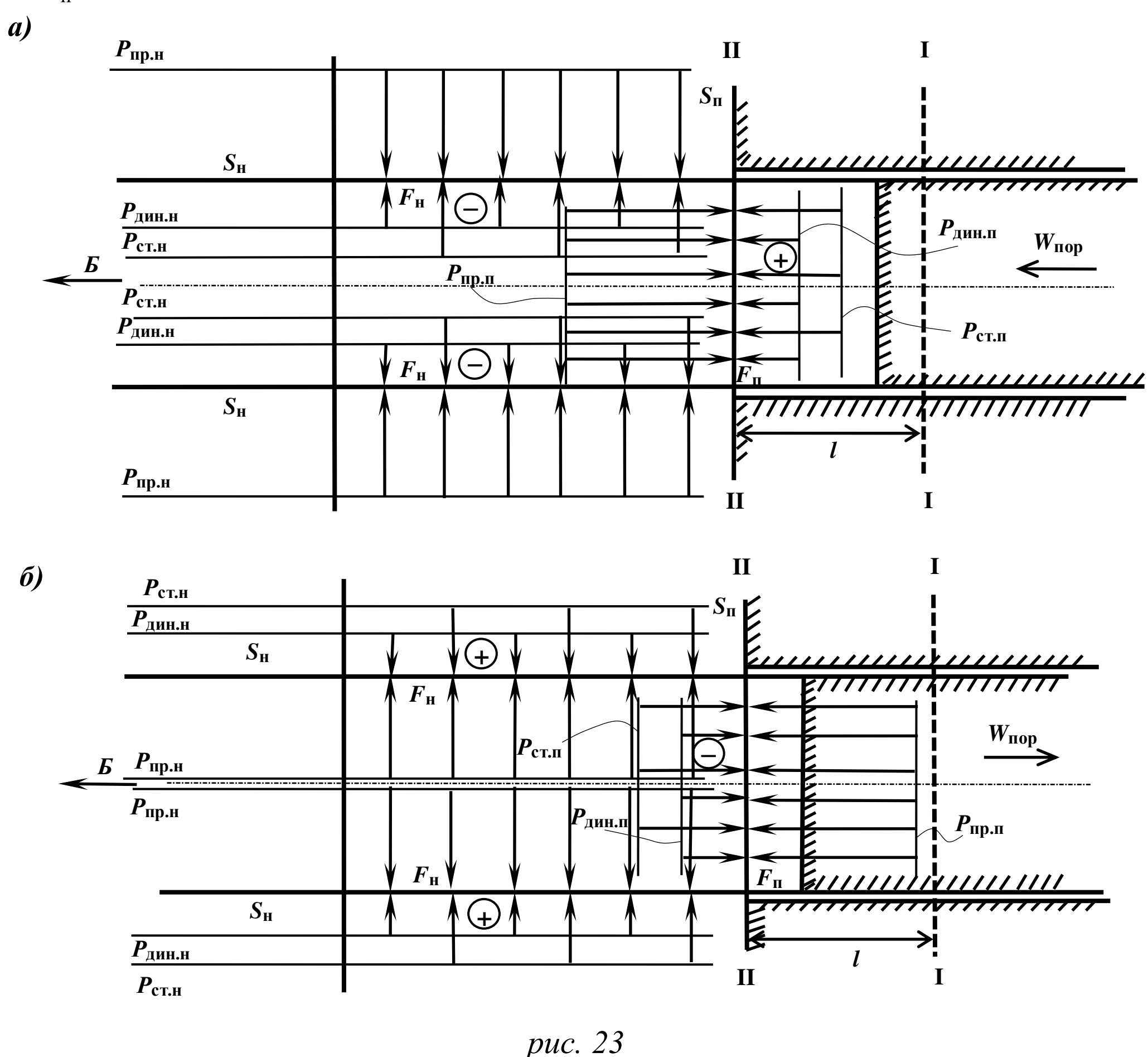

*рис. 23*

Со стороны потока на нормальную площадь $F_{н}$ будут действовать нормальные статические силы давления $P_{ст.н}$, изобразим их на рис. 23,*а*, и нормальные динамические силы давления $P_{дин.н}$, которые тоже покажем на этом рисунке. Чтобы нормальная поверхность $S_{н}$ не перемещалась от действия этих сил, с противоположной стороны её приложим нормальные силы давления $P_{пр.н}$. Теперь мы можем записать условие равновесия для площади потока $F_{н}$. Оно будет иметь вид:

$$F_{н} P_{пр.н.} = F_{н} P_{ст.н.} + F_{н} P_{дин.н.}. \tag{24}$$

В уравнении (24) нормальные динамические силы давления $P_{дин.н}$ выразим через характеристики потока с помощью уравнения сил расходного вида движения, получим:

$$F_{н} P_{ст.н.} = F_{н} P_{ст.н.} + M_{н} W_{н}. \tag{25}$$

Уравнение (25) является уравнением сил нормального движения жидкости на первом этапе движения поршня.

В связи с тем, что при движении поршня на первом этапе нормальные скорости имеют отрицательный знак, то в уравнении (25) для динамических составляющих сил давления мы должны заменить положительный знак на отрицательный. Тогда оно примет вид:

$$F_{\text{н}} P_{\text{пр.н.}} = F_{\text{н}} P_{\text{ст.н.}} - M_{\text{н}} W_{\text{н}}. \qquad (26)$$

Уравнения (23) и (26) являются окончательной формой уравнений сил поступательного и нормального движения жидкости на первом этапе движения поршня.

Эти уравнения можно отнести к единице площади и заменить расход массы в единицу времени по уравнению движения установившегося вида движения ( $M = \rho F W$ ). Тогда они примут вид:

$$P_{\text{пр.п.}} = P_{\text{ст.п.}} + \rho W_{\text{п}}^2. \qquad (27)$$

$$P_{\text{пр.н.}} = P_{\text{ст.н.}} - \rho W_{\text{н}}^2. \qquad (28)$$

При необходимости их надо решать на промежутке времени первого этапа движения поршня. При этом его либо разбивают на части, как это делалось в разделе кинематики, либо решают аналитическим путём.

Отметим, что в настоящий момент мы можем проследить только изменение динамических составляющих сил давления, которые связаны с изменением характера расхода массы в единицу времени и скоростью движения. Статические составляющие зависят от первоначальных сил давления потока жидкости. Единственно, что можно отметить по статическим силам давления, что нормальные статические силы давления $P_{\text{ст.н}}$ в любой момент времени равны поступательным статическим силам давления $P_{\text{ст.п}}$, то есть $P_{\text{ст.н}} = P_{\text{ст.п}}$ . Это связано с тем, что во всех точках потока, относящихся к первоначальным плоскостям исследования, происходит одинаковое нормальное и поступательное движение, которое соответствует записанным зависимостям.

**Рассмотрим движение жидкости на втором этапе движения поршня - из положения II в положение I.** Для нормальной и поступательной плоскостей первоначального исследования изобразим силы давления потока и принятые силы давления, как показано на рисунке 23, *б*. Их расположение будет отличаться от расположения сил давления первого этапа тем, что динамические силы давления будут направлены в противоположную сторону. Поэтому уравнения сил нормальной и поступательной плоскостей исследования будут иметь одинаковый вид с уравнениями сил первого этапа движения поршня, с той разницей, что динамические составляющие сил давления второго этапа движения поршня имеют разные знаки с динамическими составляющими первого этапа. Поэтому запишем их без вывода:

$$F_{\text{п}} P_{\text{пр.п.}} = F_{\text{п}} P_{\text{ст.п}} - M_{\text{п}} W_{\text{п}}, \qquad (29)$$

$$F_{\text{н}} P_{\text{пр.п.}} = F_{\text{н}} P_{\text{ст.н.}} + M_{\text{н}} W_{\text{н}}. \qquad (30)$$

Уравнения (29) и (30) являются уравнениями сил поступательного и нормального движения жидкости при акустическом виде движения.

$$P_{\text{пр.п.}} = P_{\text{ст.п.}} - \rho W_{\text{п}}^2, \qquad (31)$$

$$P_{\text{пр.п.}} = P_{\text{ст.н}} + \rho W_{\text{н}}^2. \qquad (32)$$

Уравнения (31) и (32) являются уравнениями сил, отнесёнными к единице площади. Статические силы давления для этих уравнений тоже будут равны. Характер распределения динамических нормальных и поступательных сил давления по этапам движения пластины аналогичен характеру распределения нормальных и поступательных скоростей, с той разницей, что силы давления пропорциональны квадрату скорости движения. Пример распределения скоростей на этапах движения поршня см. на рис. 15.

***Действие четвёртое.*** Здесь мы должны разместить полярную систему координат относительно первоначальных плоскостей исследования и провести исследование всего потока жидкости.

В предыдущей главе мы убедились, что линия тока акустического потока является, прежде всего, прямой линией. По отношению к прямой линии радиус полярной системы координат является бесконечно большой величиной. Поэтому нет смысла размещать эту систему в полном её виде. В этом случае применим упрощенный вариант этой системы, то есть разместим необходимое число поступательных плоскостей исследования $S_п$ на различных расстояниях друг от друга по линии тока, а нормальную поверхность исследования $S_н$ продолжим на всю длину потока (рис. 18). Затем проведём исследование изменения нормальных и поступательных сил давления в принятых плоскостях и поверхностях с помощью плоскости фиксации, которая может двигаться со скоростью возмущения *C*.

В определённый момент времени *t* произведем двухэтапное движение пластины. С этого времени плоскость фиксации, передней границей которой служит плоскость фронта волны, начнёт перемещаться со скоростью возмущения *C* в направлении *Б*. Перемещаясь таким образом, она достигнет первой поступательной плоскости исследования $S_{п1}$ за время $t_1$, которое будет равным частному от деления расстояния $l_1$ на скорость возмущения *C*. Следующих плоскостей исследования она достигнет за соответствующее время, которое для каждой плоскости будет равным частному от деления расстояния $l_i$ на скорость возмущения *C*. При этом на плоскости фиксации для каждой плоскости исследования зарегистрируется картина изменения нормальных и поступательных сил давления. Для каждой плоскости исследования она примерно будет соответствовать картине, показанной на рис.17, с той разницей, что силы давления пропорциональны квадрату скорости движения, а не скорости движения. Это значит, что для каждой плоскости исследования фиксируется картина изменения сил давления для первоначальных плоскостей и поверхностей исследования для двухэтапного движения пластины возмущения.

Общая же картина движения акустического потока представляется, как непрерывное движение волны возмущения от пластины возмущения по длине всего потока в направлении *Б*. Поэтому изменение сил давления для любой плоскости и поверхности исследования должно быть записано с помощью уравнений сил поэтапного движения пластины возмущения (26), (27), (30) и (31) или (28), (29), (32) и (33). Следовательно, уравнения сил для первоначальных плоскостей исследования являются общими уравнениями сил для всего акустического потока.

На плоскости фиксации для каждой плоскости исследования картина изменения сил давления для двух этапов движения пластины расположится на некоторой длине, которая называется длиной волны λ и записывается уравнением (III-55). Предположим, что пластина колеблется непрерывно, тогда акустический поток превратится в перемещающуюся цепь волн возмущения.

Отметим, что в поступательных плоскостях исследования $S_п$ и в связанных с ними участках нормальной плоскости $S_н$ в любой фиксированный момент времени характеристики сил давления будут одинаковыми, если между этими поступательными плоскостями по линии тока укладывается целое число длин волн λ при условии, что характер колебаний пластины остаётся неизменным.

Таким образом, мы получили общее уравнение сил акустического потока и выявили общую картину его движения.

***Действие пятое.*** Здесь мы должны сопоставить уравнения сил каждой поверхности исследования относительно друг друга, чтобы выявить общие уравнения сил акустического вида движения.

Исследования, связанные с данным пунктом, мы провели в пункте 4, то есть мы получили там общие уравнения сил акустического потока жидкости. Поэтому здесь мы делаем ссылку на четвёртое действие.

## IV. 5. ОСНОВНЫЕ ПОЛОЖЕНИЯ СУЩЕСТВУЮЩЕЙ МЕХАНИКИ ЖИДКОСТИ И ГАЗА ПО РАЗДЕЛУ ДИНАМИКИ

В существующей механике жидкости и газа действующие силы определяются так называемыми уравнениями движения. Выпишем их из книги [11]:

$$\left.\begin{aligned}
&\frac{\partial v_x}{\partial t}+v_x\frac{\partial v_x}{\partial x}+v_y\frac{\partial v_x}{\partial y}+v_z\frac{\partial v_x}{\partial z}=F_x-\frac{1}{\rho}\frac{\partial P}{\partial x}\\
&\frac{\partial v_y}{\partial t}+v_x\frac{\partial v_y}{\partial x}+v_y\frac{\partial v_y}{\partial x}+v_z\frac{\partial v_y}{\partial z}=F_y-\frac{1}{\rho}\frac{\partial P}{\partial y}\\
&\frac{\partial v_z}{\partial t}+v_x\frac{\partial v_z}{\partial x}+v_y\frac{\partial v_z}{\partial y}+v_z\frac{\partial v_z}{\partial z}=F_z-\frac{1}{\rho}\frac{\partial P}{\partial z}
\end{aligned}\right\},$$

где $\partial v_x, \partial v_y, \partial v_z$ – составляющие скорости по соответствующим осям координат, $F_x, F_y, F_z$ – составляющие объёмных сил по соответствующим осям координат, $t$ – время, $\rho$ – плотность, $P$ – давление.

Эти общие уравнения выражают силы, действующие при движении жидкости. Основные противоречия этих уравнений с уравнениями сил четырёх видов движения заключаются в следующем:

а) что эти уравнения получены на основе инерциальной системы координат;

б) что вывод этих уравнений получен на основе второго закона Ньютона, где сила определяется как произведение массы на ускорение;

в) что действие сил рассматривается на выделенном параллелепипеде потока жидкости, а не на поверхности или плоскости;

г) что эти уравнения допускают движение жидкости одновременно в трёх направлениях, хотя жидкость может двигаться одновременно не более чем в двух взаимно перпендикулярных направлениях;

д) что эти уравнения допускают бесчисленное множество различных видов движения, а положения механики безынертной массы ограничивают их количество четырьмя видами;

е) что объёмные силы и силы давления выделены здесь как разные силы, хотя они имеют одинаковое назначение.

При наличии таких противоречий можно сравнительно легко проверить правильность тех и других уравнений сил. В настоящее время имеется уже достаточное количество практических примеров, подтверждающих уравнения сил четырёх видов движения. На них мы остановимся ниже. [12]

# Г Л А В А V. РАБОТА И ЭНЕРГИЯ

В предыдущих разделах мы ознакомились с понятиями состояния покоя и движения жидкости. Состояние жидкости определяют действующие на неё силы давления, которые проявляются в результате взаимодействия массы жидкости с силовым полем. Практически силы проявляются как эффект в виде либо определённого состояния жидкости, либо изменения её состояния. Как эффект сила существует некоторое мгновение, так как этой силе соответствует определённое мгновение в движения жидкости. Последовательное чередование мгновений сил давления во времени создаёт определённое движение жидкости. Это движение выражается в том, что через некую неподвижную плоскость или поверхность потока проталкивается определённый расход массы соответствующий силам, действующим в тот момент времени. За время действий сил давления через эту поверхность потока проталкивается определённый объём жидкости, который мгновения действия сил давления превращает в реально существующее пространственное восприятие. Количественная сторона этого его определяется понятием работы $L$. Отсюда следует определение понятия «работа»:

---

[12] См. *Приложение* «Примеры применения зависимостей механики безынертной массы»

***работой L сил давления P на неподвижной поверхности или плоскости потока называется произведение объёма жидкости V на силы давления:***

$$L = V P. \qquad (1)$$

***При этом жидкости V является тем объёмом, который был вытолкнут через неподвижную поверхность этими силами давления.***

Работа считается положительной, если расход массы поступает в рассматриваемый объём, и – отрицательной, если расход массы вытекает из него. Размерность работы здесь сохраняется, т. к. работой $L$ в механике называется произведение силы на длину пути.

Понятие работы в механике жидкости и газа выражается объёмом жидкости, который либо уже был вытолкнут действующими силами давления, либо выталкивается ими в настоящий момент времени. Поэтому работа как бы определяет результат действующих или уже действовавших в течение определённого времени сил давления, то есть конкретные результаты действия сил давления.

Поэтому работа выражает **количество** *(выделено автором. – ред.)* затраченных сил давления на вытолкнутый ими объём, а не движение жидкости или действующие силы давления.

В другом случае жидкость может находиться в состоянии покоя или перемещаться в определённом объёме, не совершая работы. Тогда силы давления мгновение своего действия на неподвижной плоскости или поверхности как бы превращают в неопределённо долгое время, то есть в состоянии покоя или установившегося движения силы давления в любой неподвижной плоскости или поверхности остаются неизменными сколь угодно долгое время. В состоянии покоя в жидкости регистрируется силы давления, которые выражают застывшее движение, так как нет выталкиваемого объёма жидкости. Поэтому жидкость в этом случае обладает только располагаемой работой, которая может быть совершена, если дать ей возможность двигаться. Располагаемую работу называют энергией.

При установившемся виде движения силы давления в любой неподвижной плоскости потока выталкивают определённый расход массы жидкости. Здесь мы имеем и действующие силы, и выталкиваемый объём жидкости, но в этом случае выталкивание расхода массы происходит в объёме потока, то есть в целом объём потока остаётся неизменным. Поэтому фактически мы не имеем вытолкнутого объёма жидкости, а значит, и работы тоже. Следовательно, установившийся вид движения тоже имеет только располагаемую работу, или энергию.

Энергия отличается от работы не только временным фактором, но и тем, что при переходе энергии в работу она может иметь множество способов этого перехода, которые будут различаться по расходу массы и по силам давления в широком диапазоне. Поэтому из общих рассуждений невозможно определить количественные зависимости для энергии.

В общих положениях работы и энергии к настоящему моменту мы получили определение работы, энергии и получили общую количественную зависимость для работы (1). Для энергии мы располагаем пока теми знаниями, что размерность её одинакова с работой, и что она определяет количество располагаемой работы.

Дальнейшее изучение работы и энергии будет связано непосредственно с видами движения жидкости и газа.

## V. 1. РАСХОДНОЕ ДВИЖЕНИЕ ИДЕАЛЬНОЙ ЖИДКОСТИ (РАБОТА)

При расходном движении жидкость непрерывно проталкивается через поверхность потока под действием сил давления, тем самым она увеличивает или уменьшает объём потока. В самом общем случае силы давления на поверхности потока непрерывно изменяются во времени. Поэтому работа в каждый промежуток времени при этом виде движения может быть различной.

В этом случае необходимо определять работу в интересующий нас период движения жидкости. Чтобы получить необходимые для нас зависимости работы для этого случая, поступим следующим образом.

Полагаем, что проталкиваемый объём жидкости *V* через поверхность потока *F* есть величина переменная. В этом случае для вычисления работы разобьём интервал изменения сил давления *P* на большое число *n* очень маленьких участков $\Delta P_1, \Delta P_2, \Delta P_3 ... \Delta P_n$ (рис. 24, *а*).

Чтобы определить интервал изменения сил давления *P* воспользуемся плоскостью фиксации. Расположим эту плоскость перпендикулярно поверхности потока. Затем полагаем, что она движется со скоростью жидкости, которая движется на поверхности потока. Тогда на плоскости фиксации будут регистрироваться силы давления, которые в этот момент действовали на поверхности потока (рис 24,*б* ). Используя диапазон изменения давления, зафиксированный на плоскости фиксации, мы получим необходимый интервал изменения сил давления, как это показано на рисунках 24, *а* и 24, *б*. Дальше разбиваем этот интервал на ряд маленьких участков $\Delta P_i$, как это было сказано выше.

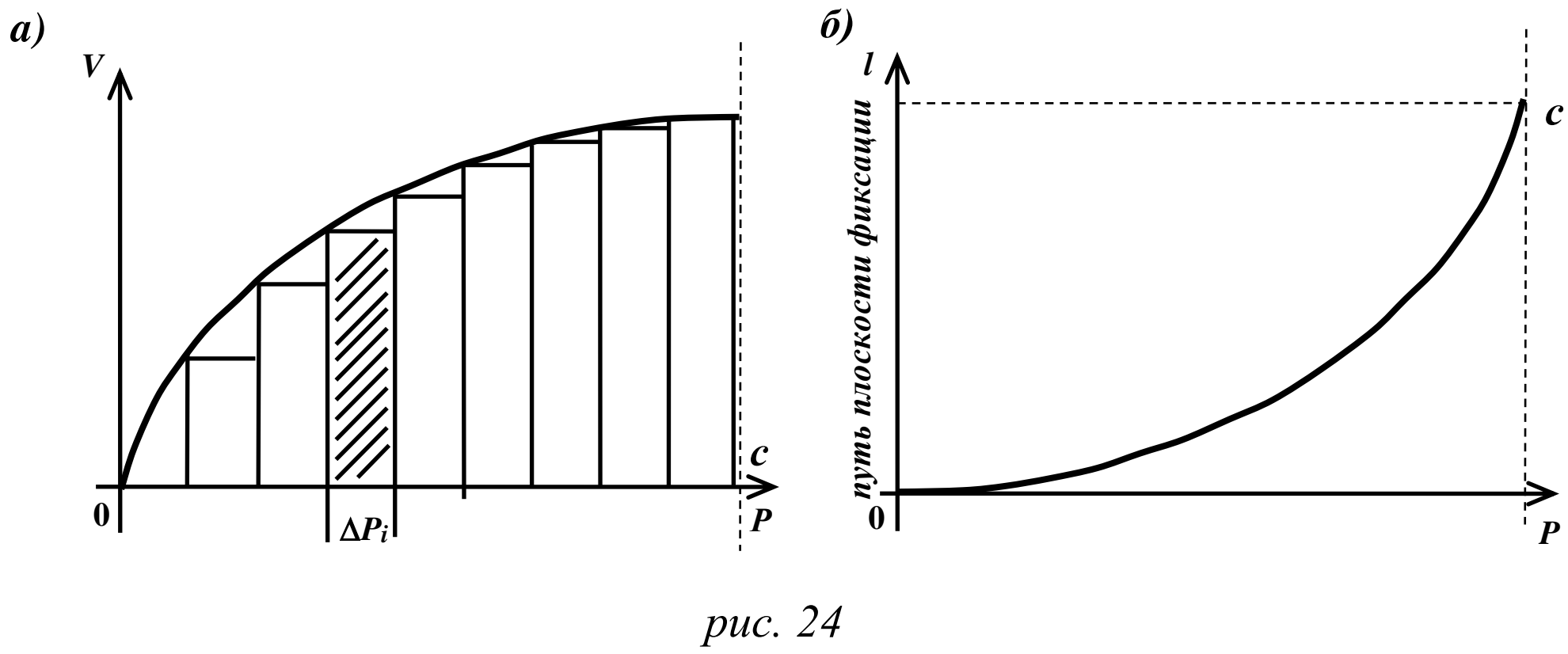

*рис. 24*

Значения переменного объёма в начале каждого из этих участков обозначим соответственно через $V_1, V_2, V_3 ... V_i$. Так как участок $\Delta P_i$ очень мал, то величину $V_i$ на этом участке приближенно можно считать постоянной. Тогда элементарная работа на участке $\Delta P_i$, согласно равенству (1), будет равна:

$$\Delta L = V_i \Delta P_i \ .$$

Взяв сумму этих элементарных работ и переходя к пределу при $\Delta P_i \to 0$ и при $n \to \infty$, получим работу переменного давления на конечном интервале сил давления. В пределе сумма этих элементарных работ выразится определённым интегралом, и мы получим:

$$L = \lim \sum_{i=1}^{n} V_i \Delta P_i = \int_{P_1}^{P_2} V dP \ . \qquad (2)$$

Уравнение (2) является уравнением работ в интегральной форме.

Дальше его можно преобразовать. Например, можно выразить силы давления в этом уравнении через уравнение сил расходного вида движения, тогда получим:

$$dP = \frac{M}{F} dW \ . \qquad (3)$$

Уравнение сил расходного вида движения продифференцировали по скорости потому, что для работы переменной характеристикой является объём.

Подставим в уравнение (2) уравнение (3) и заменим пределы, получим:

$$L = \int_{W_1}^{W_2} V \frac{M}{F} dW \ . \qquad (4)$$

Уравнение (4) тоже является уравнением работ расходного вида движения. Это уравнение можно преобразовать ещё, заменив в нём расход массы по уравнению движения установившегося вида движения ( $M = F\rho W$ ). Тогда оно примет вид:

$$L = \int_{W_1}^{W_2} V\rho W dW \ . \qquad (5)$$

Уравнение (5) тоже является уравнением работ расходного вида движения жидкости.

Таким образом, мы получили уравнение работ расходного вида движения в нескольких вариантах. Все эти уравнения можно отнести к единице объёма, тогда уравнение (2) примет вид:

$$L = \int_{P_1}^{P_2} dP \ ,$$

уравнение (4) примет вид:

$$L = \int_{W_1}^{W_2} \frac{M}{F} dW \ ,$$

уравнение (5) примет вид:

$$L = \int_{W_1}^{W_2} \rho W dW \ .$$

## V. 2. УСТАНОВИВШЕЕСЯ ДВИЖЕНИЕ ИДЕАЛЬНОЙ ЖИДКОСТИ (ЭНЕРГИЯ)

При установившемся движении жидкость движется прямолинейно или по окружности. Границы потока расположены симметрично относительно линии тока. В зависимости от площади сечения потока $F$ жидкость движется либо быстро, либо медленно по отношению к скорости движения. Расход массы потока в единицу времени остаётся неизменным с течением времени. Изобразим такой поток на рисунке 25.

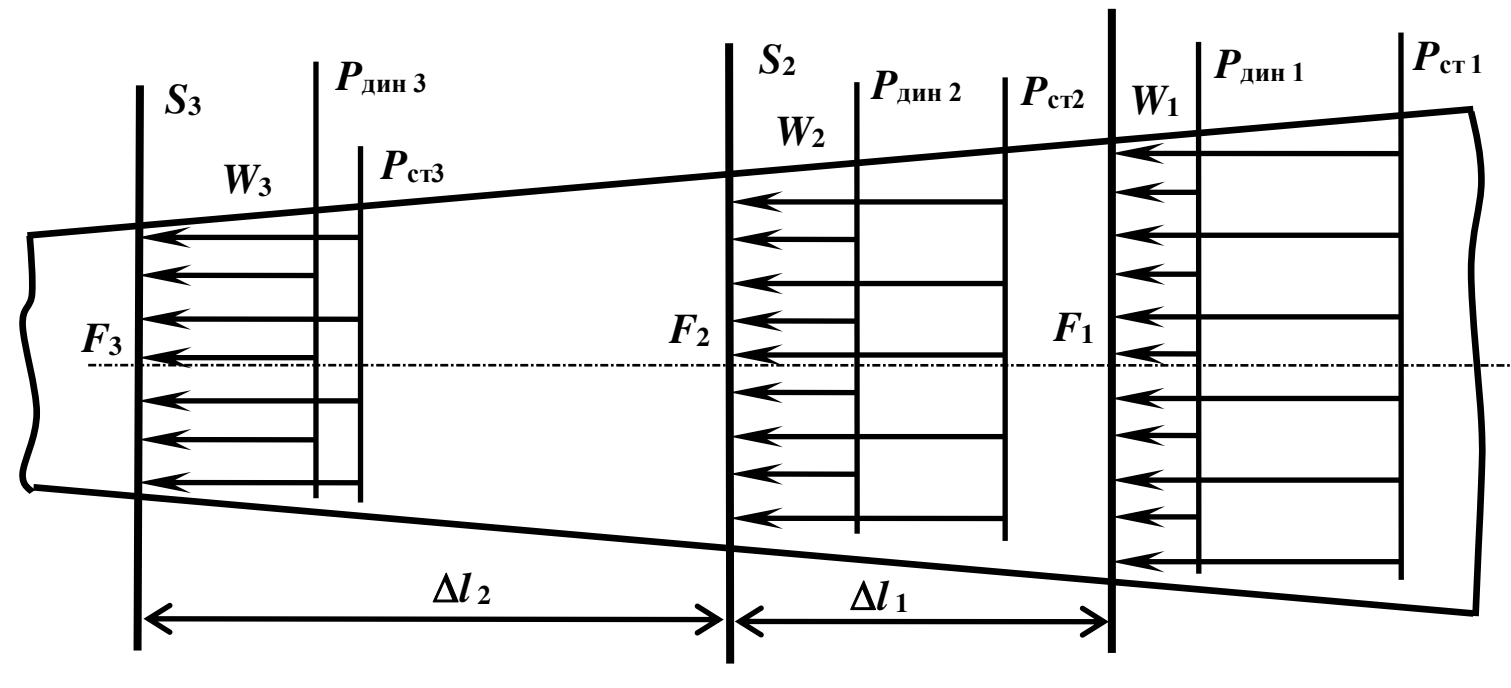

*Рис.25*

Для вывода уравнений энергии расположим по длине потока три плоскости исследования $S_1, S_2, S_3$, соответственно площади сечения потока будут $F_1, F_2, F_3$. На каждую площадь сечения потока будут действовать статические и динамические силы давления. Покажем их на рис.25.

В связи с тем, что мы имеем дело с динамическими и статическими силами давления, то в данном потоке должны быть два вида энергии: динамическая $U_{дин}$ и статическая $U_{ст}$. Чтобы получить для них необходимые зависимости, воспользуемся законом сохранения энергии и уравнениями работ. Согласно определению работы, статическая энергия $U_{ст}$ для каждого сечения потока будет равна произведению вытесненного объёма жидкости $V_{ст.}$ на статическую силу давления $P_{ст}$, тогда получим:

для площади сечения $F_1$: $$U_{ст1} = V_{вт1} P_{ст1},\qquad(6)$$

для площади сечения $F_2$: $$U_{ст2} = V_{вт2} P_{ст2},\qquad(7)$$

для площади сечения $F_3$: $$U_{ст3} = V_{вт3} P_{ст3}.\qquad(8)$$

Статические силы давления определяются по уравнению сил установившегося вида движения жидкости. Чтобы определить вытесненный объём жидкости $V_{вт.}$, мы засекаем определённое время $t$ и с этого времени производим замер вытесненной через каждую площадь сечения потока жидкости. Затем, в следующий момент времени $t_1$ фиксируем вытесненный объём жидкости $V_{вт}$, через каждую из этих площадей исследования. Все вытесненные объёмы жидкости будут равны друг другу, т. к. расход массы в единицу времени является постоянной величиной, а идеальная жидкость не сжимаема.

Отметим, что работа и энергия различаются временным фактором, который определяется именно объёмом жидкости. Работа связана с объёмом жидкости, вытесненным из объёма потока, а энергия связана с тем объёмом, который может быть вытеснен из объёма потока. Поэтому для энергии им является обычно полный объём потока. В нашем случае вытесненный объём мы принимаем по отношению к плоскостям исследования, а по отношению ко всему объёму потока он является частью его объёма. Следовательно, уравнения (6), (7), (8) являются уравнениями *статической энергии*.

Принимаем, что плоскости исследования расположены на бесконечно малом расстоянии друг от друга ($\Delta l_1$ и $\Delta l_2$). Теперь обратимся к величинам статической энергии в плоскостях исследования $S_1, S_2, S_3$, полученных с помощью уравнений (6), (7) и (8). Вытесненный объём жидкости для всех трёх уравнений одинаков, а величины статических давлений $P_{ст1}$, $P_{ст2}$, $P_{ст3}$ – разные. Поэтому величины статических энергий плоскостей исследования тоже будут разными.

Вычтем из статической энергии второй плоскости исследования статическую энергию первой плоскости исследования и из статической энергии третьей плоскости исследования – статическую энергию второй плоскости исследования, получим:

$$\Delta U_{ст1} = U_{ст2} - U_{ст1},\qquad(9)$$

$$\Delta U_{ст2} = U_{ст3} - U_{ст2}.\qquad(10)$$

Согласно закону сохранения энергии разность статических энергий пошла на вытеснение объёма жидкости $V_{вт}$. Тогда работу, затраченную на вытеснение этого объёма, можно определить с помощью уравнения работ (2), получим:

$$\Delta U_{ст1} = \int_{P_{ст1}}^{P_{ст2}} V_{вт}\, dP,\qquad(11)$$

$$\Delta U_{ст2} = \int_{P_{ст2}}^{P_{ст3}} V_{вт}\, dP\qquad(12)$$

Либо вместо уравнения работ (2) можно воспользоваться уравнением работ (5), тогда получим:

$$\Delta U_{ст1} = \int_{W_1}^{W_2} V_{вт}\, \rho W dW,\qquad(13)$$

$$\Delta U_{ст2} = \int_{W_2}^{W_3} V_{вт}\, \rho W dW.\qquad(14)$$

Заменив в уравнениях (9) и (10) величины статических энергий их значениями с помощью уравнений (6), (7), (8), (13) и (14), получим:

$$\int_{W_1}^{W_2} V_{\text{вт}} \rho W dW = V_{\text{вт}} P_{\text{ст}2} - V_{\text{вт}} P_{\text{ст}1}, \tag{15}$$

$$\int_{W_2}^{W_3} V_{\text{вт}} \rho W dW = V_{\text{вт}} P_{\text{ст}3} - V_{\text{вт}} P_{\text{ст}2}. \tag{16}$$

Так как величины расхода массы, сил давления и скорости для каждой плоскости исследования потока являются величинами постоянными, то определённые интегралы левой части уравнений (15) и (16) можно записать с помощью их значений, тогда получим:

$$V_{\text{вт}} \frac{\rho W_1^2}{2} - V_{\text{вт}} \frac{\rho W_2^2}{2} = V_{\text{вт}} P_{\text{ст}2} - V_{\text{вт}} P_{\text{ст}1}, \tag{17}$$

$$V_{\text{вт}} \frac{\rho W_2^2}{2} - V_{\text{вт}} \frac{\rho W_3^2}{2} = V_{\text{вт}} P_{\text{ст}3} - V_{\text{вт}} P_{\text{ст}2}. \tag{18}$$

Уравнения (17) и (18) являются уравнениями энергий. Левая часть этих уравнений является разностью кинетических энергий плоскостей исследования, а правая – разностью статических энергий плоскостей исследования.

Преобразуем уравнения (17) и (18) по плоскостям исследования, то есть перенесём члены уравнений, относящиеся к одной плоскости исследования, в правую часть уравнения, а относящиеся к другой плоскости – в левую часть уравнений, получим:

$$V_{\text{вт}} P_{\text{ст}1} + V_{\text{вт}} \frac{\rho W_1^2}{2} = V_{\text{вт}} P_{\text{ст}2} + V_{\text{вт}} \frac{\rho W_2^2}{2}, \tag{19}$$

$$V_{\text{вт}} P_{\text{ст}2} + V_{\text{вт}} \frac{\rho W_2^2}{2} = V_{\text{вт}} P_{\text{ст}3} + V_{\text{вт}} \frac{\rho W_3^2}{2}. \tag{20}$$

Сопоставив уравнения, получим:

$$V_{\text{вт}} P_{\text{ст}1} + V_{\text{вт}} \frac{\rho W_1^2}{2} = V_{\text{вт}} P_{\text{ст}2} + V_{\text{вт}} \frac{\rho W_2^2}{2} = V_{\text{вт}} P_{\text{ст}3} + V_{\text{вт}} \frac{\rho W_3^2}{2} = \text{const}. \tag{21}$$

Из уравнения (21) следует, что сумма кинетических и статических энергий для каждой плоскости исследования есть величина постоянная.

Отсюда следует, что для любой плоскости исследования потока эта величина тоже является постоянной и равной величине энергии этих плоскостей исследования. Тогда общее уравнение энергии для установившегося вида движения запишется так:

$$U = V_{\text{вт}} P_{\text{ст}} + V_{\text{вт}} \frac{\rho W^2}{2} = \text{const}. \tag{22}$$

Это уравнение можно отнести к единице объёма, то есть разделить левую и правую части уравнения (22) на величину вытесненного объёма жидкости $V_{\text{вт}}$, получим:

$$U = P_{\text{ст}} + \frac{\rho W^2}{2} = \text{const}. \tag{23}$$

В таком виде уравнение энергии более удобно в применении, т. к. оно применимо к любому объёму жидкости. В то же время нельзя забывать, что уравнения (22) и (23) относятся к площади сечения потока.

Отметим, что эти уравнения выражают полную энергию установившегося потока жидкости.

В механике твёрдого тела статическую энергию принято называть потенциальной энергией. По аналогии с механикой твёрдого тела назовём статическую энергию ($V_{вт}P_{ст}$) тоже потенциальной.

Тогда полную энергию потока $U$ так и назовём – полной энергией потока, а кинетическую энергию потока ($U_{кин.} = \frac{1}{2}V_{вт}\rho W^2$) – кинетической энергией потока $U_к$.

В уравнении (23) тоже будем называть соответственно: полная энергия потока для единицы объёма $U$,

потенциальная энергия потока для единицы объёма $|U_{ст}| = |P_{ст}|$,

кинетическая энергия потока для единицы объёма ($U_к = \rho W^2 / 2$).

Впредь будем придерживаться этой терминологии.

С энергетической точки зрения движение установившегося потока жидкости характеризуется непрерывным от сечения к сечению потока переходом потенциальной энергии в кинетическую и кинетической энергии - в потенциальную при постоянстве полной энергии потока.

Датчиками манометрического типа в установившемся потоке жидкости замеряется полная энергия потока и потенциальная энергия потока единицы объёма. В то же время этими же датчиками замеряются статические силы давления потока $P_{ст}$. Поэтому шкала датчиков манометрического типа может быть проградуирована как для замера статических сил давления, так и для замера полной и потенциальной энергии потока для единицы объёма.

Проведём исследование уравнений энергии (22) и (23), которое будет заключаться в том, чтобы определить максимальные и минимальные значения потенциальной и кинетической энергии в установившемся потоке жидкости.

1.Так как установившееся движение характеризует поток жидкости, то в уравнениях энергии потока обязательно должны быть члены, которые характеризуют и потенциальную, и кинетическую энергию потока одновременно. Поэтому в этом потоке обязательно должна быть скорость движения жидкости, которая может быть по величине близкой к нулю, но никогда она не будет равна ему, то есть скорость в своём пределе может стремиться к нулю. Тогда кинетическая энергия потока тоже будет стремиться к нулю, а потенциальная энергия будет стремиться к полной энергии потока. Это значит, что минимальное значение кинетической энергии практически может иметь нулевое значение, а максимальное значение потенциальной энергии практически может быть равно полной энергии потока. В этом случае абсолютная величина потенциальной энергии для единицы объёма равна абсолютной величине статических сил давления для единицы площади $P_{ст}$, и эта величина будет максимальной величиной статических сил давления потока.

2. Теперь предположим, что на плоскости исследования $S_3$ для площади сечения потока $F_3$ (рис. 25) динамические силы давления $P_{дин}$ по абсолютной величине достигнут максимальной величины статических сил давления $P_{ст}$. В этом случае скорость жидкости достигнет максимальной величины, и динамические силы давления тоже будут иметь максимальное значение ($P_{дин.} = \rho W^2$), т. к. прирост скорости происходит за счёт статических сил давления.

Тогда в этом сечении потока динамические и статические силы давления (максимальные) будут равны по величине ($P_{ст.(макс.)} = P_{дин.(макс.)} = \rho W^2_{(макс.)}$), а уравнение энергии (23) будет иметь

такие значения:

$$U_{\text{п}} = \frac{U}{2} = P_{\text{ст}}$$
$$U_{\text{к}} = \frac{U}{2} = \frac{\rho W_{\max}^2}{2}. \qquad (24)$$

Из уравнения (24) следует, что максимальная величина кинетической энергии потока достигает половины величины полной энергии потока *U*. Из уравнения (24) можно также определить возможную максимальную скорость жидкости в установившемся потоке жидкости.

Таким образом, мы определили максимально возможные величины потенциальной и кинетической энергии в установившемся потоке жидкости:

потенциальная энергия потока изменяется от половины полной энергии потока $U/2$ до величины полной энергии потока *U*, а кинетическая энергия изменяется от нуля до половины полной энергии потока, то есть

$$U/2 \le U_{\text{п}} < U,$$
$$0 < U_{\text{к}} \le U/2. \qquad (25)$$

Посмотрим теперь положения существующей механики жидкости и газа по вопросу механической энергии.

Уравнение энергии, типа уравнения энергии (23), для единицы объёма в сечении установившегося потока впервые было получено Д. Бернулли. В механике жидкости и газа оно известно под названием механической энергии потока, или просто уравнения Бернулли. В прикладных науках, таких как гидравлика [11], газовая динамика [7], с этим уравнением связаны следующие противоречия:

когда идёт речь об энергии в потоке жидкости, то вывод и объяснение к уравнению Бернулли даются в соответствии с положениями об энергии. В то же время, когда речь идёт о силах давления в потоке жидкости, то уравнение Бернулли превращается в уравнение сил, то есть полная *энергия* потока превращается в полное *давление* потока: потенциальная энергия – в статическое давление, а кинетическая энергия – в динамическое давление. Так что, например, в книгах [11] и [7] уравнение Бернулли выполняет две функции одновременно: оно является и уравнением механической энергии, и уравнением сил давления потока.

Пожалуй, такое явное противоречие связано с тем, что пока не научились измерять динамические силы давления, т. к. «динамическое давление», вычисленное по уравнению Бернулли, ровно в два раза меньше действительных динамических сил давления, действующих в потоке жидкости. С помощью замеров этих давлений можно было бы просто выявить их количественное различие.[13]

## V. 3. ПЛОСКИЙ УСТАНОВИВШИЙСЯ ВИД ДВИЖЕНИЯ ИДЕАЛЬНОЙ ЖИДКОСТИ (ПЕРЕХОД ЭНЕРГИИ ПОТОКА С БОЛЬШИМ ПОТЕНЦИАЛОМ В ПОТОК С МЕНЬШИМ ПОТЕНЦИАЛОМ)

Поток плоского установившегося вида движения жидкости представляет собой цилиндрический объём, где жидкость движется одновременно в двух направлениях: тангенциальном и радиальном. Радиальные и тангенциальные расходы масс в единицу времени как для всего потока, так и для любой его точки есть величина постоянная во времени. Объём потока составляют плоскости тока, прилегающие друг к другу. В его поперечном разрезе они образуют линии тока, которые имеют форму логарифмической спирали.

Уравнения движения и сил для этого вида движения были получены в предыдущих разделах. Теперь остаётся получить для него соответствующие уравнения энергии. Как для остальных разделов механики жидкости и газа, так и для данного раздела принимаем общий метод исследования. Постараемся воспользоваться им в несколько упрощенной форме, то есть не будем рассматривать исследование по действиям.

[13] в главе VI (см. Часть II) приводится конструктивная схема прибора для измерения динамических сил давления

Дальше изобразим поток плоского установившегося вида движения (рис. 26, *а*). На одной из линий тока примем точку *А* и проведём в ней исследование. Отметим, что ось цилиндрического объёма потока совпадает с осью полярной системы координат.

Затем выделим элемент линии тока с точкой *А* в увеличенном масштабе и разместим в этой точке радиальную $S_r$ и тангенциальную $S_{tg}$ плоскости исследования, изобразим на них действующие силы (рис. 26, *б*).

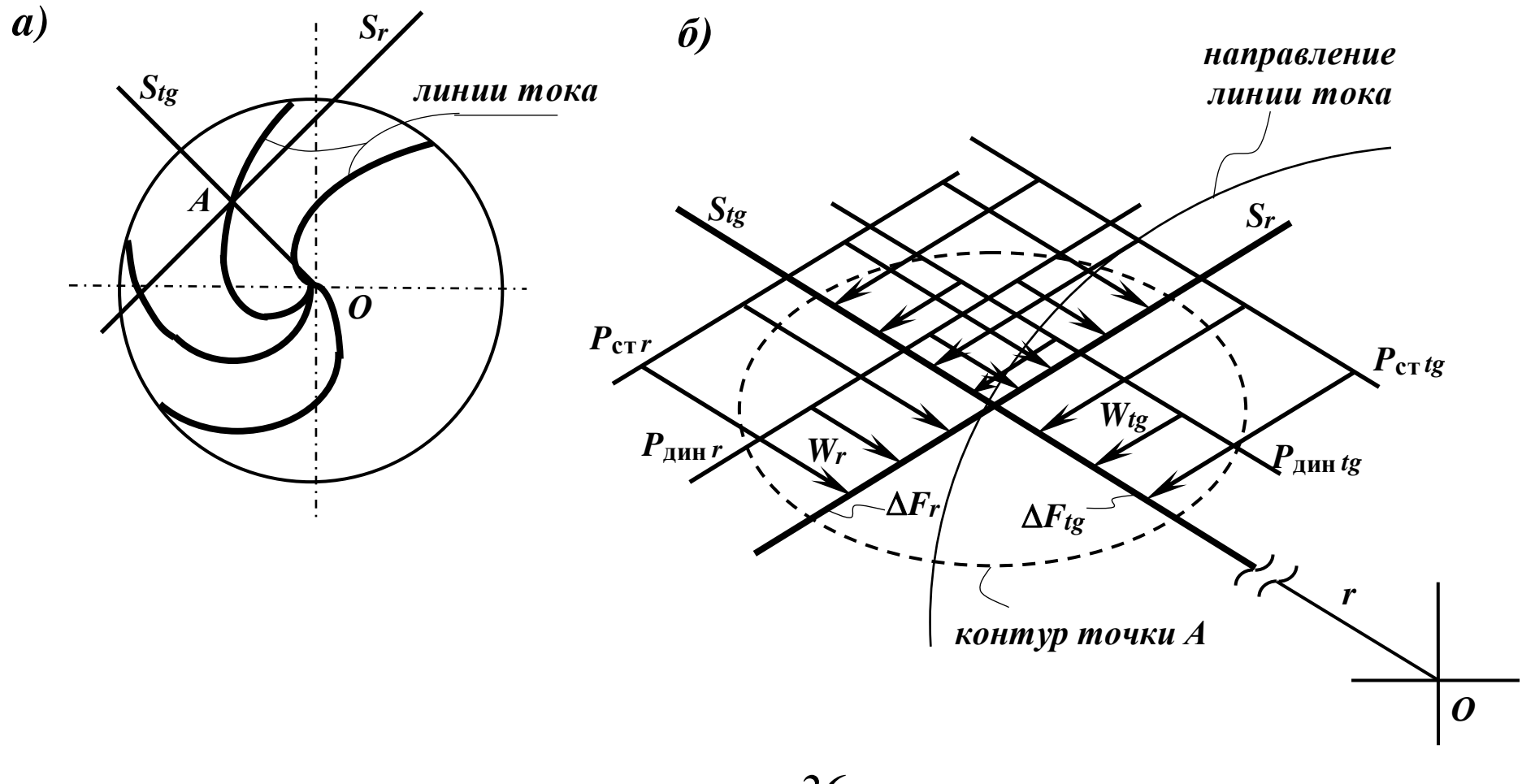

*рис. 26*

На тангенциальной и радиальной плоскостях исследования будут действовать соответствующие динамические и статические силы давления, которые записываются уравнениями сил (IV - 9), (IV - 10), (IV - 11), (IV - 12). Эти уравнения аналогичны уравнениям сил установившегося вида движения жидкости, для которого мы уже получили уравнения энергии (22) и (23).

В плоском установившемся виде движения мы тоже имеем дело с энергией. Поэтому вывод уравнений энергии плоского установившегося вида движения будет аналогичен выводу уравнений энергии установившегося вида движения жидкости.

В связи с этим воспользуемся здесь аналогией уравнений энергии плоского установившегося вида движения и установившегося вида движения и запишем их без вывода.

В соответствии с уравнением энергии (22) и уравнениями сил (IV-9) и (IV-10), уравнения энергии плоского установившегося вида движения в исследуемой точке *А* линии тока будут иметь вид:

*в радиальном направлении*:
$$U_r = V_{вт.r} P_{ст.r} \pm V_{вт.r} P_{дин.r}, \qquad (26)$$

*в тангенциальном направлении*:
$$U_{tg} = V_{вт.tg} P_{ст.tg} \pm V_{вт.tg} P_{дин.tg}. \qquad (27)$$

Уравнения (26) и (27) можно отнести к единице объёма выталкиваемой жидкости, то есть разделить эти уравнения соответственно на выталкиваемые объёмы жидкости. Тогда по аналогии с уравнением (23) они примут вид:

*в радиальном направлении*:
$$U_r = P_{ст.r} \pm \frac{\rho W_r^2}{2}, \qquad (28)$$

*в тангенциальном направлении*:
$$U_{tg} = P_{ст.tg} \pm \frac{\rho W_{tg}^2}{2}. \qquad (29)$$

Пользоваться уравнениями (28) и (29) более удобно, чем уравнениями (26) и (27). Для любой точки потока уравнения энергий будут иметь вид уравнений (26), (27) или (28), (29). Следовательно, эти уравнения являются общими уравнениями энергии для каждой точки объёма плоского установившегося потока жидкости.

Теперь нам необходимо определить их общность в отношении всего потока и найти максимальные и минимальные пределы величин кинетической и потенциальной энергий в радиальном и тангенциальном направлении.

***В отношении общности***

Уравнения энергии точки потока плоского установившегося вида движения не включают в себя непосредственно площадь сечения потока, но они, как и уравнения энергии установившегося вида движения (22) и (23), относятся к определённой плоскости исследования потока. В то же время они, как и уравнения (22) и (23), записывают полную энергию потока в его точке, а полная энергия одинакова для всех точек потока. По этой причине полная энергия любой точки потока будет полной энергией и для всего потока в целом. Следовательно, уравнения энергии (26), (27) или (28), (29) являются общими уравнениями энергии для всего потока в целом.

***В отношении распределения кинетической и потенциальной энергий***

Для этого проведём исследование распределения энергии по одной из линий тока плоского установившегося потока. Изобразим этот поток и линию тока *на рисунке 27*. Поток примем со сравнительно конкретными условиями движения.

Полагаем, что поток находится в среде, полная энергия которой равна некоторой величине $U_1$. Жидкость движется по направлению к оси потока и переходит в среду, полная энергия которой равна $U_2$, при этом $U_1 \rangle U_2$. Возьмём точку $A$ на линии тока, как показано на рис. 27, и запишем для неё уравнения энергий в соответствии с конкретными условиями движения.

Тангенциальный расход массы в единицу времени в точке $A$ будет втекать в объём точки $A$. Поэтому тангенциальные динамические силы давления $P_{\text{дин..}tg}$ будут совершать положительную работу по отношению к объёму точки. Поэтому динамическая составляющая энергии будет иметь положительный знак. Тангенциальное уравнение энергии примет вид:

$$U_{tg} = P_{\text{ст.}tg} + \frac{\rho W_{tg}^2}{2}. \qquad (30)$$

Радиальный расход массы $M_r$ будет вытекать из объёма точки $A$. Поэтому радиальные силы давления будут совершать отрицательную работу по отношению к объёму точки, а динамическая составляющая энергии тоже будет иметь отрицательный знак. Радиальное уравнение энергии примет вид:

$$\Delta U_r = P_{cm.r} - \frac{\rho W_r^2}{2}. \qquad (31)$$

В этом случае уравнение (31) выражает не полную энергию потока, а разность между потенциальной и кинетической энергией потока.

Отметим, что в любых уравнениях плоского установившегося вида движения одноимённые члены радиальных и тангенциальных составляющих равны между собой по абсолютной величине, например, динамические составляющие уравнений сил, уравнений движения и уравнений энергии. Это значит, что в уравнениях (30) и (31) тоже соответствующие члены уравнений, такие как потенциальная энергия, и как кинетическая энергия тоже равны между собой по абсолютной величине.

Теперь с помощью уравнений (30) и (31), которые учитывают конкретные условия движения жидкости, проведём исследование распределения энергии на длине линии тока.

Для этого полагаем, что точка $A$ теперь расположена на границе потока и занимает положение точки $A_2$ (рис. 27, *а*). В этом случае тангенциальные и радиальные скорости стремятся к нулю. Тогда тангенциальные и радиальные потенциальные энергии и разность энергий в уравнении (31) будут стремиться к полной энергии потока, которая равна полной энергии среды $U_1$, то есть

$$\lim_{W_{tg} \to 0} \frac{\rho W_{tg}^2}{2} = 0,$$

$$\lim_{W_{tg}\to 0} P_{ст.tg} = U_{tg} = U_1,$$

$$\lim_{W_r\to 0} \frac{\rho W_r^2}{2} = 0,$$

$$\lim_{W_r\to 0} P_{ст.r} = \Delta U_r = U_1.$$

Таким образом, мы получили первое крайнее положение для распределения кинетической и потенциальной энергии.

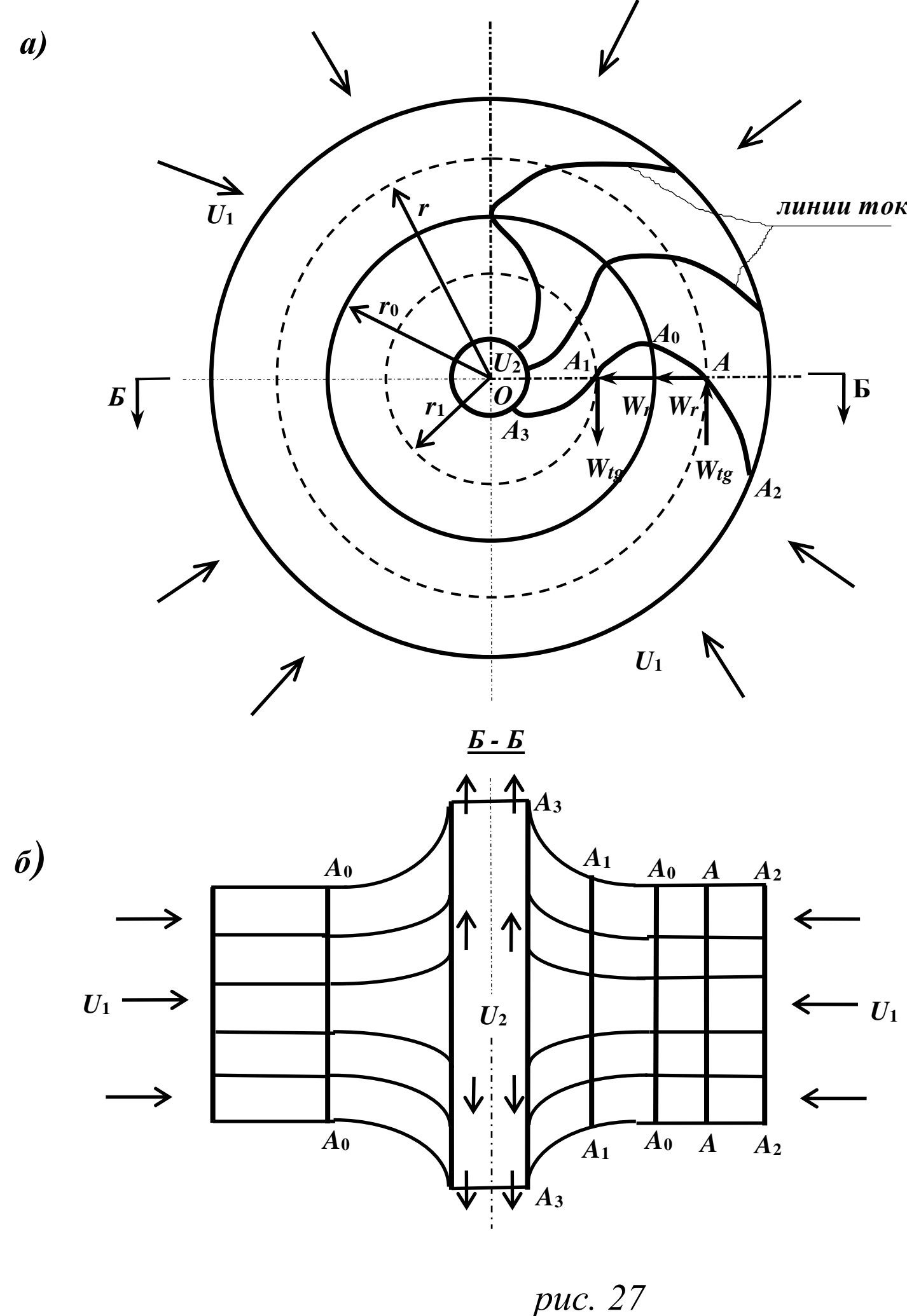

*рис. 27*

Теперь полагаем, что точка $A$ заняла положение точки $A_0$, в котором потенциальная и кинетическая составляющие обоих направлений равны между собой, т.е.

$$\left|P_{ст.tg}\right| = \left|\frac{\rho W_{tg}^2}{2}\right| = \left|\frac{U_1}{2}\right|,$$

$$\left|P_{ст.r}\right| = \left|\frac{\rho W_r^2}{2}\right| = \left|\frac{U_1}{2}\right|,$$

В связи с тем, что в этом случае и потенциальная, и кинетическая составляющие равны половине полной энергии потока, то разность энергий в уравнении (31) будет равна нулю:

$$\Delta U_r = P_{ст.r} - \frac{\rho W_r^2}{2} = 0. \quad (32)$$

Уравнение (30) так и будет выражать полную энергию потока, то есть

$$U_{tg} = P_{cm.tg} + \frac{\rho W_{tg}^2}{2} = U_1. \qquad (33)$$

Равенство (32) означает, что вся потенциальная энергия потока перешла в кинетическую, а равенство (33) показывает, что полная энергия потока в предельном случае распределяется по пятьдесят процентов между потенциальной и кинетической энергиями потока. При этом максимальная величина статических сил давления равна максимальной величине динамических сил давления в соответствии с положениями установившегося вида движения, которые мы рассматривали в предыдущем пункте. Это значит, что в потоке точки $A_0$ отсутствуют статические силы давления и присутствуют только динамические силы давления.

Точка $A_0$ одновременно является границей раздела между двумя потоками с разными уровнями энергии. При движении из точки $A_2$ в точку $A_0$ был поток с энергией $U_1$, а за точкой $A_0$ начинается поток с энергией $U_2$. Поэтому величина полной энергии потока, записанная за точкой $A_0$ уравнением (30), становится недействительной и само уравнение (30) тоже неприемлемо для данной области потока.

В то же время за точку $A_0$, согласно уравнениям (31) и (32), поступает поток жидкости с кинетической энергией равной половине величины полной энергии потока среды $U_1$. Следовательно, эта энергия должна быть использована в другой части плоского установившегося потока жидкости, т. к. связь между потоками с разными уровнями энергии осуществляется за счёт потока жидкости с кинетической энергией. Поэтому продолжим свои исследования.

Обратимся опять к рисунку 27. Мы видим, что, начиная от точки $A_0$, линия тока резко меняет своё направление, но, в то же время, остаётся логарифмической спиралью. Возьмём на этой линии точку $A_1$. Радиальный расход массы в единицу времени будет втекать в объём точки $A_1$, как показано на рис. 27. Поэтому работа радиальных сил давления будет положительной относительно объёма точки и кинетическая энергия радиального движения жидкости тоже будет положительной. Тогда уравнение энергии радиального движения жидкости будет иметь вид:

$$U_r = P_{\text{ст}.r} + \frac{\rho W_r^2}{2}. \qquad (34)$$

Тангенциальный расход массы в единицу времени в точке $A_1$ будет вытекать из а этой точки. Поэтому работа тангенциальных сил давления относительно объёма этой точки будет отрицательной и кинетическая энергия тангенциального движения жидкости тоже будет отрицательной. Уравнение энергии тангенциального движения жидкости будет иметь вид:

$$\Delta U_{tg} = P_{cm.tg} - \frac{\rho W_{tg}^2}{2}. \qquad (35)$$

Как видим, уравнение (34) радиального движения жидкости здесь выражает полную энергию потока, а уравнение энергии (35) тангенциального движения жидкости − разность потенциальной и кинетической энергий. С помощью уравнений (34) и (35) проведём исследование на линии тока от точки $A_0$ до точки $A_3$ – границы плоского установившегося потока жидкости.

Теперь считаем, что точка $A_0$ переместилась в точку $A_1$ и заняла её место. В этой точке потенциальная энергия радиальной и тангенциальной составляющих движения будут равны нулю. Тогда полная энергия потока и разность энергий будет равна радиальной кинетической энергии, то есть при $P_{cm.r} = P_{cm.tg} = 0$,

$$U_r = \Delta U_{tg} = \frac{\rho W_r^2}{2}. \qquad (36)$$

Мы определили третье крайнее положение для линии тока.

Затем переместим точку $A_1$ на границу объёма потока в точку $A_3$ (рис. 27, *а*). В точке $A_3$ радиальные и тангенциальные скорости будут стремиться к нулю. Тогда в уравнении (34) радиальная кинетическая энергия тоже будет стремиться к нулю, а радиальная потенциальная энергия будет стремиться к полной энергии потока, которая равна энергии среды с новым энергетическим уровнем $U_2$. Запишем всё это в математической форме:

$$\lim_{W_r \to 0} \frac{\rho W_r^2}{2} = 0,$$

$$\lim_{W_r \to 0} P_{\text{ст}\,r} = U_r = U_2.$$

В уравнении (35) тангенциальная кинетическая энергия тоже будет стремиться к нулю, а потенциальная энергия и разность энергий будут стремиться к полной энергии потока, которая будет равна энергии среды с новым энергетическим уровнем $U_2$, запишем:

$$\lim_{W_{tg} \to 0} \frac{\rho W_{tg}^2}{2} = 0,$$

$$\lim_{W_{tg} \to 0} P_{\text{ст}\,tg} = \Delta U_{tg} = U_2.$$

Мы получили предельные значения энергии на второй границе плоского установившегося потока жидкости.

Рассматривая все четыре предельных перехода потенциальной и кинетической энергий потока, мы сможем теперь определить переход среды с одного энергетического уровня на другой.

Вернёмся опять к рисунку 27. Если теперь применить для анализа скоростей в точках $A$ и $A_1$ инерциальную систему координат, то мы должны были бы записать, что радиальные скорости этих точек имеют одинаковое направление. Поэтому они должны иметь одинаковый знак, а у нас они имеют разные знаки. Действительно, радиальные скорости имеют одно направление, хотя они имеют разные знаки.

Тангенциальные скорости в точках $A$ и $A_1$ с точки зрения инерциальной системы координат имеют противоположные направления. Действительно, жидкость движется в этом потоке в двух направлениях. Поэтому они должны иметь разные знаки. У нас эти скорости тоже имеют разные знаки для этих точек. Следовательно, при определении положительных и отрицательных величин при помощи инерциальной и неинерциальной систем координат есть существенное различие, которое делает эти системы независимыми друг от друга.

Мы рассмотрели распределение потенциальной и кинетической энергии по одной из линий тока плоского установившегося потока жидкости. Теперь рассмотрим это распределение по отношению ко всему объёму потока. Распределение энергии в потоке тоже будет происходить в соответствии с тем распределением, которое происходит по линии тока. Разница будет заключаться лишь в том, что для потока оно рассматривается на цилиндрической поверхности с постоянным радиусом $r$, который определяется точкой на линии тока. Значит, определив распределение кинетической и потенциальной энергий, например, в точке $A$, которая определяет радиус $r$, мы должны перенести это распределение на всю цилиндрическую поверхность потока с этим радиусом $r$ (рис. 27). Следовательно, сделав распределение кинетической и потенциальной энергии по линии тока, мы тем самым делаем распределение энергии по всему потоку в целом.

Отметим, что площадью сечения потока являются цилиндрические поверхности, т. к. для плоского установившегося потока определяющим является радиальный расход массы в единицу времени. Так как этот расход является постоянным для этого потока, то, как было показано выше, величина тангенциальной и радиальной скоростей движения зависит от площади сечения потока. От точки $A_2$ до точки $A_0$ происходит уменьшение площади сечения потока за счёт уменьшения радиуса этих цилиндрических поверхностей. Поэтому происходит увеличение скорости движения жидкости, и в точке $A_0$ она достигает своей максимальной величины. В связи с этим происходит увеличение радиальных и тангенциальных динамических сил давления, которые в точке $A_0$ достигают своей максимальной величины.

При движении от точки $A_0$ до точки $A_3$ происходит уменьшение скорости движения жидкости. Поэтому происходит уменьшение динамических сил давления и кинетической энергии, которые в точке $A_3$ превращаются в ноль. Следовательно, площади сечения потока должны возрастать от точки $A_0$ до точки $A_3$, и в точке $A_3$ они достигнут своей максимальной величины.

Площади сечения потока и в этом случае являются цилиндрическими поверхностями, но радиус их при движении от точки $A_0$ до точки $A_3$ всё время уменьшается. Поэтому увеличение площади сечения возможно за счёт увеличения высоты этих цилиндрических поверхностей. На рисунке 27, *б* так и показано: от точки $A_2$ до точки $A_0$ высота цилиндрической поверхности сечения потока является постоянной величиной, а от точки $A_0$ до точки $A_3$ она непрерывно возрастает и в точке $A_3$ имеет максимальную высоту.

С точки зрения площадей сечения плоский установившийся поток жидкости можно охарактеризовать следующим образом: в точке $A_0$ поток имеет минимальную площадь сечения, а при движении от точки $A_0$ к точке $A_2$ или $A_3$ происходит увеличение площади сечения потока[14].

## V. 4. ЭНЕРГИЯ СОСТОЯНИЯ ПОКОЯ ИДЕАЛЬНОЙ ЖИДКОСТИ

Состояние покоя жидкости определяется двумя силовыми полями: скалярным и векторным. При действии скалярного поля сил на жидкость силы давления в ней определяются законом Паскаля как действие внешних сил. При этом в жидкости существуют только статические силы давления $P_{ст}$. Её состояние покоя рассматривается в некотором объёме $V$, то есть жидкость в скалярном силовом поле находится под действием статических сил давления, которые в любой точке этого объёма равны по величине.

Работа и энергия определяются произведением сил давления на объём жидкости. Для работы это уже вытолкнутый объём жидкости, а для энергии это объём, который может быть вытолкнут силами давления. Поэтому энергия запишется следующим равенством:

$$U = VP_{ст}, \qquad (37)$$

где $V$ – полный объём жидкости, которая находится в состоянии покоя, $P_{ст}$ – статические силы давления в этом объёме.

Эту энергию можно отнести к единице объёма, то есть разделить левую и правую части уравнения (37) на объём, тогда получим:

$$|U| = |P_{ст}| \qquad (38)$$

В таком виде уравнение энергии более удобно в пользовании, т. к. её можно отнести, например, к какой-то части объёма. При этом статические силы давления $P_{ст}$ и энергия единицы объёма равны по абсолютной величине, что является большим удобством при практическом измерении энергии.

Энергия и статические силы давления измеряются датчиками манометрического типа. Поэтому достаточно провести градуировку шкалы этого датчика для замера энергии и для замера статических сил давления. Следовательно, уравнения (37) и (38) являются общими уравнениями энергии для объёма жидкости, находящейся в состоянии покоя в скалярном силовом поле.

Выше мы назвали энергию, которая образуется от действия статических сил давления, потенциальной. Поэтому уравнения (37) и (38) также выражают потенциальную энергию объёма жидкости, которая находится в скалярном силовом поле.

В векторном силовом поле жидкость находится в состоянии покоя под действием направленных сил давления. Примером такого состояния может служить жидкость, которая находится в состоянии покоя в поле земного тяготения. Величина сил давления, действующих в этой жидкости, зависит от высоты $h$, которая определяется в направлении действия силового поля (рис. 6). Зависимость величины сил давления от высоты $h$ определяется уравнением (II-15). Состояние покоя жидкости является состоянием застывшего движения, поэтому объём жидкости определяется как объём, который может быть вытолкнут действием сил векторного силового поля.

---

[14] данное положение уточняется в работе «Строение Солнца и планет солнечной системы с точки зрения механики безынертной массы»

Обозначим этот возможный объём жидкости как $V_{вт}$. Тогда энергию этого объёма можно выразить определённым интегралом, как:

$$U_h = \int_{P_1}^{P_2} V_{вт}\, dP. \qquad (39)$$

Выразим силы давления с помощью уравнения (II-15), которое определяет силы давления для жидкости, находящейся в состоянии покоя в векторном силовом поле. Тогда получим:

$$dP = \rho w^2 dh, \qquad (40)$$

Где $\rho$ – плотность есть постоянная величина,
$w$ – скорость, которая тоже есть постоянная величина. В поле земного тяготения по абсолютной величине она равна корню квадратному из ускорения $g$ в поле земного тяготения $|w| = \left|\sqrt{g}\right|$;.

Так как переменной величиной у нас является высота $h$, то нам придётся поменять пределы в определённом интеграле. Подставим равенство (40) в уравнение (39), получим:

$$U_h = V_{вт} \rho w^2 \int_{h_1}^{h_2} dh. \qquad (41)$$

Подставив пределы в интеграл, получим:

$$U_h = V_{вт} \rho w^2 (h_2 - h_1). \qquad (42)$$

Отнесём это уравнение к единице объёма, то есть разделим левую и правую части уравнения (42) на объём $V_{вт}$, получим:

$$U_h = \rho w^2 (h_2 - h_1). \qquad (43)$$

Уравнения (42) и (43)являются уравнениями энергии для жидкости, находящейся в состоянии покоя в векторном силовом поле. Замеры, как сил давления, так и потенциальной энергии, производятся на соответствующих высотах датчиками манометрического типа.

## V. 5. АКУСТИЧЕСКИЙ ВИД ДВИЖЕНИЯ ИДЕАЛЬНОЙ ЖИДКОСТИ (РАБОТА И ЭНЕРГИЯ)

Акустический поток жидкости образуется в результате возвратно-поступательного, или колебательного, движения пластины. Под действием этой пластины происходит поступательное и нормальное движение жидкости. Это движение образует волну возмущения, которая перемещается в потоке жидкости со скоростью возмущения $C$.

Скорость возмущения определяет общую направленность движения акустического потока. По отношению к направлению движения волны, возмущения образуют фронт возмущения, который определяется плоскостью, перпендикулярной направлению движения акустического потока. Фронт возмущения движется со скоростью возмущения ($C$). Непосредственное движение жидкости как бы складывается из плоского установившегося и расходного вида движения.

Вспомнив общие характеристики акустического потока, перейдём к его исследованию с точки зрения работ и энергии. Опять воспользуемся общим методом исследования и опытом предыдущих разделов механики жидкости и газа.

Исследование движения жидкости начнём с изучения движения пластины в два этапа. Опять заменим пластину возмущения поршнем, как показано на рис. 14. Границы выделенного участка потока определим из условия, что поступательная площадь сечения потока $F_п$, умноженная на величину хода поршня $l$, должна равняться единице объёма $v$. Это условие в дальнейшем облегчит нам исследование. Нормальную и поступательную плоскости исследования разместим тоже, как показано на рис. 14.

За время первого этапа поршень переместится из положения I в положение II. Перед поршнем находится жидкость в состоянии покоя, которая обладает определённой величиной потенциальной энергии $U$. Переместившись из положения I в положение II, поршень вытеснит единицу объёма $v$. Теперь полагаем, что поршень начал своё движение из положения I в положение II. При этом он будет выталкивать через поступательную плоскость исследования $S_п$ определённый объём жидкости $v$, то есть будет совершать работу, и за время первого этапа вытолкнет единицу объёма $v$. Воспользуемся уравнением работ (2) и запишем эту работу:

$$L_п = v\int_{P_1}^{P_2} dP. \quad (44)$$

Выразим силы давления через плотность $\rho$ и поступательную скорость $W_п$, то есть воспользуемся уравнением работ (5), получим:

$$L_п = v\rho\int_{W_{п1}}^{W_{п2}} W_п dW. \quad (45)$$

Своё движение жидкость начинает с определённой поступательной скорости $W_{п1}$, а в конечном положении II поршня поступательная скорость будет равна нулю, т. к. поршень в этот момент имеет нулевую скорость. Тогда поступательная скорость $W_{п2}$ тоже будет равна нулю, то есть $W_{п2} = 0$. Верхний предел интеграла (45) тоже будет равен нулю. Перепишем его:

$$L_п = v\rho\int_{W_{п1}}^{0} W_п dW. \quad (46)$$

В то же время выталкиваемая жидкость поступает в среду, где жидкость в состоянии покоя обладает определённым уровнем потенциальной энергии $U$. Поэтому к потенциальной энергии среды в поступательной плоскости исследования $S_п$ добавится работа вытесненной единицы объёма жидкости. Следовательно, чтобы получить полную работу и энергию в поступательной плоскости $S_п$ в момент первого этапа движения поршня, мы должны сложить потенциальную энергию и поступательную работу, получим:

$$\sum U_п = U + v\rho\int_{W_{п1}}^{0} W_п dW. \quad (47)$$

Уравнение (47) выражает сумму потенциальной энергии и поступательной работы в возмущённом объёме жидкости за время первого этапа движения поршня.

За это же время через нормальную поверхность исследования $S_н$ будет вытолкнута тоже единица объёма $v$, то есть работа будет иметь отрицательный знак. Нормальная скорость движения в начальный момент времени тоже будет иметь некоторую величину $W_{н1}$, которая в конце этапа превращается в ноль. Тогда работа для единицы объёма будет иметь вид:

$$L_н = -v\rho\int_{W_{н1}}^{0} W_н dW. \quad (48)$$

Так как нормальная работа имеет отрицательный знак, то нам придётся её вычесть из потенциальной энергии жидкости $U$, получим:

$$\Delta U_н = U - v\rho\int_{W_{н1}}^{0} W_н dW. \quad (49)$$

Уравнение (49) выражает разницу потенциальной энергии и нормальной работы в возмущённом объёме жидкости за время первого этапа движения поршня.

Теперь рассмотрим второй этап движения поршня - из положения II в положение I (рис. 14). После аналогичных рассуждений мы получим уравнения такого же типа, что и уравнения (47) и (49) для поступательного и нормального движения жидкости. С той лишь разницей, что работа поступательного движения будет иметь отрицательный знак, а работа нормального движения будет иметь положительный знак, т. к. поступательный расход массы будет вытекать из объёма потока, а нормальный – втекать. Запишем эти уравнения по аналогии:

*для поступательного движения:*

$$\Delta U_{\text{п}} = U - v\rho \int\limits_{W_{\text{п}1}}^{0} W_{\text{п}} dW \,, \tag{50}$$

*для нормального движения:*

$$\sum U_{\text{н}} = U + v\rho \int\limits_{W_{\text{н}1}}^{0} W_{\text{н}} dW . \tag{51}$$

Уравнения (50) и (51) выражают работу и энергию поступательного и нормального движения жидкости в возмущённом объёме жидкости за время второго этапа движения поршня.

В уравнениях данного пункта пятой главы везде проставлена величина единицы объёма *v*. Это сделано потому, что она определяет нам величину участка выделенного из акустического потока объёма жидкости. Единица объёма здесь равна произведению длины *l* хода пластины возмущения на поступательную площадь сечения выделенного объёма потока.

Уравнениями (47), (49) и (50), (51) записан непрерывный переход положительной работы в отрицательную на участке объёма потока равным длине волны λ.

Работа каждого этапа движения поршня определяется тем же способом, что и скорости движения, то есть аналитическим или графическим способом. Исследование скоростей, а следовательно, и работ производится с помощью плоскостей фиксации, которые перемещаются вместе с фронтом волны со скоростью возмущения (*C*), и на которых фиксируются скорости и силы давления.

Определяющим в этих уравнениях является работа. Поэтому можно сказать, что уравнения энергий (47), (49), (50) и (51) акустического потока жидкости являются уравнениями работ, то есть с помощью акустического вида движения работа передаётся на расстояние.

Отметим, несколько забегая вперёд, что с помощью акустических зависимостей записывается не только непосредственно акустическое движение жидкости, но и ударные волны, волны, возникающие на воде при возмущении её поверхности, и так далее.

***